\documentclass[oneside]{aa}

\usepackage[varg]{txfonts}
\usepackage{natbib}
\usepackage{graphicx}
\usepackage{subcaption}
\usepackage{wasysym} 
\usepackage{rotating}
\usepackage{multirow} 
\usepackage{xcolor} 
\usepackage{amsmath}
\usepackage{siunitx}
\usepackage{ulem} 
\usepackage{blindtext}
\usepackage{color}
\usepackage[title,toc]{appendix}
\usepackage{threeparttable}
\usepackage{amssymb}
\usepackage{longtable}
\graphicspath{{Figure/}}
\usepackage{booktabs}
\usepackage{pdflscape}
\usepackage{placeins}

\DeclareMathAlphabet{\mathsc}{OT1}{cmr}{m}{sc}
\def\testbx{bx}
\DeclareRobustCommand{\ion}[2]{
\relax\ifmmode
\ifx\testbx\f@series
{\mathbf{#1\,\mathsc{#2}}}\else
{\it{#1\,\mathsc{#2}}}\fi
\else\textup{#1\,{\mdseries\textsc{#2}}}
\fi}

\begin{document}

\title{Supernovae at Distances $<$ 40 Mpc: I.}
\subtitle{Catalogues and fractions of Supernovae in a Complete Sample}

\author{
Xiaoran Ma \inst{\ref{inst1} 
}
\and Xiaofeng Wang \inst{\ref{inst1},\ref{inst2} \thanks{E-mail: wang\_xf@mail.tsinghua.edu.cn}}
\and Jun Mo \inst{\ref{inst1}}
\and D. Andrew Howell \inst{\ref{inst3},\ref{inst4}}
\and Craig Pellegrino \inst{\ref{inst3},\ref{inst4}}
\and Jujia Zhang \inst{\ref{inst5},\ref{inst6}}
\and Shengyu Yan \inst{\ref{inst1}}
\and Iair Arcavi \inst{\ref{inst7}}
\and Zhihao Chen \inst{\ref{inst1}}
\and Joseph Farah \inst{\ref{inst3},\ref{inst4}}
\and Estefania Padilla Gonzalez \inst{\ref{inst3},\ref{inst4}}
\and Fangzhou Guo \inst{\ref{inst1}}
\and Daichi Hiramatsu \inst{\ref{inst8},\ref{inst9}}
\and Gaici Li \inst{\ref{inst1}}
\and Han Lin \inst{\ref{inst5},\ref{inst6}}
\and Jialian Liu \inst{\ref{inst1}}
\and Curtis McCully \inst{\ref{inst3},\ref{inst4}}
\and Megan Newsome \inst{\ref{inst3},\ref{inst4}}
\and Hanna Sai \inst{\ref{inst1}}
\and Giacomo Terreran \inst{\ref{inst3},\ref{inst4}}
\and Danfeng Xiang \inst{\ref{inst1}}
\and Xinhan Zhang \inst{\ref{inst10}}
\and Tianmeng Zhang \inst{\ref{inst11},\ref{inst12}}
}

\institute{
Department of Physics, Tsinghua University, Beijing 100084, China \label{inst1}
\and Purple Mountain Observatory, Chinese Academy of Sciences, Nanjing 210023, China  \label{inst2}
\and Las Cumbres Observatory, 6740 Cortona Drive Suite 102, Goleta, CA 93117-5575, USA \label{inst3}
\and Department of Physics, University of California, Santa Barbara, CA 93106-9530, USA \label{inst4}
\and Yunnan Observatories, Chinese Academy of Sciences, Kunming 650216, China \label{inst5}
\and International Centre of Supernovae, Yunnan Key Laboratory, Kunming 650216, P. R. China \label{inst6}
\and School of Physics and Astronomy, Tel Aviv University, Tel Aviv 69978, Israel \label{inst7}
\and Center for Astrophysics \textbar{} Harvard \& Smithsonian, 60 Garden Street, Cambridge, MA 02138-1516, USA \label{inst8}
\and The NSF AI Institute for Artificial Intelligence and Fundamental Interactions, USA \label{inst9}
\and School of Physics and Information Engineering, Jiangsu Second Normal University, Nanjing 211200, China \label{inst10}
\and Key Laboratory of Space Astronomy and Technology, National Astronomical Observatories, Chinese Academy of Sciences, 20A Datun Road, Beijing 100101, China \label{inst11}
\and School of Astronomy and Space Science, University of Chinese Academy of Sciences, Beijing 101408, China \label{inst12}
}
 
\date{Received Month XX, 2022 / Accepted Month XX, 2022}

\abstract{This is the first paper of a series aiming to determine the fractions and birth rates of various types of supernovae (SNe) in the local Universe.}{In this paper, we aim to construct a complete sample of SNe in the nearby universe and provide more precise measurement of subtype fractions.}{We carefully selected our SN sample at a distance of $<$ 40 Mpc mainly from wide-field surveys conducted over the years from 2016 to 2023.}{The sample contains a total of 211 SNe, including 109 SNe II, 69 SNe Ia, and 33 SNe Ibc. With the aid of sufficient spectra, we can obtain relatively accurate subtype classifications for all SNe in this sample. After corrections for the Malmquist bias, this volume-limited sample gives fractions of SNe Ia, SNe Ibc, and SNe II as $30.4^{+3.7}_{-11.5}\%$, $16.3^{+3.7}_{-7.4}\%$, and $53.3^{+9.5}_{-18.7}\%$, respectively.
In the SN Ia sample, the fraction of the 91T-like subtype becomes relatively low ($\sim$5.4\%), while that of the 02cx-like subtype shows a moderate increase ($\sim$6.8\%). In the SN Ibc sample, we find significant fractions of broadlined SNe Ic ($\sim$18.0\%) and SNe Ibn ($\sim$8.8\%). The fraction of 87A-like subtype is determined as $\sim$2.3\% for the first time, indicating rare explosions from blue supergiant stars. We find that SNe Ia show a double peak number distribution in S0- and Sc-type host galaxies, which may serve as a straightforward evidence for the presence of "prompt" and "delayed" progenitor components giving rise to SN Ia explosions. Several subtypes of SNe such as 02cx-like SNe Ia, broadlined SNe Ic, SNe IIn (and perhaps SNe Ibn) are found to occur preferentially in less massive spiral galaxies (i.e., with stellar mass $<$ 0.5$\times$10$^{10}$ M$_{\odot}$), favoring their associations with young stellar progenitors. Moreover, the 02cx-like subtype shows a trend of exploding in the outer skirt of their hosts, suggestive of metal-poor progenitors.} {}
\authorrunning{Ma et al.} 
\titlerunning{A Nearby Supernova Sample}
\maketitle

\section{Introduction}

Supernovae (SNe) represent catastrophic explosions of certain types of stars evolving towards their final stages, which are one of the most spectacular phenomena in the universe. Based on spectroscopic features such as hydrogen, helium, and silicon, SNe can be observationally divided into four subtypes of Ia, Ib, Ic, and II \citep{Filippenko1997}. Balmer lines are usually absent in the former three types, while they are main spectral features of SNe II. Among the H-poor subtypes, SNe Ib are characterized by helium absorption features, while such features are generally invisible in both SNe Ia and SNe Ic. Although the latter two subtypes both show Si II $\lambda$ 6355 absorptions in their spectra, SNe Ia show relatively bluer continuum and "w-shaped" S II doublet near 5400-\SI{5600}{\angstrom} compared to SNe Ic. 
Theoretically, SNe Ia are believed to arise from thermonuclear explosion of carbon-oxygen white dwarves, while SNe Ib, Ic, and II (dubbed as CCSNe) are produced due to gravitational collapse of central iron core in massive stars. By studying the observational properties of various kinds of SNe, we can probe the extreme physics process in the explosion, gain new insights into stellar evolution, and understand the chemical and dynamic evolution of the universe.  

SNe Ia play an important role in observational cosmology, measuring the expansion history of the universe, due to their high and relatively uniform luminosity around the peak \citep{Riess1998,Perlmutter1999,Filippenko2005}. On the other hand, SNe Ia have been found to show increasing diversities over the past two decades \citep{Filippenko1997,Maoz2014}, such as the discoveries of luminous 91T-like \citep{Filippenko1992a,Phillips1992} and overluminous 03fg-like \citep{Howell2006,Hsiao2020} subclasses, subluminous 91bg-like \citep{Filippenko1992b,Leibundgut1993}, 02cx-like \citep{Li2003,Foley2013}, and 02es-like SNe Ia \citep{Ganeshalingam2012,Xi2024}. In addition to those spectroscopically peculiar SNe Ia, different subtypes have been identified even for spectroscopically normal SNe Ia \citep{Branch1993,Benetti2005,Branch2009,Wang2009}. In particular, those SNe Ia characteristics of high photospheric velocities (HV) are recognized to have different local environments \citep{Wang2013,Wang2019,Pan2020}, with a higher progenitor metallicity compared to their normal-velocity (NV) counterparts. 
A reliable constraint of the SN Ia rate of different subtypes as well as its evolution with redshift would allow us to better understand their progenitor scenarios, the single-degenerate channel \citep[SD,][]{Whelan1973, Nomoto1982} and double-degenerate channel \citep[DD,][]{Tutukov1981, Iben1984, Webbink1984}. For the SD scenario, the system consists of a CO WD and a main sequence, sub-giant, or giant star companion, while for the DD scenario, the companion star is another WD. 

Among all types of SN explosions, SNe II represent the most common stellar explosions in the local universe \citep{Li2011}, which can be further classified into subtypes of IIP, IIL, IIb according to their light-curve shapes and spectral features \citep{Filippenko1997}. Pre-explosion images have established their origins from yellow (IIb) or red (IIP/IIL) supergiants \citep{Smartt2009}. A group of peculiar 1987A-like type II SNe is believed to originate from blue supergiants. For SN Ibc, theoretical studies suggest that their progenitors should be massive Wolf-Rayet (WR) stars or of binary origin with relatively low masses \citep[i.e.,][]{Yoon2015}. Possible WR progenitors of SNe Ibc are rarely detected, with only two detections for SN Ib, that is, SN 2013bvn \citep{Cao2013} and SN 2019yvr \citep{Kilpatrick2021}, and one possible detection for SN Ic \citep{Van2018, Xiang2019}. In recent years, numerous subtypes have also been identified for CCSNe. 
For example, the appearances of narrow emission features such as hydrogen, helium and carbon in the spectra lead to identifications of subtypes of SNe IIn \citep{Schlegel1990}, SNe Ibn \citep{Matheson2000,Pastorello2008, Hosseinzadeh2017}, and even SNe Icn \citep{Gal-Yam2022,Perley2022}, respectively.  
Over the past decade, wide-field surveys have led to the discovery of numerous superluminous supernovae \citep[SLSNe,][]{Quimby2007,Gal-Yam2012, Chen2023a}, which are 10 to 100 times luminous than normal CCSNe but are rarely detected in the local universe, likely due to their very massive, metal-poor progenitor properties. 

In recent years, with the development of wide-field rolling search programs such as the Zwicky Transient Facility \citep[ZTF,][]{Masci2019, Bellm2019}, the Asteroid Terrestrial-impact Last Alert System \citep[ATLAS,][]{Tonry2018, Smith2020}, and the All-Sky Automated Survey for Supernovae \citep[ASAS-SN,][]{Shappee2014, Kochanek2017}, the completeness for the discovery and classification of nearby SNe has improved significantly. This would allow us to construct a more improved supernova sample and derive birth rates of different types of SNe more precisely.   
	
In this series of papers, we report our construction of a nearby SN sample and determinations of subtype fractions and rates from this sample. 
This is paper I of a series of papers and is organized as follows.
In Sect.~\ref{sample}, we introduce this relatively complete nearby SN sample. 
We then calculate the subtype fractions of SNe in Sect.~\ref{fraction}. We describe the host galaxies of our SN sample and examine the SNe distribution in galaxies of different Hubble types in Sect.~\ref{host}. We summarize in Sect.~\ref{sum}.  
	
\section{The supernova sample} \label{sample}

In this section, we describe the procedures for selecting our SN sample and the ways of collecting their main parameters, including photometric properties, spectroscopic classification, extinction, and the application of a distance cut. Throughout our study, we adopt H$_0$ = 70 km s$^{-1}$ Mpc$^{-1}$, $\Omega_m = 0.3$, and $\Omega_\Lambda = 0.7$, and the magnitudes are in the AB system. 

\subsection{Sample selection and classification} \label{classification}
 
We collected our initial SN sample from WISeREP \footnote{\url{https://www.wiserep.org/}} (formerly known as WISEASS, the Weizmann Institute of Science Experimental Astrophysics Spectroscopy System) and the Transient Name Server \footnote{\url{https://www.wis-tns.org/}} (TNS) with the following two criteria: 1) exploded during the period from 2016 to 2023; 2) at a distance $<$ 60 Mpc \footnote{The reason why we first chose 60 Mpc is that the luminosity function sample (the volume-limited sample which was later corrected to 100\% completeness) constructed by \citep{Li2011} has a cutoff distance of 60 Mpc. As we aim to construct a relatively complete nearby SN sample, the cutoff distance should be: 1) not too large so that the collected SN sample is not severely biased; 2) not too small so that the sample size is large enough to avoid severe statistical uncertainty.}.
The first criterion is adopted because several main wide-field surveys such as ZTF, ATLAS, and ASAS-SN have been regularly scanning the sky since 2015. Given their detection limits (see Table~\ref{surveys} for details) and the brightness of nearby SNe, the sample collected after 2015 should suffer less effect due to Malmquist bias. The basic parameters, including SN name, discoverer, discovery date, right ascension and declination for all supernovae, were collected from both WISeREP and TNS. Classification of each SN given by these two catalogs is adopted as a reference before obtaining a more precise classification. Redshift, distance, and host information were obtained from two large galaxy databases, NASA/IPAC Extragalactic Database \footnote{\url{https://ned.ipac.caltech.edu/}} (NED) and Set of Identifications, Measurements and Bibliography for Astronomical Data  \footnote{\url{http://simbad.u-strasbg.fr/Simbad}} (SIMBAD).

Distances to our SNe Ia were taken from the NED in the following two manners: 1) Tully-fisher distance whenever possible, but normalized to the assumed Hubble constant (i.e., 70 km s$^{-1}$ Mpc$^{-1}$); 2) Hubble flow distance when the former distance is not available, and it is corrected for the infall of the local group toward the Virgo Cluster \citep{Mould2000}.

The spectra used for the SN classifications were mainly taken from WISeREP, TNS, the Global Supernova Project \citep[GSP,][]{Howell2019} using 2~m telescopes of Las Cumbres Observatory \citep[LCO,][]{Shporer2011,Brown2013}, the Tsinghua Supernova Group using the Xinglong 2.16~m telescope \citep[XLT,][]{Fan2016} of NAOC, and the Lijiang 2.4~m telescope \citep[LJT,][]{Fan2015} of Yunnan Observatories. The numbers of spectra collected from different sources are listed in Table~\ref{sources}. The light curve data were obtained from the literature whenever possible (detailed references are given in Table~\ref{reference}). For those without published light curve data, we turned to the LCO light curves and forced photometry of ZTF \citep{Smith2019} and ATLAS \citep{Shingles2021}. Our classifications of the SN sample were mainly based on spectroscopic characteristics. 
We used the Supernova Identification Library \citep[SNID,][]{Blondin2007} to fit every spectrum to obtain a best fit subtype, and it was then double checked by eyes to make sure their classifications are reliable. When it was difficult to make a clear classification solely from the spectra, we examined the light curves for a certain classification, in particular for SNe IIP and SNe IIL which show quite similar spectral features. 

        \begin{table}
		\centering
		\caption{Number of spectra collected from different sources.}\label{sources}
		\begin{threeparttable}
			
			\begin{tabular}{ccccc}
				\toprule
				Type & WISeREP & LCO & XLT & LJT\\
				\midrule
				Ia &  25 & 28 & 13 & 2 \\
				Ibc &  15 & 11 & 6 & 2 \\
				II &  60 & 23 & 20 & 5 \\
                Total &  100 & 62 & 39 & 9 \\
				\bottomrule
			\end{tabular}
			
		\end{threeparttable}
	\end{table}

In our study, we classify supernovae into three main categories: Ia, Ibc, and II. Each main category is then classified into a few subtypes in the following manner. SNe Ia are grouped into 5 subtypes: 
	(1) normal Ia (HV and normal subtypes), 
	(2) 91bg-like,
	(3) 91T-like,
	(4) 02cx-like (Iax),
	(5) other peculiar Ia, such as Super-Chandra SNe Ia (Ia-SC), Ia-CSM, etc.
	SNe Ibc are grouped into 5 subtypes:
	(1) normal Ib,
	(2) Ibn,
	(3) normal Ic,
	(4) Ic-BL,
	(5) Ib/c, which means that we can only verify that they are of type Ibc, or a transition in between Ib and Ic. 
	SNe II are grouped into 6 subtypes:
	(1) IIP,
	(2) IIL,
	(3) IIb,
	(4) IIn,
	(5) peculiar II, such as 87A-like objects,
	(6) II, which are SNe II with an uncertain subclassification. The classifications of some peculiar SNe in our sample are highlighted below.

Among SNe Ia, SN 2021fcg is classified as an 02cx-like object according to \citet{Karambelkar2021} as we do not have any spectra for it, while SN 2022pul is classified as an SN Ia characteristic of superchandrasekhar explosion (Ia-SC), which showed a strong C II absorption feature in the early spectra and emission of [O I] in the nebular ones \citep{Kwok2024, Siebert2024}.

SNe 2021foa, SN 2022qhy, and SN 2023fyq can be classified as SNe Ibn. Among them, the classification of SN 2021foa is somewhat complicated. \citet{Reguitti2022} presented spectroscopic data for SN 2021foa. They found that in the early phase, its spectra showed blue continuum superposed with narrow H emission lines, putting it into the subgroup of SNe IIn. However, around 2 weeks after the peak, its He I lines became much stronger than those of SNe Ibn. Thus, they argued that SN 2021foa may represent a transitional object linking H-rich SNe IIn with He-rich SNe Ibn\footnote{In our later statistics, we count this SN as one Ibn and one IIn when calculating the fractions of subtypes of SNe Ibc and II, when calculating the fractions of Ibc and II in the total SN sample in Sect.~\ref{fraction} and the SN rates in Paper II of the series, we count it as 0.5 Ibc and 0.5 II.}.
SN 2022qhy presents prominent He I $\lambda$ 5876 and 6678 in its spectrum, a SNID fit to the spectrum indicates that it is a typical type Ibn supernova.
SN 2023fyq is classified as a SN Ibn according to the prominent narrow P-Cygni helium lines in its early time spectra, and it is reported to show an exponential rise in flux prior to core collapse \citep{Brennan2024}. 

Among our sample, the fraction of the subtype of SNe Ib/c seems a bit high. SN 2020zgl and SN 2022frl can be classified as transitional objects between SNe Ib and SNe Ic, as both show close resemblances to SN 2005bf and SN 2009er \citep{Folatelli2006,Foley2009} according to the SNID fit. Thus, we assigned them the subtype of SNe Ib/c. For SN 2016dsl, SN 2017gax, and SN 2021bhd, the SNID fits to their late-time spectra give the best results of type Ib/c, so they are also grouped as SNe Ib/c.

The spectra of SNe 2016afa, 2018hna and 2020mmz show characteristics of SN 1987A according to SNID fitting, we further checked their light curve shapes to confirm the classification. In particular, for SN 2018hna, \citet{Singh2019} and \citet{Xiang2023} identified evident features of Ba II (i.e., 4554 and \SI{6142}{\angstrom}).
	
\subsection{Light curve fitting}

Apart from the light-curve shapes that aid in the subclassification, we also count on peak magnitudes for later bias corrections as introduced in Sect.~\ref{biascor}. The photometric data were not always well sampled to allow a direct measurement of this parameter. To determine the r-band peak magnitudes for those without photometric results from the literature, we fitted the light curves collected from the sources introduced in Sect.~\ref{classification} in the following manner. We performed a spline fit to directly measure the peak magnitudes when near-maximum light photometric data were available and well sampled. However, when the photometric data were available only in pre- or post-maximum, we used a $\chi^2$-minimizing method to fit the photometric data with the light curve templates from \citet{Li2011}. For SNe IIP, we adopted the maximum brightness measured during the plateau phase.
	
We then converted the apparent peak magnitudes to the absolute magnitudes using distances collected in Sect.~\ref{classification} and corrected for extinctions estimated in Sect.~\ref{extinction}. The final results for the absolute peak magnitudes in the r band are listed in Tables~\ref{Ia_info},~\ref{Ibc_info}, and ~\ref{II_info}.
	
\subsection{Extinction} \label{extinction}
	
Extinction values are important for estimates of peak magnitudes. Galactic reddening towards each SN in our sample was estimated according to \citet{Schlafly2011}. The host-galaxy extinction value is mainly taken from the literature of each individual SN, with the extinction value given in Tables~\ref{Ia_info},~\ref{Ibc_info},~\ref{II_info}, and the corresponding references given in Table C.1. The rest were estimated using the empirical relation $\mathrm{E(B-V) = 0.16 \times EW(Na \: I \: D)}$ \citep{Turatto2003} when the absorptions of Na ID are visible in the spectra. Note that $\mathrm{EW(Na \: I \: D)}$ represents the equivalent width of Na I D $\lambda \lambda$ 5890,5896 absorption lines. 

\subsection{Distance cut} \label{distance_cut}
	
We aimed to construct a relatively complete sample of nearby SNe to obtain a more accurate estimate of subtype fractions and rates. The detection of SNe at larger distances is expected to be more underestimated because they become fainter with distances and are then likely to be beyond the detection limit. Thus, a proper distance cut is necessary so that the resultant sample size is not too small and the selected SN sample is close to being complete within this volume. Such a selected SN sample could be regarded as an approximate representation of the true supernova distribution after a slight but careful correction. 
	
	\begin{figure*}
		\includegraphics[width=\textwidth]{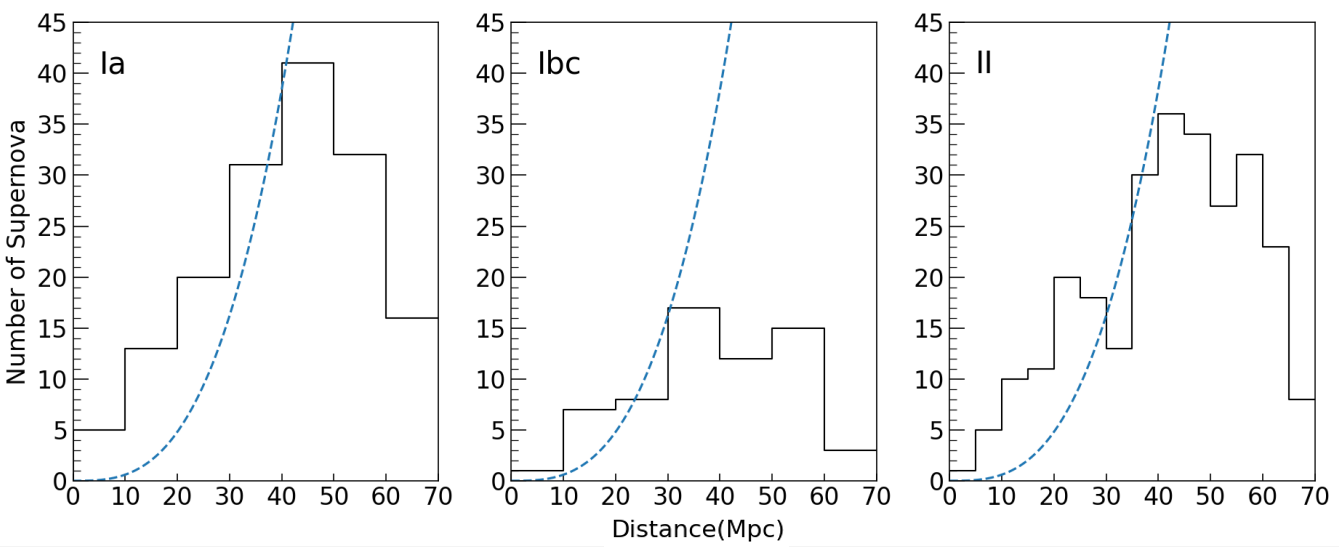}
		\caption{The number distribution of our SN sample as a function of distance for Ia (left), Ibc (middle), and II (right) samples, respectively. We adjust the distance bin from 10 Mpc to 5 Mpc for SNe II because its sample size at around 40 Mpc is rather large for this type. The dashed lines represent the expected number distribution of SNe $\propto$ d$^3$.}\label{distance}
	\end{figure*}
	
Fig.~\ref{distance} shows the number of SNe as a function of distance for our 
SN sample. The dashed lines represent the number distribution $\propto$ d$^3$ assuming a constant SNe rate in the nearby universe. The numbers of all three SN types show a similar distribution, with an initial increase with distance until around 40-50 Mpc, followed by a significant decline. The SN population is expected not to change much within a distance of 60 Mpc, so the number of SNe detected beyond the turnover point, 40 Mpc, is significantly smaller than it should be according to the dashed lines in Fig.~\ref{distance}. This suggests that a severe observation bias exists beyond 40 Mpc.

    \begin{figure}
		\includegraphics[width=\columnwidth]{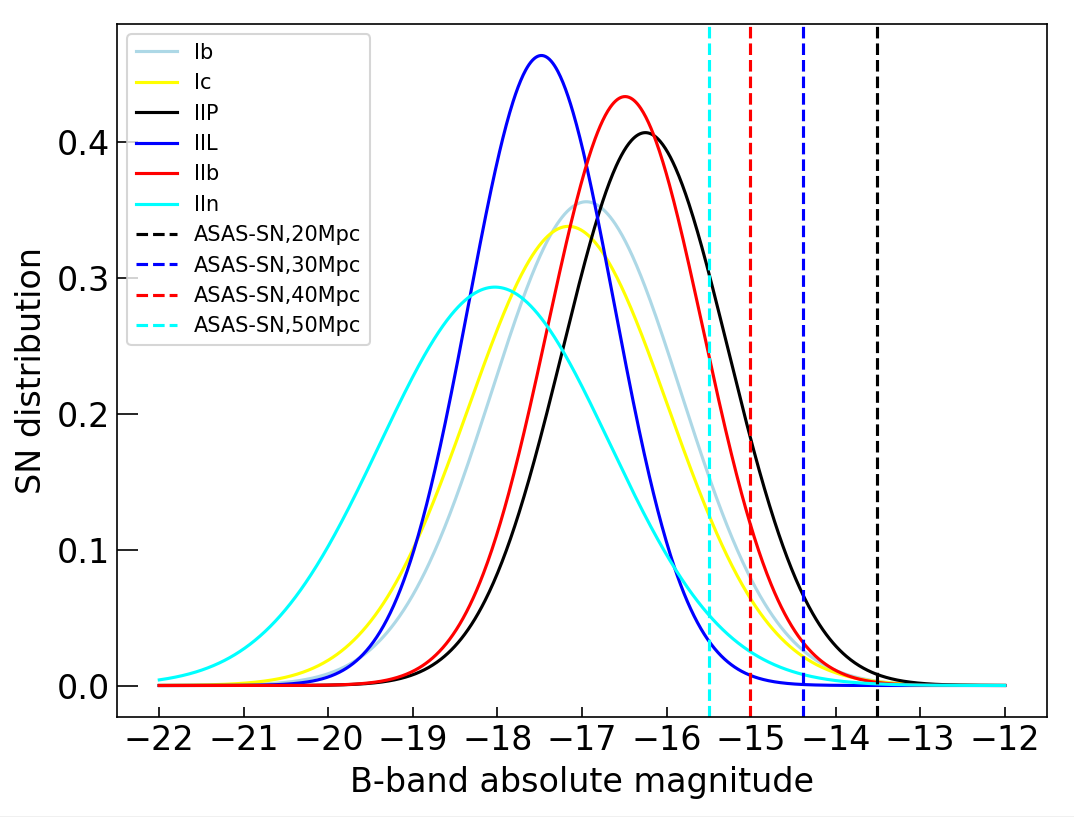}
		\caption{Peak absolute magnitude distributions for different types of CCSNe. The dashed lines represent the corresponding detection limits of ASAS-SN at different distances.}\label{distance_choice}
    \end{figure}

Nevertheless, as shown in Fig.~\ref{distance}, the detection sensitivity of some surveys for CCSNe might drop at distances $<$ 40Mpc. To examine such an effect on CCSNe sample, we present the distribution of B-band peak absolute magnitudes \citep{Richardson2014} for different types of CCSNe in Fig.~\ref{distance_choice}. At a distance of 40 Mpc, one can see that the missing fraction is only about 10\% for SNe IIP and SNe IIb, and even smaller for other subtypes of CCSNe. While this missing fraction increases rapidly at a distance $\sim$ 50 Mpc. Thus, the choice of 40 Mpc distance limit should be reasonable for the purpose of SN rate study.

\begin{figure} [h!]
		  \includegraphics[width=\columnwidth]{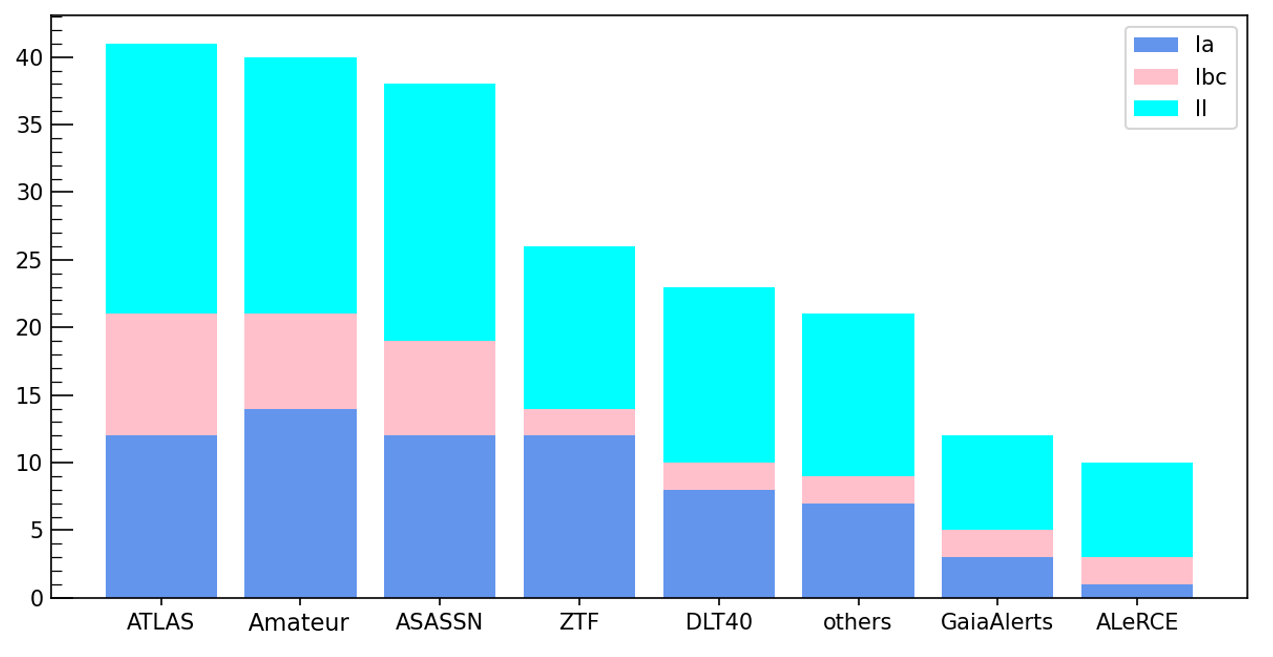}
		  \caption{Number of SNe discovered by different surveys in our sample. SNe Ia, Ibc and II are represented by blue, pink, and cyan colors, respectively. Apart from amateurs, most SNe were discovered by ASAS-SN, ATLAS, ZTF, and DLT40 project.}\label{discoverer}
		\end{figure}

After applying a 40-Mpc distance cut, our final sample contains 211 SNe in total, and the locations of all these SNe in the sky are shown in Fig.~\ref{sky}(a). The average distances for the total SN sample, the subsamples of SNe Ia, SNe Ibc, and SNe II are 26.84 Mpc, 25.95 Mpc, 28.17 Mpc, and 26.25 Mpc, respectively. As shown in Fig.~\ref{discoverer}, most of this sample is discovered by several wide-field surveys such as ATLAS, ASAS-SN, and ZTF and those dedicated to searching for nearby SNe such as DLT40 \citep[Distance Less Than 40 Mpc,][]{Tartaglia2018}. Nevertheless, the contribution by amateur astronomers like K. Itagaki is still an indispensable part, accounting for about 11\% of the total sample. The detection limits of the main telescopes used for the SN search are all deeper than 17.0 mag, and are summarized in Table~\ref{surveys}.

    \begin{table}[h!]
		\centering
		\caption{Detection limits of main surveys.}\label{surveys}
		\begin{threeparttable}
		  \begin{tabular}{cccc}
				\toprule 
				Survey & Detection limits/mag & Band & Reference \\
				\midrule 
				ATLAS & $\sim$19.5 & o & (6),(8)\\
				ASASSN & $\sim$17-18 & V & (3),(5)\\
				ZTF & $\sim$20.8 & g & (1),(4)\\
                DLT40 & $\sim$19-20 & r & (7)\\
                GaiaAlerts & 20.7 & G & (2) \\
                K. Itagaki & $\sim$17.8 &  &    \\
                \bottomrule
          
            \end{tabular}

		\end{threeparttable}

        \tablebib{(1) \citet{Bellm2019}; (2) \citet{Hodgkin2021}; (3) \citet{Kochanek2017}; (4) \citet{Masci2019}; (5) \citet{Shappee2014}; (6) \citet{Smith2020}; (7) \citet{Tartaglia2018}; (8) \citet{Tonry2018}.}
        
    \end{table} 
 
\subsection{Type Ia supernovae sample}
	
Our final SN Ia sample consists of 69 objects, with detailed information about this sample listed in Table~\ref{Ia_info}. For each SN in our sample, the following parameters are given in Table~\ref{Ia_info}: the SN name, subtype classification, discoverer, discovery date, spectrum ID in WISeREP of the spectrum used for classification, the r-band peak magnitude after extinction correction, the Galactic and host extinction values, the right ascension and declination, redshift, distance, host galaxy, and offset from host galaxy nucleus (in units of arcsec). Only the first 6 entries are shown; full table would be available online. 

\begin{figure}
  \centering
  \subcaptionbox{All\label{fig:All}}
    {\includegraphics[width=\columnwidth]{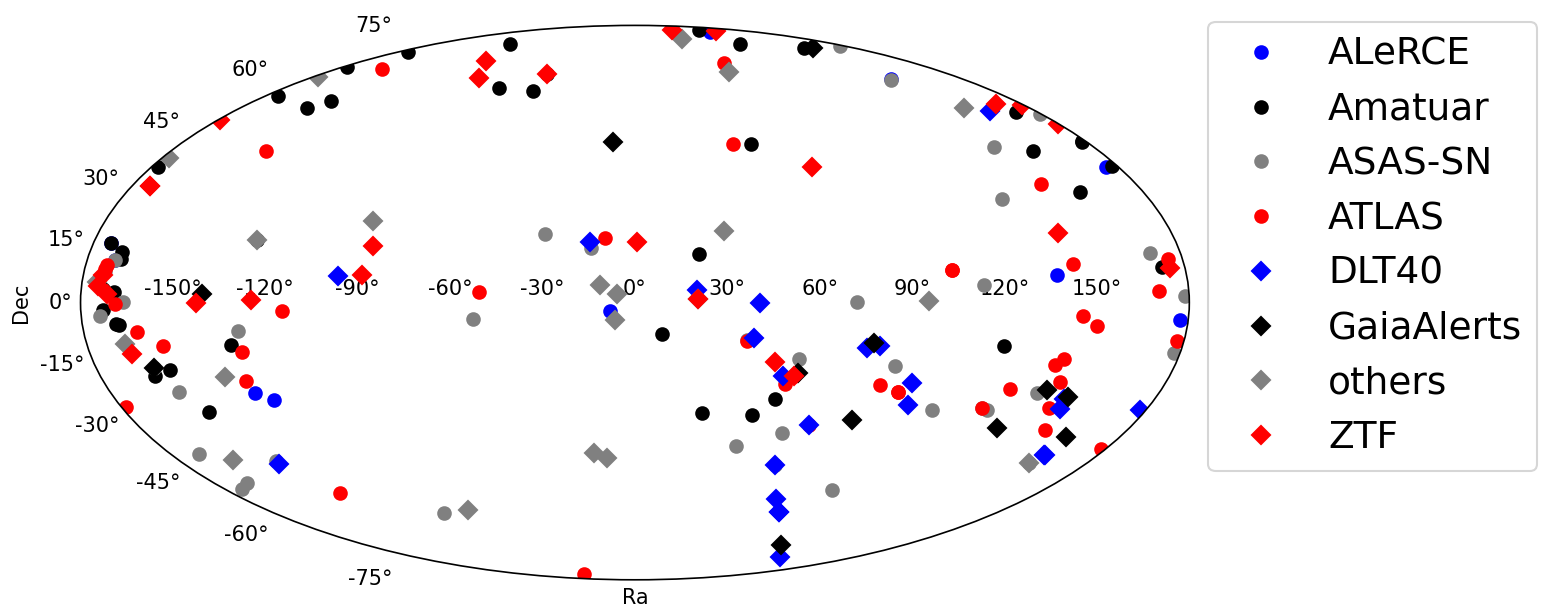}}
  \subcaptionbox{Ia\label{fig:Ia}}
    {\includegraphics[width=\columnwidth]{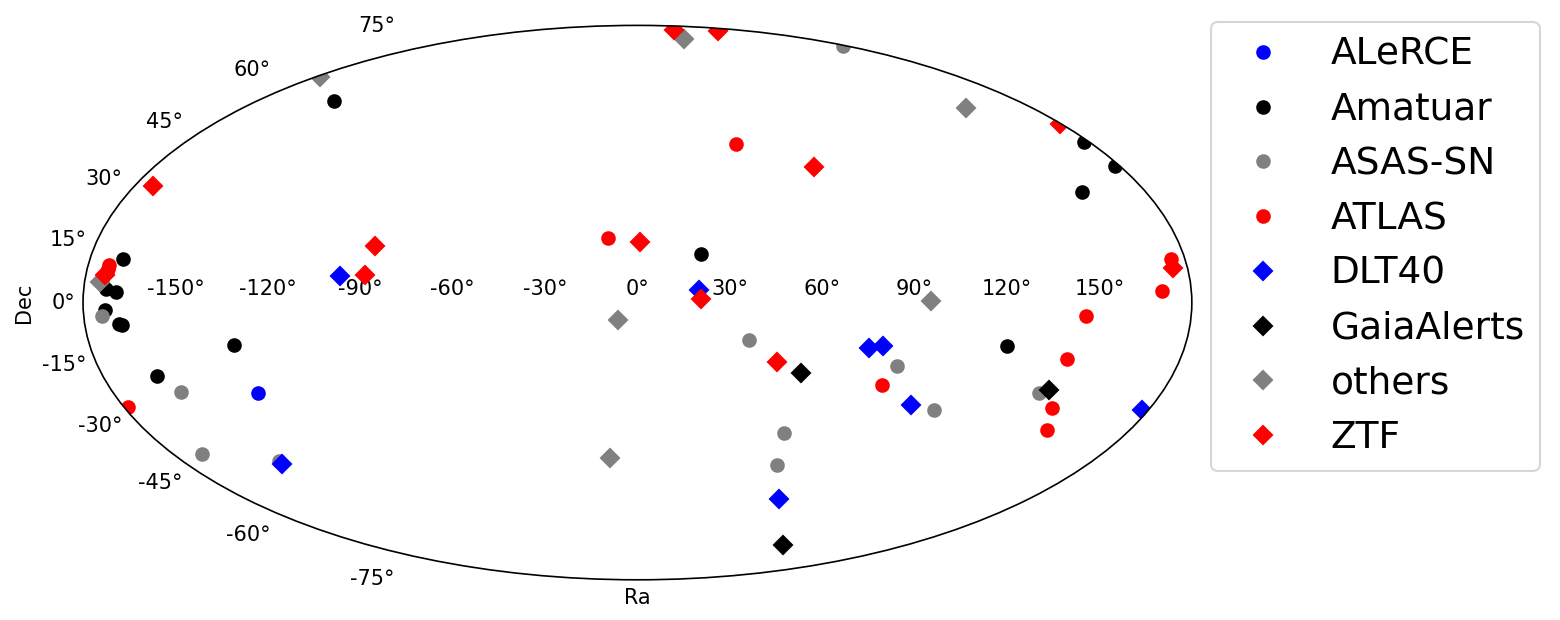}}
  \subcaptionbox{Ibc\label{fig:Ibc}}
    {\includegraphics[width=\columnwidth]{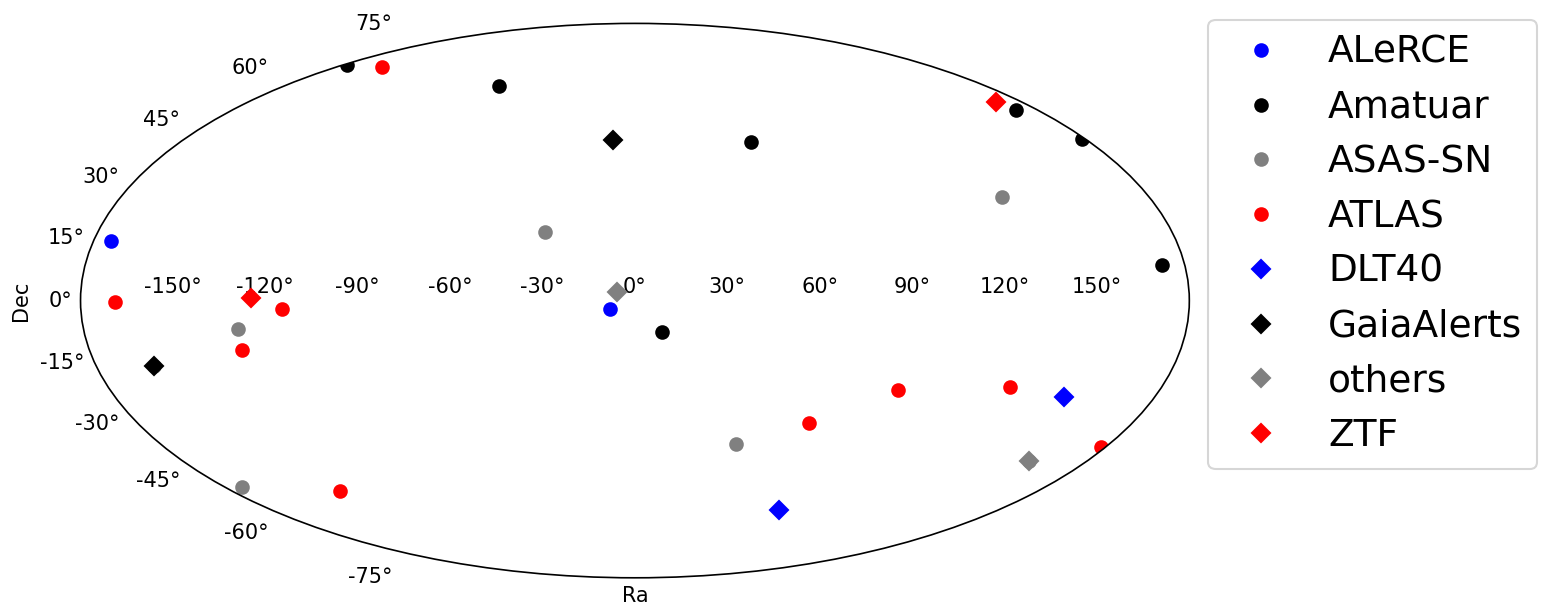}}
    \subcaptionbox{II\label{fig:II}}
    {\includegraphics[width=\columnwidth]{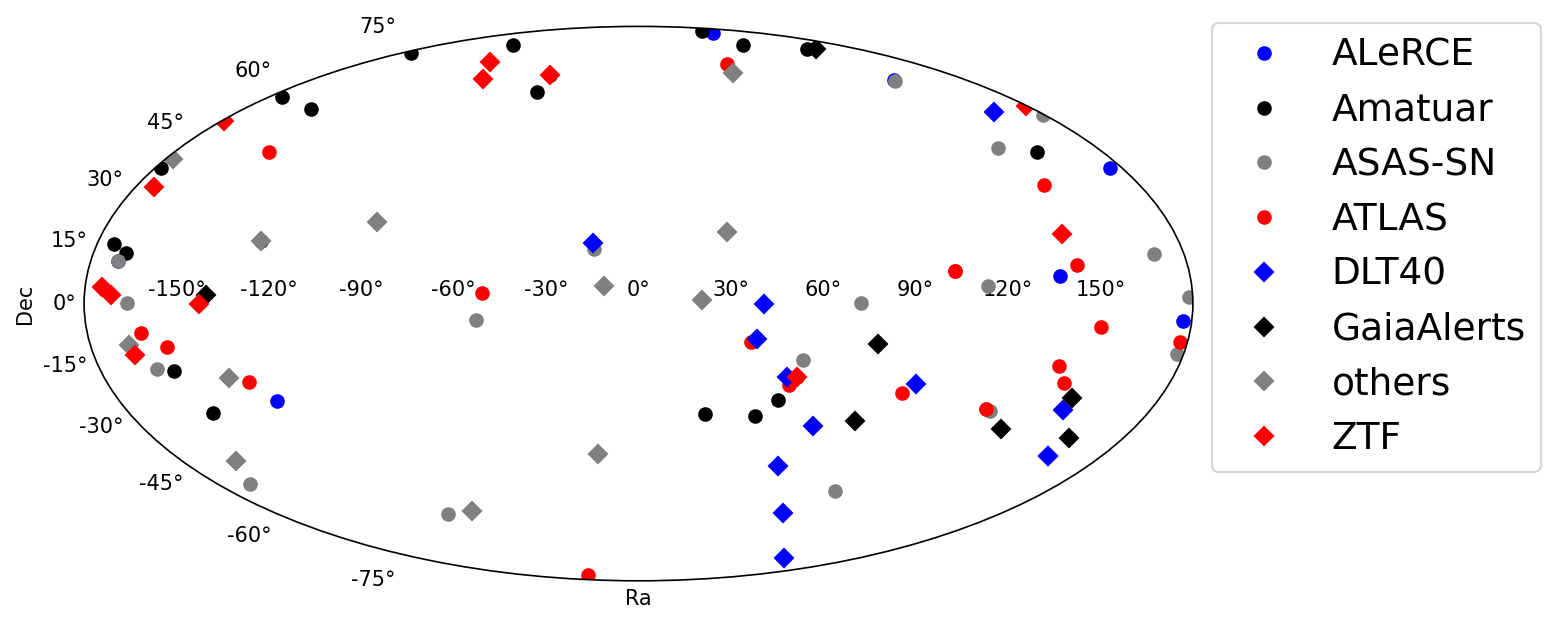}}
  \caption{Sky location of our SN sample.}
  \label{sky}
\end{figure}
    
The 69 SNe Ia in our sample include 49 normal, 10 91bg-like, 4 91T-like, 5 02cx-like, and 1 Ia-SC ones. The observed fractions of the SN Ia subtypes are shown in the second panel of Fig.~\ref{obs_frac}. The locations of these SNe Ia and the corresponding discoverers are shown in Fig.~\ref{fig:Ia}.

We collected a total of 68 spectra for our SN Ia sample, with 1 for each SN Ia except for SN 2021fcg mentioned in Sect.~\ref{classification}. Of these 68 spectra, 53 were obtained within one week of the maximum. Pre-maximum spectra are available for 60 SNe Ia.
Part of the SN Ia spectra are shown in Fig.~\ref{Iaspectra1}.
Peak magnitudes have been estimated for 66 of the 69 SNe Ia, with 25 taken from the literature (details seen in Table~\ref{reference}) and the rest estimated from the LCO photometric database (25), the ZTF database (11), and the ATLAS forced photometry (5), respectively.

\subsection{Core-collapse supernovae sample}
	
\begin{figure*}[t]
		\includegraphics[width=\textwidth]{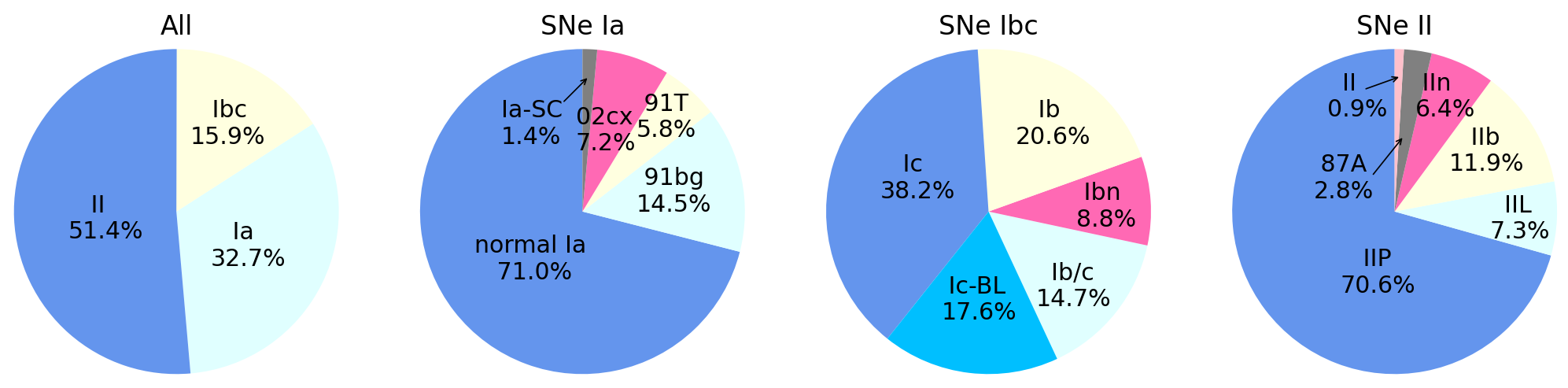}
		\caption{The observed fractions of supernova sub-types.}\label{obs_frac}
	\end{figure*}

Our final core-collapse SN sample contains 142 CCSNe, including 34 SNe Ibc and 109 SNe II. We note that SN 2021foa was counted for both SNe Ibc and SNe II.	
Detailed parameters for our SN Ibc and II samples are listed in Tables~\ref{Ibc_info} and ~\ref{II_info}, respectively, with the full tables available online. Among the 34 SNe Ibc, 6 are normal Ib, 3 are Ibn, 13 are normal Ic, 7 are Ic-BL, and 5 are classified as Ib/c which cannot be clearly assigned to type Ib or Ic. Among the 109 SNe II, 77 are IIP, 8 are IIL, 13 are IIb, 7 are IIn, 3 are 87A-like, and 1 can only be classified as type II. The observed fractions of the SNe Ibc and II subtypes are shown in the third and fourth panels of Fig.~\ref{obs_frac}. The locations of our sample of SNe Ibc and SNe II are shown in Figs.~\ref{fig:Ibc} and ~\ref{fig:II}, respectively.
	
A total of 34 spectra were collected for 34 SNe Ibc. Of these spectra, 26 were taken near the maximum light according to the photometric data or the SNID fit. 
Peak magnitudes are estimated for 31 of our 34 SNe Ibc, with 2 taken from the literature (see Table~\ref{reference} for the references), 17 estimated from LCO data, 3 from ZTF data, and 9 from ATLAS forced photometry.

A total of 109 spectra were collected for 109 SNe II, among which 102 were obtained near the maximum light or during the plateau phase. 
Photometric information is derived for 95 of the 109 SNe II, with 11 taken from the literature (see Table~\ref{reference} for the references), 44 estimated from LCO data, 16 from ZTF data, and 24 from ATLAS forced photometry.

\section{Fractions of supernova subbypes}\label{fraction}
	
In this section we describe the fractions of supernova sub-types estimated from our sample. We present Miller diagrams of our sample and introduce the way to correct for the bias seen in the diagram.  
    
\subsection{Observed factions and the Malmquist bias}
	
The observed fractions of different subtypes are given in Fig.~\ref{obs_frac} and the second column of Table~\ref{obs_table}. SNe II are the most abundant type (51.4\%) in our sample, while SNe Ia and SNe Ibc take up 32.7\% and 15.9\%, respectively. No SLSNe were found in our sample, indicating their rarity in local universe. The pie chart of SNe Ia shows that most of them belong to normal ones (71.0\%) and other commonly seen subtypes include 91bg-like (14.5\%), 91T-like (5.8\%), 02cx-like (7.2\%), and Ia-Sc (1.4\%). No 02es-like events are found in current sample, implying an intrinsically low rate for this peculiar low-luminosity SNe Ia. The pie chart of SNe Ibc indicates that normal Ic (38.2\%) and Ib (20.6\%) are two most abundant subtypes, with significant fractions of SNe Ic-BL (17.7\%), SNe Ibn (8.8\%), and SNe Ib/c (14.7\%), respectively. Among SNe II, the main subtype is SN IIP which takes up to 70.6\%, and the remaining subtypes include SN IIL (7.3\%), SNe IIb (11.9\%), SNe IIn (6.4\%), and 87A-like (2.8\%), respectively. There is still a small fraction (0.9\%) of SNe II that cannot be accurately assigned to any of the subtypes. 

\begin{figure}
		\includegraphics[width=\columnwidth]{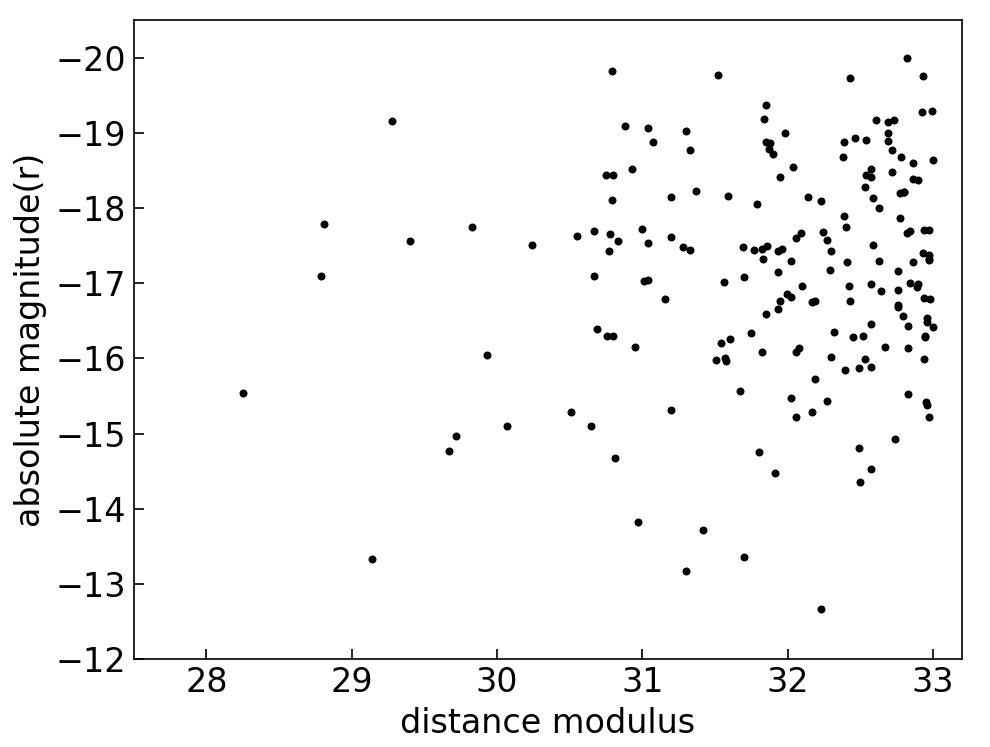}
		\caption{The absolute r-band peak magnitudes of all SNe. }\label{mag}
	\end{figure}

The observed fractions derived from the current sample may still suffer from some selection bias, even though our sample is close to being complete. To investigate this issue, we show the absolute peak magnitudes versus the corresponding distance modulus in Fig.~\ref{mag} (dubbed the "Miller diagram"). As can be seen, fainter SNe tend to be absent at larger distances, suggesting the presence of a selection effect in our sample. This selection effect is also called the Malmquist bias \citep{Teerikorpi1984,Giraud1987}. 

\subsection{Bias correction}\label{biascor}

\begin{figure}
		\includegraphics[width=\columnwidth]{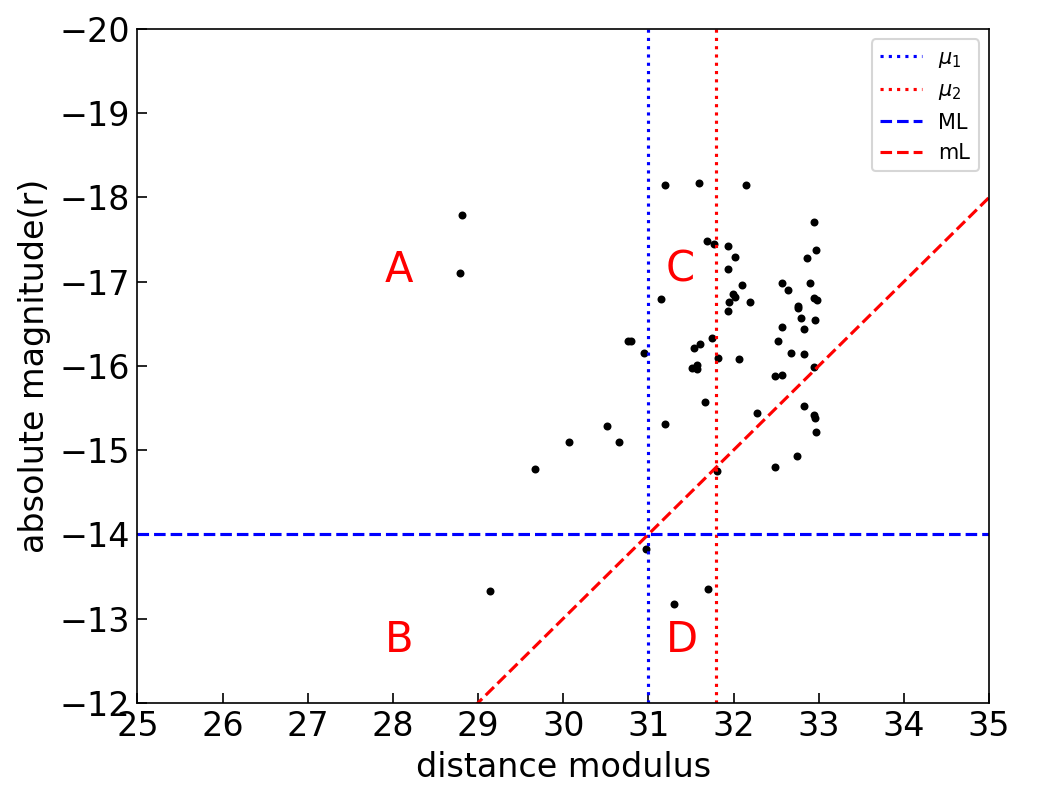} 
		\caption{Miller diagram for our SN IIP sample. The diagonal line is the apparent magnitude completeness limit (mL), while the horizontal line is the absolute magnitude completeness limit (ML). The two vertical lines at different distance modulus separate the volume within $\mu_2$ into a sphere and a concentric spherical shell. A, B, C and D represent the four regions of the volume within $\mu_2$ separated by the ML and the $\mu = \mu_1$ lines.}\label{bias}
	\end{figure}

To correct for the Malmquist bias, we followed the method adopted by \citet{Richardson2014}. The Miller diagram of our SN IIP sample is presented in Fig.~\ref{bias}, where a Malmquist bias can be clearly seen. Before performing the bias correction, we first make two reasonable assumptions: (1) The completeness limit, the sample brighter than the limit magnitude can be considered as a complete representation of the true SN distribution. Given an absolute magnitude limit M (dubbed as ML) for SNe of a certain type, the completeness limit of the apparent magnitude m (dubbed as mL) should follow a diagonal line with distance $\mu$ in the M $- ~\mu$ plane (see the solid line marked by mL in Fig.~\ref{bias}).  
From Figs.~\ref{mag} and ~\ref{bias} and detection limits of the main SN surveys involved in our study (presented in Table~\ref{surveys}), we choose the apparent-magnitude completeness limit to be about 17.0 mag for the SNe IIP sample. The sample with apparent magnitudes brighter than this limit (located above the mL line) is considered to be complete, while those fainter than the limit line are considered incomplete. 
(2) The real distribution of absolute magnitudes or luminosity function is assumed to be roughly the same for any relatively large volume of space in the local universe. If all volumes have roughly the same luminosity distribution, then the number ratio between dim and bright SNe should be the same in different volumes for a certain brightness limit.
	
As shown in Fig.~\ref{bias}, volumes A and B extend to some distance modulus $\mu_1$, the absolute magnitude/luminosity completeness limit (ML) is determined by mL and the vertical line $\mu = \mu_1$. Volume A represents the region brighter than ML, while volume B represents the region fainter than ML. The samples within volume A and volume B have already been corrected and thus are thought to be complete. We then carefully choose another distance modulus $\mu_2$ so that the sample located in volume C (the region brighter than ML) lies above the apparent magnitude limit mL and could be considered complete, while the sample in region D is incomplete and needs to be corrected. In the iteration procedure, we choose the maximum value of $\mu_2$ that satisfies the above criteria.

We started the bias-correction procedures by first picking up the faintest SNe that are brighter than mL in our sample and then setting the absolute magnitude
limit, ML, above that point. Then $\mu_1$ and $\mu_2$ were determined in a manner similar to that discussed above. According to the assumptions we made, regions A, B, and C are considered complete, while region D is incomplete, and the true distribution of SNe in volume A/B should be the same as that in volume C/D. Thus, we can randomly generate the missing SNe in volume D using Monte Carlo simulation according to the SN distribution in volumes A, B, and C.
Then we took $\mu_2$ as the new $\mu_1$ and continued the process until we reached a distance with $\mu$ = 33.01 mag (40 Mpc). For subtypes with small sample sizes (i.e., type Ia-SC and 91T-like, 02cx-like, 87A-like SNe), we did not apply the bias corrections due to that such corrections would bring uncertainties that are apparently larger than the possible number changes. Instead, possible errors due to a small number of statistics are estimated for these subtypes in Sect.~\ref{Uncertainty}. Since the total number of SNe Ibc in our sample is small, we combined Ib-normal and Ibn as Ib, Ic-normal and Ic-BL as Ic when correcting the Mamquist bias for SNe Ibc. The added SNe for each subtype are shown in the Miller diagrams in Appendix~\ref{Miller}.

\subsection{Uncertainty estimation}\label{Uncertainty}

Taking into account the statistical error as Poisson noise, the 1$\sigma$ scatter of the fractions measured from the 10,000 different versions generated of the SN sample is taken as the statistical uncertainty. For subtypes with small sample sizes mentioned in Section~\ref{biascor}, we estimate the error due to  small number of statistics. For the systematic error, we considered the following three sources:
(1) SNe with distances around 40 Mpc: Due to the uncertainties in distance measurements, it is possible for an SN with measured distance smaller than 40 Mpc to be actually positioned outside, and vice versa; (2) SNe with uncertain redshifts: Our SN sample is composed of very nearby SNe, and their host galaxies are well observed in most cases. So, the number of SNe with uncertain redshifts is rather small, thus uncertainty caused by redshift is smaller compared to the first case; (3) Dust extinction/obscuration of SNe: This is a non-negligible effect especially for CCSNe, even in the nearby universe. 
	 
The contributions of the first two sources were estimated by calculating the probability that each SN is inside 40 Mpc. For each SN, we assumed a Gaussian distribution centered on the measured distance, with 1$\sigma$ uncertainty inferred from that in the distance measurement. Then the probability of an SN being inside 40 Mpc was calculated by performing a Monte Carlo simulation, generating 10,000 distances according to the assumed distribution and counting the ones smaller than 40 Mpc. The difference between the sum of all the probabilities and the observed number was taken as the uncertainty in the SN number, and then it transferred to the uncertainty in the fraction. For the third source, \citet{Mattila2012} and \citet{Jencson2019} estimated that $18.9^{+19.2}_{-9.5}\%$ and $38.5^{+26.0}_{-21.9}\%$ of CCSNe were, respectively, missed by optical surveys even in the local universe. We adopted the average value of the missed fractions and estimated the number of CCSNe missed due to dust extinction/obscuration as $57.16^{+92.42}_{-35.94}$, then assigned this number to SNe Ibc and II according to their relative proportions.
The total uncertainty is calculated by adding up the statistical and systematic errors, which is shown in the third column of Table~\ref{obs_table}.
	
\subsection{Corrected fractions}
	
\begin{figure*}
\includegraphics[width=\textwidth]{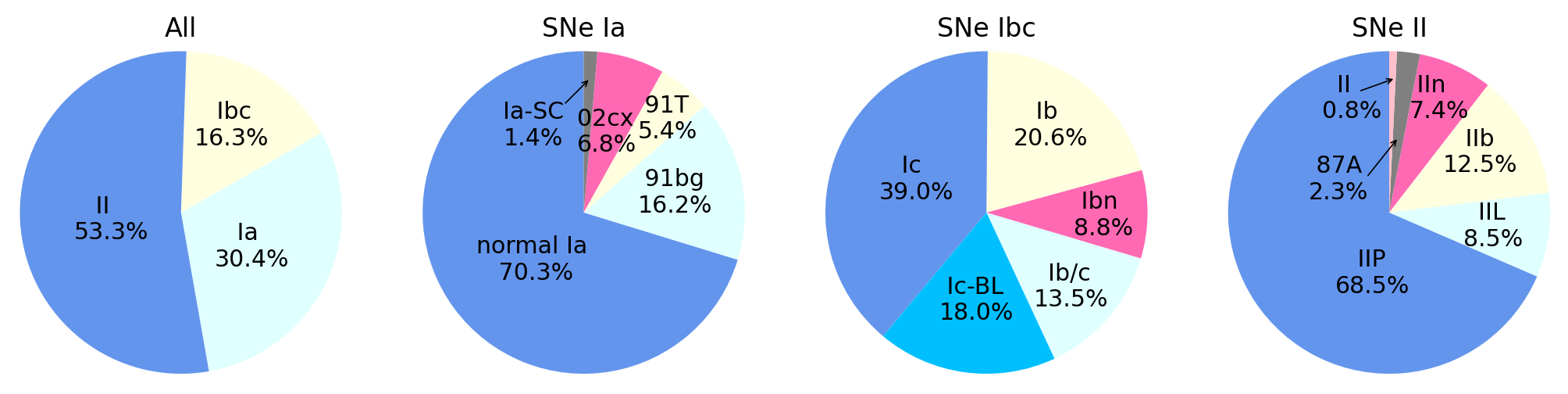}
\caption{The bias-corrected fractions of supernova sub-types.}\label{fraction_cor}
\end{figure*}

The bias-corrected SN fractions are shown in Fig.~\ref{fraction_cor} and the third column of Table~\ref{obs_table}, respectively. The corrected fractions are $30.4^{+3.7}_{-11.5}\%$ for SNe Ia, $16.3^{+3.7}_{-7.4}\%$ for SNe Ibc, and $53.3^{+9.5}_{-18.7}\%$ for SNe II, respectively. In comparison, \citet{Li2011} gave fractions of $24.1^{+3.7}_{-3.5}\%$, $18.7^{+3.5}_{-3.3}\%$, and $57.2^{+4.3}_{-4.1}\%$ for Ia, Ibc, and II, respectively. Our SN Ia fraction is larger than that given by \citet{Li2011} at a confidence level of about 2$\sigma$, while the changes in the fractions of both SN Ibc and SN II are approximately 1$\sigma$.

\subsubsection{Type Ia supernovae}
	
For SNe Ia, the bias-corrected fractions are $70.3^{+6.4}_{-9.4}\%$ for normal Ia, $16.2^{+5.1}_{-5.6}\%$ for 91bg-like, $5.4^{+3.6}_{-3.8}\%$ for 91T-like, $6.8^{+3.8}_{-3.2}\%$ for 02cx-like, and $1.4^{+2.6}_{-2.7}\%$ for Ia-SC, respectively. We could also give an upper limit of $1.3^{+1.3}_{-1.3}\%$ for the undetected 02es-like events, for that the detection number of 02es-like events within 40 Mpc is less than one. Compared to the results of \citet{Li2011}, which are $70.0^{+16.7}_{-15.0}\%$ for Ia normal, $15.2^{+6.8}_{-5.9}\%$ for 91bg-like, $9.4^{+5.9}_{-4.7}\%$ for 91T-like, and $5.4^{+4.7}_{-3.3}\%$ for 02cx-like, the fraction of the 91bg-like subtype is well consistent, while the fraction of 02cx-like subtypes becomes slightly larger and that of 91T-like shows a significant decrease (i.e., $\sim$5.4\% versus $\sim$9.4\%).  
\citet{Desai2024} recently gave a fraction of around 3.7\% for the 91T-like events, which is similar to our result. We caution that the fraction of 91T-like SNe Ia given by \citet{Li2011} could be overestimated because their sample was constructed using the SNe Ia up to a distance of 80 Mpc. At such a large distance, the 91T-like objects should be more easily detected than other subtypes, resulting in a relatively higher fraction. 
 
Following the classification scheme introduced by \citet{Wang2009},  we measured the Si~II $\lambda$ 6355 velocity for Branch-normal SNe Ia of our sample and identified 13 high-velocity (HV) SNe Ia and 36 normal-velocity (NV) SNe Ia, with the HV:NV ratio of 1:2.77 being consistent with the ratio of 1:3 given by \citet{Wang2009}. The classification information is presented in the second column of Table~\ref{Ia_info}.

\subsubsection{Core-collapse supernovae}
The bias-corrected fractions of SNe Ibc are $20.6^{+6.5}_{-10.3}\%$ for normal Ib, $8.8^{+2.5}_{-4.8}\%$ for Ibn, $39.0^{+10.2}_{-8.2}\%$ for Ic, $18.0^{+6.8}_{-9.1}\%$ for Ic-BL and $13.5^{+7.1}_{-7.3}\%$ for Ib/c, respectively. In comparison, \citet{Li2011} gave $21.2^{+8.4}_{-7.7}\%$ for Ib, $54.2^{+9.8}_{-9.8}\%$ for Ic, and $24.5^{+9.0}_{-8.4}\%$ for Ibc-pec. In our study, we give more detailed classifications, making the comparison with theirs difficult. However, the overall trend is roughly the same, with the fraction of SNe Ic being larger than that of SNe Ib. The subtype of SNe Ibn was not recognized in the sample from \citet{Li2011}, probably related to the rapid evolution of its light curve.  
	
The bias-corrected fractions of SNe II are $68.5^{+4.5}_{-7.5}\%$ for IIP, $8.5^{+3.0}_{-3.3}\%$ for IIL, $12.5^{+2.6}_{-3.5}\%$ for IIb, $7.4^{+2.6}_{-2.8}\%$ for IIn, and $2.3^{+2.1}_{-2.1}\%$ for 87A-like. And there is a fraction of SNe II ($0.8^{+1.3}_{-0.8}\%$) that cannot be accurately subclassified. The corresponding fractions given by \citet{Li2011} are $69.9^{+5.1}_{-5.8}\%$ for IIP, $9.7^{+4.0}_{-3.2}\%$ for IIL, $11.9^{+3.9}_{-3.6}\%$ for IIb, and $8.6^{+3.3}_{-3.2}\%$ for IIn. Our results are overall consistent with those of \citet{Li2011}, although the fractions of IIP and IIn are slightly lower and those of IIL and IIb are slightly larger. We include 87A-like events in our classification scheme, which were not involved in the sample of \citet{Li2011}. 
	
	\begin{table}
		\centering
		\caption{Fractions of sub-types of supernovae}\label{obs_table}
		\resizebox{\columnwidth}{!}{
			\begin{tabular}{ccc}
				\hline 
				Type & Observed fraction & Bias-corrected fraction \\
				\hline 
				Ia & 32.7\% & $30.4^{+3.7}_{-11.5}\%$  \\
				Ibc & 15.9\% & $16.3^{+3.7}_{-7.4}\%$  \\
				II & 51.4\% & $53.3^{+9.5}_{-18.7}\%$  \\
                SLSN & 0\% & $< 0.4^{+0.4}_{-0.4}\%$ \\
				\hline
				normal Ia & 71.0\% & $70.3^{+6.4}_{-9.4}\%$  \\
				91bg-like & 14.5\% & $16.2^{+5.1}_{-5.6}\%$  \\
				91T-like & 5.8\% & $5.4^{+3.6}_{-3.8}\%$  \\
				02cx-like & 7.3\% & $6.8^{+4.0}_{-3.4}\%$  \\
				Ia-SC & 1.5\% & $1.4^{+2.6}_{-2.7}\%$  \\
                02es-like & 0\% & $< 1.3^{+1.3}_{-1.3}\%$ \\
				\hline
				Ib & 20.6\% & $20.6^{+6.5}_{-10.3}\%$ \\
				Ibn & 8.8\% & $8.8^{+2.5}_{-4.8}\%$ \\
				Ic & 38.2\% & $39.0^{+10.2}_{-8.2}\%$ \\
				Ic-BL & 17.7\% & $18.0^{+6.8}_{-9.1}\%$ \\
				Ib/c & 14.7\% & $13.5^{+7.1}_{-7.3}\%$ \\
				\hline
				IIP & 70.6\% & $68.5^{+4.5}_{-7.5}\%$ \\
				IIL & 7.3\% & $8.5^{+3.0}_{-3.3}\%$ \\
				IIb & 11.9\% & $12.5^{+2.6}_{-3.5}\%$ \\
				IIn & 6.4\% & $7.4^{+2.6}_{-2.8}\%$ \\
				87A-like & 2.8\% & $2.3^{+2.1}_{-2.1}\%$ \\
				II & 0.9\% & $0.8^{+1.3}_{-0.8}\%$ \\
				\hline
			\end{tabular}
		}    
		
	\end{table} 

\section{The host galaxy sample} \label{host}

We constructed two galaxy samples in our study, one representing the host galaxies of our SN sample and the other including a full sample of nearby galaxies within 40 Mpc.

\subsection{Host galaxies of our supernova sample}

For each supernova in our sample, we examined the galaxy images near the supernova positions through the SIMBAD database to clarify its host galaxy, and then collected information of each host galaxy from the NED and Hyperlinked Extragalactic Databases and Archives \footnote{\url{http://leda.univ-lyon1.fr/}} \citep[HyperLEDA,][]{Makarov2014}. 
	
The host galaxy sample, hereafter the host sample, contains 191 galaxies. This is due to that 2 SNe (1 Ibc and 1 II) of our sample do not show clear host galaxies and 17 galaxies are found to host multiple SNe:
	ESO 430-20 (SN 2019ejj and SN 2019fcn), 
	IC 863 (SN 2016dsl and SN 2021foa), 
	M 100 (SN 2019ehk and SN 2020oi)
	NGC 0493 (SN 2016hgm and SN 2022ywf), 
	NGC 1359 (SN 2021aess and SN 2022yvw), 
	NGC 1448 (SN 2020zbv and SN 2021pit), 
	NGC 1532 (SN 2016iae and SN 2016ija), 
	NGC 1672 (SN 2017gax and SN 2022aau), 
        NGC 2139 (SN 2022qhy and SN 2023zcu),
	NGC 3256 (SN 2018ec and SN 2021afuq),
	NGC 3318 (SN 2017ahn and SN 2020aze), 
	NGC 3938 (SN 2017ein and SN 2022xlp), 
        NGC 4414 (SN 2021J and SN 2023hlf),
        NGC 4568 (SN 2020fqv and SN 2023ijd),
	NGC 5962 (SN 2016afa and SN 2017ivu), 
	NGC 6951 (SN 2020dpw and SN 2021sjt), 
	PGC 136560 (SN 2018afb and SN 2022xus).

The first 6 entries in the host-galaxy sample for our SNe Ia, SNe Ibc, and SNe II are listed in Tables~\ref{Ia_host}, ~\ref{Ibc_host}, and ~\ref{II_host}, respectively. Full tables are available online, including the SN name, the host galaxy name, the coordinates, the decimal logarithm of the major axis measured at the isophotal level of 25 mag arcsec$^{-2}$ in the B band\footnote{The major axis is in unit of arcsecond, and its measurement has been corrected for the galactic extinction and inclination effect}, the relative radial position of the SN in its host galaxy (the ratio between the projected offset of SN from its host-galaxy nucleus and half of the host major axis), the morphological type, type code, and stellar mass of the galaxy.  
We specially mention that, for E-S0 galaxies, we directly use the angle between the SN position and the host center as the offset angle to calculate the relative position of SN, while for other galaxies, we consider the effect of position angle and inclination. The stellar mass of host galaxy is estimated by fitting to its spectral energy distribution constructed with photometry in ultraviolet, optical, and near-infrared bands (see discussions in Sect.~\ref{prospectorfitting}).

        \begin{table}
		\centering
		\caption{Hubble type bins.}\label{Hubbletypebin}
		\begin{threeparttable}
			
			\begin{tabular}{ccccc}
				\toprule
				Symbol & Type & Host\tablefootmark{a} & SNe\tablefootmark{b} & GLADE+\tablefootmark{c} \\
				\midrule
				E &  E, E-S0 & 12 & 12 & 1150 \\
				S0 &  S0, S0-a & 24 & 26 & 814 \\
			  Sa &  Sa, Sab & 10 & 10 & 401 \\
                Sb &  Sb & 31 & 35 & 423 \\
                Sbc &  Sbc & 21 & 24 & 430 \\
                Sc &  Sc & 56 & 62 & 1149 \\
                Scd &  Scd, Sd, Sm & 19 & 20 & 1671 \\
                Irr &  Irr & 10 & 11 & 2355 \\
				\bottomrule
			\end{tabular}

		\end{threeparttable}

            \tablefoot{
            \tablefoottext{a}{The number of galaxies of each Hubble type in the Host sample.}
            \tablefoottext{b}{The number of SNe hosted by galaxies of each Hubble type.}
            \tablefoottext{c}{The number of galaxies of each Hubble type in the GLADE+ sample.}
            }
            
	\end{table} 
 
\begin{figure}
        \includegraphics[width=\columnwidth]{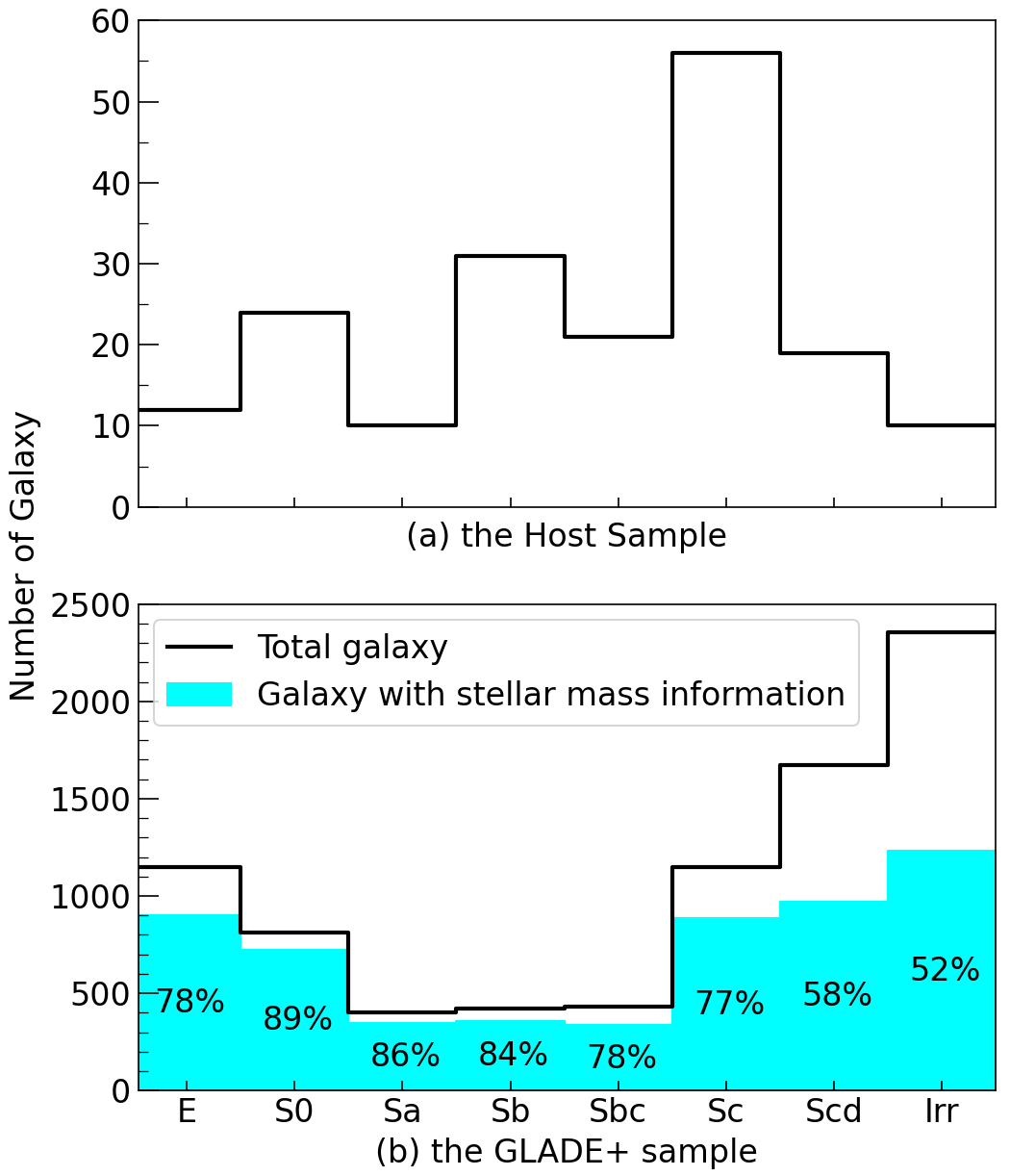}
		\caption{The Hubble-type distribution for the Host sample (upper panel) and the GLADE+ sample (lower panel). The cyan region shows the fraction of galaxies with stellar mass measurements for each Hubble type.} \label{Galaxysample}
    \end{figure}

\begin{figure}
        \includegraphics[width=\columnwidth]{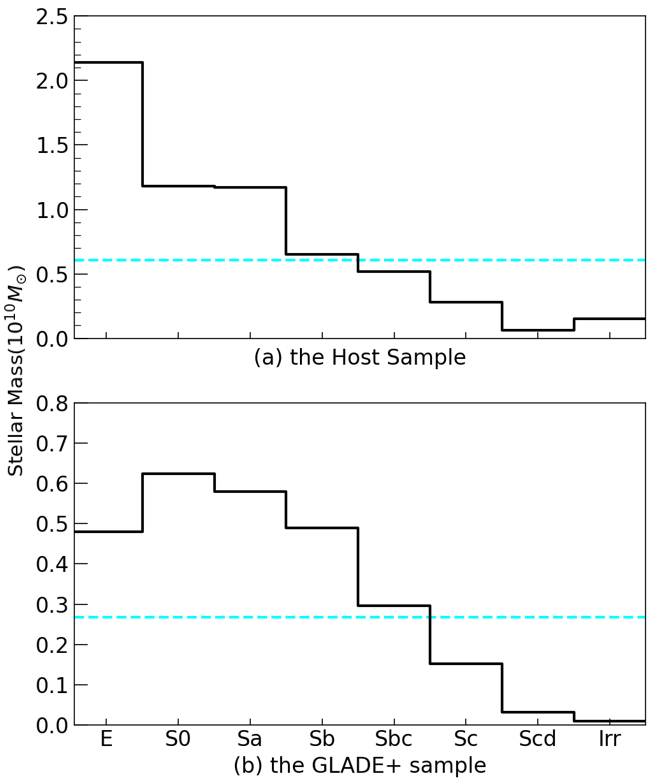}
		\caption{The average stellar mass derived for different Hubble types of galaxies in the Host sample (upper panel) and the GLADE+ sample (lower panel). The dashed cyan lines show the average stellar mass for the sample of SN hosts and the full galaxy sample from GLADE+.} \label{Galaxymass}
    \end{figure}

We group the Hubble types of host galaxies into eight bins as listed in Table~\ref{Hubbletypebin}, where the number of galaxies of each Hubble type in the host sample and GLADE+sample (details seen in Sect.~\ref{GLADE}) and the number of SNe hosted by galaxies of each Hubble type are also listed. In Fig.~\ref{Galaxysample}(a) we present the Hubble-type distribution for the host sample. Our SN sample shows a clear preference to occur in Sc galaxies but lacks in E, Sa and Irr galaxies. There is one galaxy (2MASXi J1357565-291725) labeled as S? in HyperLEDA, indicating that it can only be classified as a spiral galaxy, we did not include it in Figs.~\ref{Galaxysample}(a) and ~\ref{Galaxymass}(a).

In Fig.~\ref{Galaxymass}(a) we present the average stellar mass of galaxies of different Hubble types for the SN hosts. The average stellar mass for the Host sample is $0.605 \times 10^{10} \mathrm{M_{\odot}}$. Among this sample, the most massive ones are elliptical galaxies, with an average stellar mass of $2.141 \times 10^{10} \mathrm{M_{\odot}}$; for spirals, the average stellar mass declines from early to late types, and the mean stellar mass is only $0.065 \times 10^{10} \mathrm{M_{\odot}}$ for Scd galaxies.

\subsection{Host stellar mass estimation}\label{prospectorfitting}

To calculate the stellar mass of the host galaxies, we require broadband photometric observations covering at least two wavelength ranges in UV, optical, IR and mid-IR to model the stellar population properties, so that the spectral energy distribution (SED) can be well sampled. We collected archival photometric observations for all host galaxies via NED. Observations in the UV, optical, IR, and mid-IR bands were obtained using the Galaxy Evolution Explorer \citep[GALEX,][]{Bouquin2018}, Sloan Digital Sky Survey \citep[SDSS, ugriz,][]{Ahumada2020}, Two-Micron All Sky Survey \citep[2MASS,][]{Skrutskie2006}, and Wide-Field Infrared Survey Explorer \citep[WISE,][]{Wright2010}, respectively. A similar aperture size is required for consistency in the SED fit whenever possible. All  photometry was corrected for Galactic extinction along the line-of-sight of the SNe \citep{Schlafly2011}. 

We used Prospector \citep{Leja2017, Johnson2021}, a flexible code used to infer the parameters of stellar populations, to determine the host metallicity and stellar mass of the SNe. We fitted the observational data with Prospector through a nested sampling fitting routine, dynesty \citep{Speagle2020}. The prospector produces model SEDs with Flexible Stellar Population Synthesis \citep[FSPS,][]{Conroy2009,Conroy2010} using single stellar models through MIST \citep{Paxton2018} and the MILES spectral library \citep{Falcon2011}. For all Prospector fits, we used the \citet{Chabrier2003} initial mass function (IMF) and a parametric delayed-$\tau$ star formation history (SFH $\propto \mathrm{te}^{\mathrm{t}/\tau}$ ), defined by the e-folding time $\tau$. We also included the effects nebular emission in our simulation \citep{Byler2017}. Since we did not collect any spectra of the galaxies, we cannot measure the spectral line strengths to determine their metallicities, so we fixed the gas phase metallicity to solar in our modeling. For dust attenuation, we used the \citet{Kriek2013} model. The main fitting parameters include the total mass formed, stellar metallicity (Z$_{\ast}$), V-band optical depth, stellar age of the galaxy (the maximum value allowed is the age of the universe at a redshift of the galaxy), and the SFH parameter $\tau$. We present the best-fit SED and the corresponding parameters of the galaxy NGC 1209 (the host of SN 2018htt) in Figs.~\ref{SED} and ~\ref{prospector} as an example.

\subsection{The GLADE+ sample}\label{GLADE} 

We constructed a full sample of galaxies within 40 Mpc using the extended version of Galaxy List for the Advanced Detector Era \citep[GLADE$+$,][]{Dlya2022}. GLADE+ contains a total of $\sim$ 22.5 million galaxies and is complete up to a luminosity distance of $\sim$ 44 Mpc in terms of the cumulative B-band luminosity of galaxies. They also derived stellar masses for galaxies with WISE W1-band magnitudes. We extracted all galaxies with distances smaller than 40 Mpc from GLADE+ and checked the Hubble types of each galaxy in HyperLEDA. The final galaxy sample, hereafter the GLADE+ sample, consists of 8790 galaxies with known Hubble types and 6021 of them with known stellar masses. The stellar mass information of the host sample is also contained in the GLADE+ sample. We present the Hubble-type distribution for our GLADE+ galaxy sample and the fraction of galaxies with stellar mass measurements for each Hubble type in Fig.~\ref{Galaxysample}(a). Irr and Scd galaxies are the two most abundant types in the local universe, taking up 48\% and 34\% of the total sample, respectively. However, the stellar mass information coverage of these two Hubble types is remarkably lower than other types. For the rest Hubble types, the fractions of galaxies with stellar mass measurements are all over 75\%, while the numbers of E, S0 and Sc galaxies are notably larger than those of Sa, Sb and Sbc galaxies. A total of 393 galaxies in the GLADE+ sample are labeled as S? in HyperLEDA, and 271 of them have stellar mass measurements, with an average stellar mass of $0.05091 \times 10^{10} \mathrm{M_{\odot}}$, we do not include them in Figs.~\ref{Galaxysample}(b) and ~\ref{Galaxymass}(b).

We then calculated the average stellar mass of the 6021 GLADE+ galaxies with stellar-mass measurements. The results are shown for each Hubble type in Fig.~\ref{Galaxymass}(b) and the dashed cyan line represents the average for a total of 6021 galaxies, which is $0.257 \times 10^{10} \mathrm{M_{\odot}}$. The stellar mass distribution shows a clear difference from the galaxy number distribution. The average stellar mass peaks in E-S0 galaxies and then declines rapidly from early to late type spirals. The most abundant types of galaxies in the local universe, the Irr and Scd galaxies, turn out to have the lowest stellar mass, that is, with a fraction of $\sim$ 4.0\% and $\sim$ 12.1\% of the average value of the whole GLADE+ sample, respectively. 

\subsection{Comparison of the two galaxy samples} 

The distributions of Hubble types for the Host and GLADE+ samples show clear differences. Among the GLADE+ sample, the Irr and Scd galaxies are found to be the most abundant types in the local universe, but most of them do not host any SNe, perhaps due to their low stellar mass. Moreover, compared to the GLADE+ sample, SNe shows clear preferences to occur in Sc galaxies and Sb galaxies.  
The relative number of elliptical galaxies hosting SNe in our sample is much smaller than that of the GLADE+ sample. However, the stellar mass shows an overall similar distribution for the two galaxy samples, except that the average stellar mass of the Host sample is $\sim$ 1.4 times larger than that of the GLADE+ sample. In particular, for elliptical galaxies, the stellar mass derived from the SN host sample is $\sim$ 3-4 times larger than that from the GLADE+ sample, suggesting that the elliptical galaxies hosting SNe lie at the extremely massive end. 

\section{Host environment of different types of supernovae}

\subsection{Host Hubble types}

\begin{figure}
        \includegraphics[width=\columnwidth]{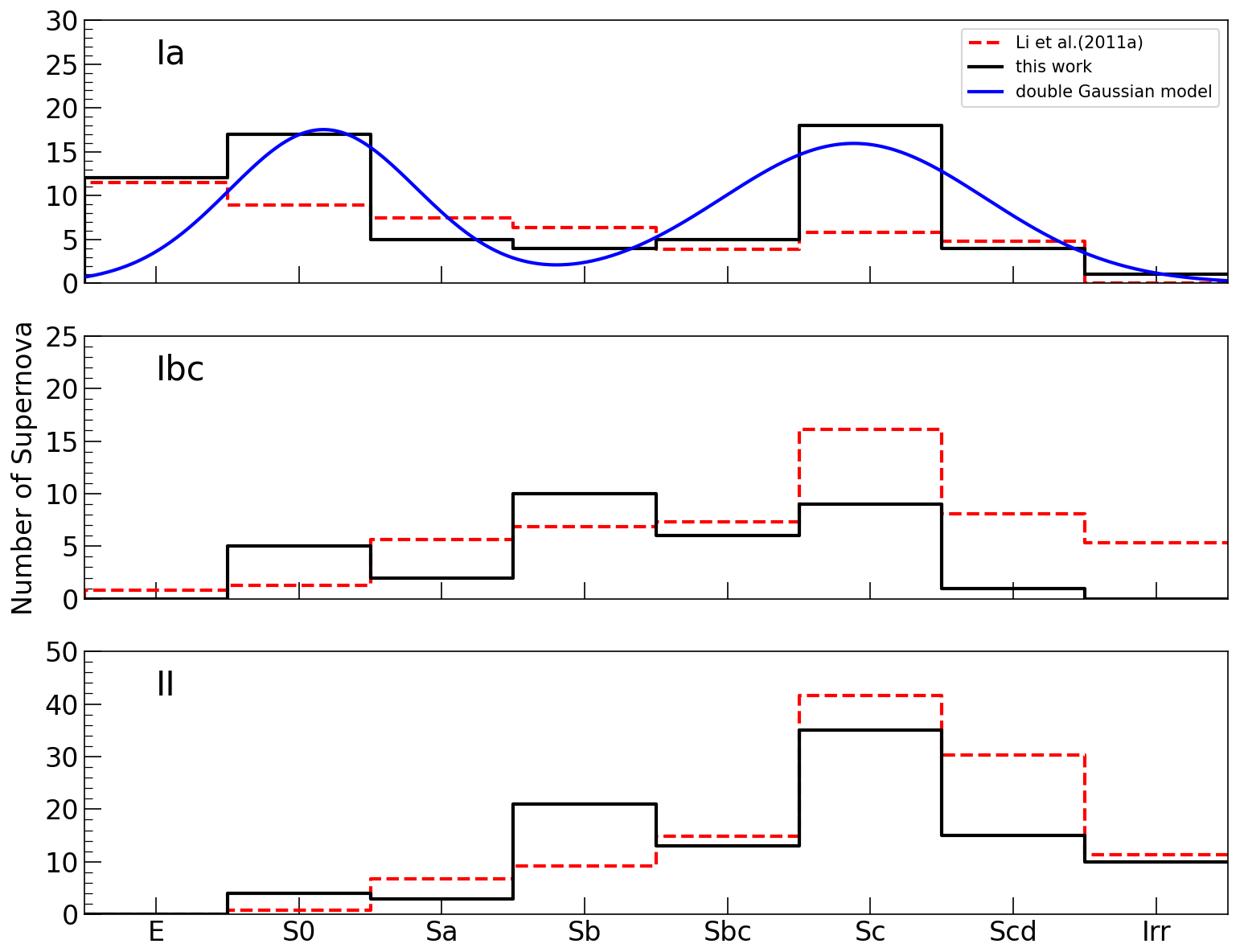}
		\caption{The Hubble-type distribution of our SN host sample compared to that of the Lick sample \citep{Li2011}. The solid black lines are for our SN sample, while the dashed red lines are interpolated from the \citet{Li2011} sample. The blue line in the upper panel represents the double Gaussian fit to the SN Ia distribution.}\label{Hubble}
    \end{figure}

We compared the host galaxy hubble types of our SN sample with those of the \citet{Li2011} sample, as shown in Fig.~\ref{Hubble}. From the stellar mass given by GLADE$+$ and the SN rate parameters given in Table 4 of \citet{Li2011b}, we are able to calculate the expected number distribution of each subtype of SNe in galaxies of different Hubble types by adopting their SN rates and rate-size relation. The results are shown as red dashed lines in Fig.~\ref{Hubble}. For galaxies marked with the "S?" type, we redistributed them into all types of spirals according to their relative fractions. One can see that CCSNe rarely occurs in early-type galaxies (E and S0) but both SNe Ibc and SNe II peak in Sc galaxies, consistent with the trend seen in the \citet{Li2011} sample. Given the massive star origin of CCSNe, this distribution is not surprising. 

However, we notice that, for our SN sample, the number of SNe Ibc discovered in Sc, Scd, and irregular galaxies, and the number of SNe II in Scd galaxies seems to be noticeably smaller than the \citet{Li2011} sample. The lack of SNe Ibc in these types of galaxies could be partly due to the bias caused by small-number statistics. If comparing the Hubble type distribution between our Host, GLADE+ sample (Fig.~\ref{Galaxysample}) and the galaxy sample from LOSS \citep{Leaman2011} (see the upper left panel of their Figure 2, hereafter the Lick sample), we find that the Hubble type distribution of the Lick sample is more similar to that of our Host sample. For instance, there are smaller numbers of E and Irr galaxies and relatively larger numbers of spirals. Moreover, the distribution of average stellar mass of the Lick sample (see the upper panel of their Fig. 3) is also more consistent with that of our host sample, rather than the GLADE+ sample. In particular for the Scd- and Irr-type galaxies, the ratio between their average stellar masses and those of other Hubble types is higher for the Lick sample than for the GLADE+ sample. Therefore, it is likely that \citet{Li2011b} overestimated the CCSNe rate in the Irr and Scd galaxies because of their preference for more massive (brighter) Scd and Irr galaxies in the Lick sample.

As can be seen in Fig.~\ref{Hubble}, the number distribution of our SN Ia sample shows a distinct double peak structure in early-type S0 and late-type Sc galaxies, with an extreme excess in Sc galaxies compared to the \citet{Li2011} sample. As the star-forming activities in Sc galaxies are not remarkably more active than other Hubble-type galaxies according to the distribution of CCSNe, this number excess in Sc galaxies could be intrinsic. On the other hand, since the elliptical and S0 galaxies are mainly composed of relatively old populations, the double peak structure could suggest two different populations of SNe Ia. One group is the "prompt" component, which originates from young stellar populations in late-type spirals and has a very short time delay when evolving from their progenitor systems to the catastrophic explosion. The other group is the so-called "delayed" component, which arises from old progenitor populations in E and S0 galaxies and has delay time as long as billions of years. For SNe Ia, the delay time measures the duration between an instantaneous burst of star formation and the final resulted SN Ia explosion. The double peak structure we discovered gives inspiration to the simulation of the delay time distribution (DTD) of SNe Ia in Paper II of our study, suggesting a two-component model for SN Ia DTD, with the prompt and delayed components corresponding to the young and old stellar population in late-type spirals and E-S0 galaxies, respectively. Performing a double Gaussian fit, we suggest a fraction of 41.9\% for the delayed component and a fraction of 58.1\% for the prompt component.  
        
\begin{figure}
        \includegraphics[width=\columnwidth]{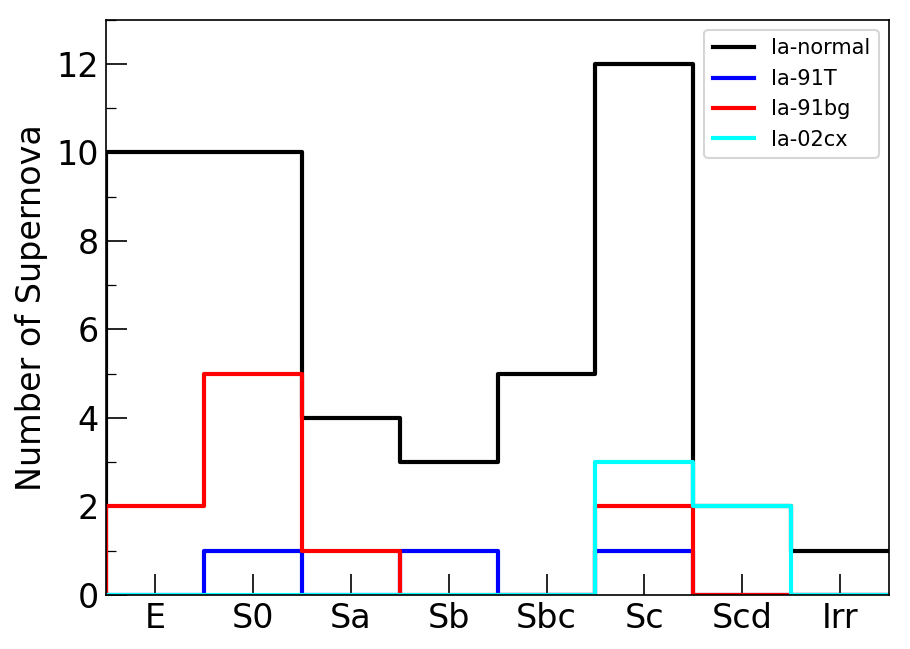}
		\caption{The number distribution of different subtypes of SNe Ia in different Hubble types of galaxies.}\label{subHubble}
        \end{figure}

We further investigated the distribution of host-galaxy Hubble types for different subtypes of SNe Ia in Fig.~\ref{subHubble}\footnote{HyperLeda does not give a Hubble type classification for the host of the 91T-like SN 2021dsz (i.e., WISEA J042253.55+363641.0).}. As can be seen, normal SNe Ia can occur in host galaxies with a wide range of Hubble types, and it also shows a prominent double peak structure. It is clear that the 02cx-like events occur only in late-type star-forming spirals; while the 91bg-like events occur preferentially in E–S0 galaxies, with a minor fraction in spirals. The 91T-like SNe are also found to explode preferentially in late-type spirals.
The distributions in host-galaxy Hubble types indicate different delay time and hence progenitor properties for different SN Ia subtypes. Thus, the delay time of 02cx-like explosions should be much shorter than that of the 91bg-like events with older stellar populations. We note, however, that two 91bg-like SNe Ia are found in Sc galaxies (i.e., SN 2017igf in NGC 3901 and SN 2022xkq in NGC 1784), both located in the spiral arms of their hosts, probably indicating the existence of a different progenitor channel for the 91bg-like objects. 

\subsection{Stellar mass} 

\begin{figure}
        \includegraphics[width=\columnwidth]{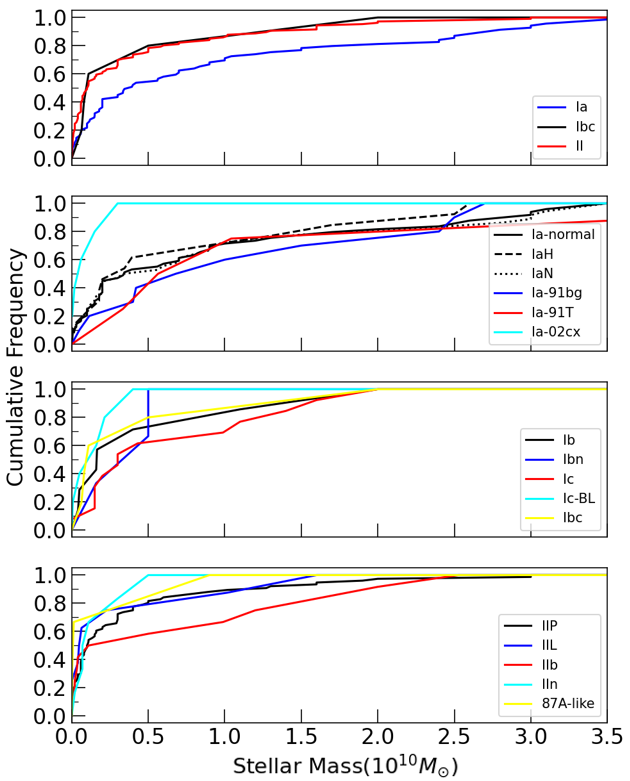}
		\caption{The cumulative fractions for the  distributions of host stellar mass for SNe Ia (blue), Ibc (black), II (red) in the upper panel and their detailed sub-types in the lower three panels, respectively.}\label{stellarmass}
    \end{figure}

The cumulative fractions for the stellar mass distribution of our SN hosts are shown in the upper panel of Fig.~\ref{stellarmass}. SNe Ia show a significant preference for residing in massive hosts, while the distributions for SNe Ibc and II are well consistent. The stellar-mass distributions of SNe Ia and CCSNe show a statistically significant difference, with p = 0.0007 from a K-S test.

For SNe Ia, no clear differences are found between subtypes of SNe Ia except for 02cx-like events, which tend to explode in low-mass hosts. We ran a K-S test between normal Ia and 02cx-like events and derived a p value of 0.109. The difference in the cumulative fraction distribution is not statistically significant according to the K-S test, but this could be due to the small sample of 02cx-like SNe Ia.  

For CCSNe sample, the most visually apparent difference occurs between SNe Ic and SNe Ic-BL, although the K-S test suggests that this difference is not that significant, i.e., with p = 0.318. This is consistent with previous results that hosts of SNe Ic-BL generally have a lower stellar mass than those of other types of CCSNe \citep{Kelly2012, Schulze2021}. The host stellar mass distributions are roughly the same for all subtypes of SNe II. The difference between SNe IIP and SNe IIb is slightly larger but not statistically significant (p = 0.262). We list the results of all the statistical K-S tests in the second column of Table~\ref{KS}.

    \begin{table}
		\centering
		\caption{P values of the K-S tests derived for stellar-mass and radial distributions between different subtypes of SNe.}\label{KS}
        \begin{threeparttable}
		  \resizebox{\columnwidth}{!}{	
			\begin{tabular}{ccc}
				\toprule 
			Samples & p-value for stellar mass & p-value for offset\\
				\midrule
				Ia vs. Ibc & 0.010 & 0.018 \\
				Ia vs. II & 0.001 & 0.272 \\
			  Ibc vs. II & 0.189 & 0.010 \\
                Ia normal vs. 91bg-like & 0.262 & 0.502  \\
                Ia normal vs. 91T-like & 0.287 & 0.027  \\
                Ia normal vs. 02cx-like & 0.109 & 0.333   \\
                91bg-like vs. 91T-like & 0.984 & 0.010  \\
                91bg-like vs. 02cx-like & 0.019 & 0.655  \\
                91T-like vs. 02cx-like & 0.016 & 0.079\\
                IaH vs. IaN & 0.907 & 0.249  \\
                Ib vs. Ibn & 0.850 & 0.400  \\
                Ib vs. Ic & 0.821 & 0.821  \\
                Ib vs. Ic-BL & 0.909 & 0.545  \\
                Ib vs. Ib/c & 0.838 & 0.737\\
                Ibn vs. Ic & 0.732 & 0.732  \\
                Ibn vs. Ic-BL & 0.286 & 0.857  \\
                Ibn vs. Ib/c & 0.464 & 0.464  \\
                Ic vs. Ic-BL & 0.319 & 0.176 \\
                Ic vs. Ib/c & 0.211 & 0.981 \\
                Ic-BL vs. Ib/c & 0.873 & 0.873 \\
                IIP vs. IIL & 0.764 & 0.520  \\
                IIP vs. IIb & 0.262 & 0.819  \\
                IIP vs. IIn & 0.877 & 0.660  \\
                IIP vs. 87A-like & 0.436 & 0.382 \\
                IIL vs. IIb & 0.897 & 0.967  \\
                IIL vs. IIn & 0.376 & 0.376  \\
                IIL vs. 87A-like & 0.709 & 0.927  \\
                IIb vs. IIn & 0.468 & 0.760  \\
                IIb vs. 87A-like & 0.338 & 0.725  \\
                IIn vs. 87A-like & 0.333 & 0.679  \\
				\bottomrule
			\end{tabular}
		  }
		\end{threeparttable}
    \end{table}

\begin{figure}
        \includegraphics[width=\columnwidth]{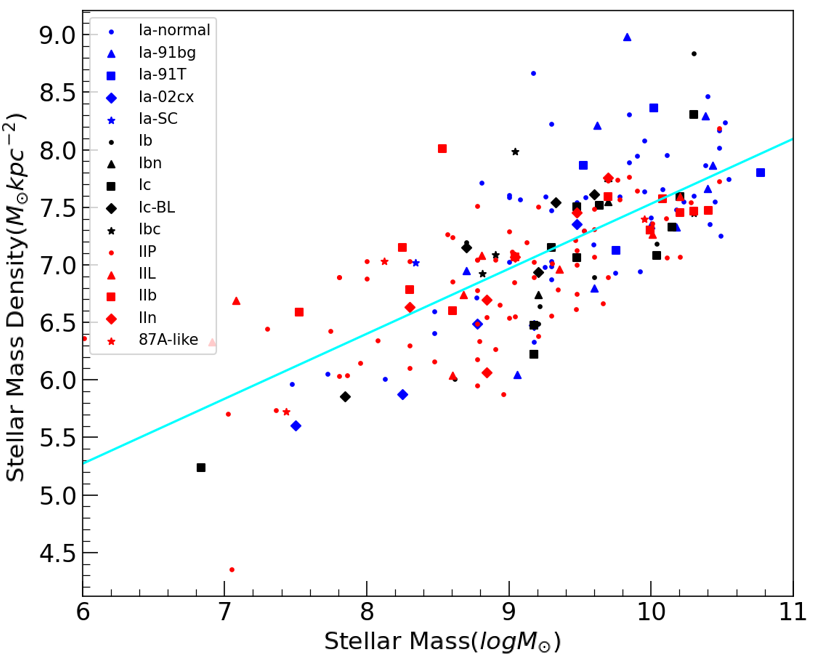}
		\caption{Host stellar mass versus the stellar mass density. SNe Ia, Ibc and II are shown by the color blue, black and red, respectively. The subtypes are displayed with different shapes. The cyan line is the linear fit of the correlation to the total SN sample.}\label{stellarmass_density}
    \end{figure}

We then calculated the projected stellar-mass density $\Sigma_{\mathrm{M}}$ using the equation $\Sigma_{\mathrm{M}} = log_{10}(\mathrm{M/\pi AB})$, where M is the stellar mass (in units of M$_{\odot}$), A and B are the semimajor and semiminor axes (in units of kpc) of the galaxy at the isophotal level 25 mag/arcsec$^2$ in B-band corrected for galactic extinction and inclination effect. The relation between host-galaxy stellar mass density and stellar mass is shown in Fig.~\ref{stellarmass_density}. The host stellar mass and stellar mass density of all types of SNe follow a positive linear correlation. For different subtypes, we find that the hosts of 02cx-like events have lower stellar mass density for their stellar mass compared to other subtypes, as most of them distribute below the cyan line, while the hosts of SNe Ic-BL show the opposite trend to locate above the cyan line. As suggested by \citet{Kelly2014}, higher stellar mass density corresponds to higher metallicity on average, we can conclude that the host environment of 02cx-like events tends to be metal-poor, while that of SNe Ic-BL prefers to be metal-rich. 

\subsection{Radial distribution} 

\begin{figure}
        \includegraphics[width=\columnwidth]{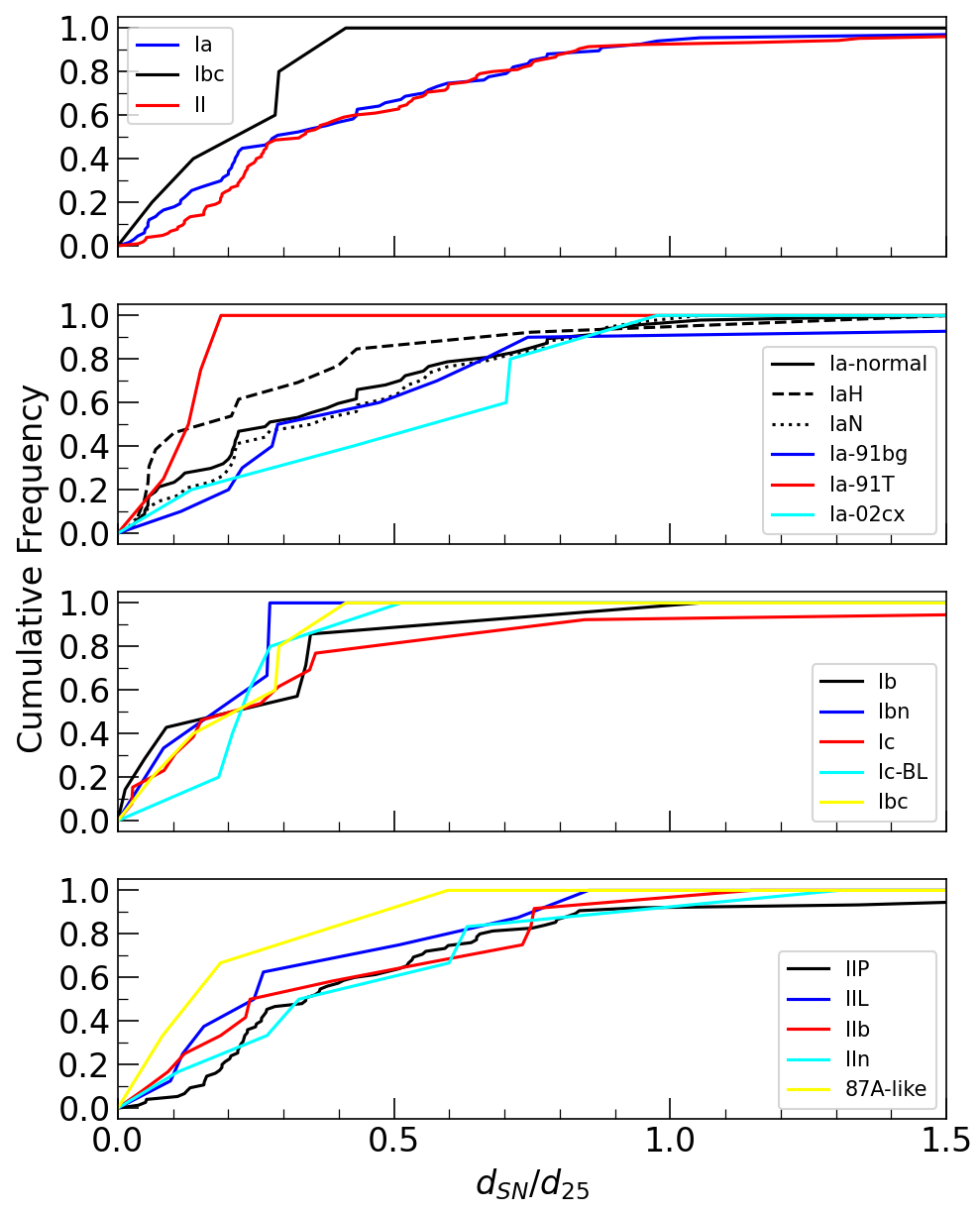}
		\caption{The cumulative fractions for the radial distributions of SNe Ia (blue), Ibc (black), II (red) in the upper panel and their detailed sub-types in the following three panels, respectively.}\label{Radial}
    \end{figure}

We show the cumulative fractions of the radial distributions derived for SNe Ia, Ibc, and II in the upper panel of Fig.~\ref{Radial}. Significant differences were found between SNe Ibc and SNe Ia/II. We ran K-S tests on the radial distribution and derived a p value of 0.018 and 0.010, respectively. It is obvious that SNe Ibc prefers to locate in more central regions in their host galaxies compared to SNe Ia and SNe II. This is consistent with the results of \citet{Anderson2009} that SNe Ibc dominates the production of CCSNe within the central parts of spiral galaxies, while in the disc SNe II become more dominant. More recently, \citet{Audcent2020} found a relative deficit of SNe Ia in central regions of galaxies, in agreement with our findings. 

Among SNe Ia, 91T-like events show a tendency to explode in the inner region of their host galaxies. In contrast, 02cx-like events prefer to occur in the outermost region of their hosts. The K-S test indicates that the radial distribution of the 91T-like and the 02cx-like subtypes is not similar, with a p value of 0.079. We also split normal Ia into the HV (IaH) and NV (IaN) subtypes and show the cumulative fractions of their radial distributions in Fig.~\ref{Radial}. The HV SNe Ia prefer to occur in the inner regions of their host galaxies compared to the NV counterparts, although  the difference is not that substantial according to the p value of the K-S test. 

Observations have shown that negative metallicity gradients are common in both the Milky Way and many external galaxies, in the sense that the heavy element abundances decrease systematically outward from the center of galaxies \citep{Henry1999}. This indicates that the progenitor populations are metal-poor for 02cx-like SNe Ia but are metal-rich for SNe Ibc, 91T-like, and HV SNe Ia. 

\begin{figure}
        \includegraphics[width=\columnwidth]{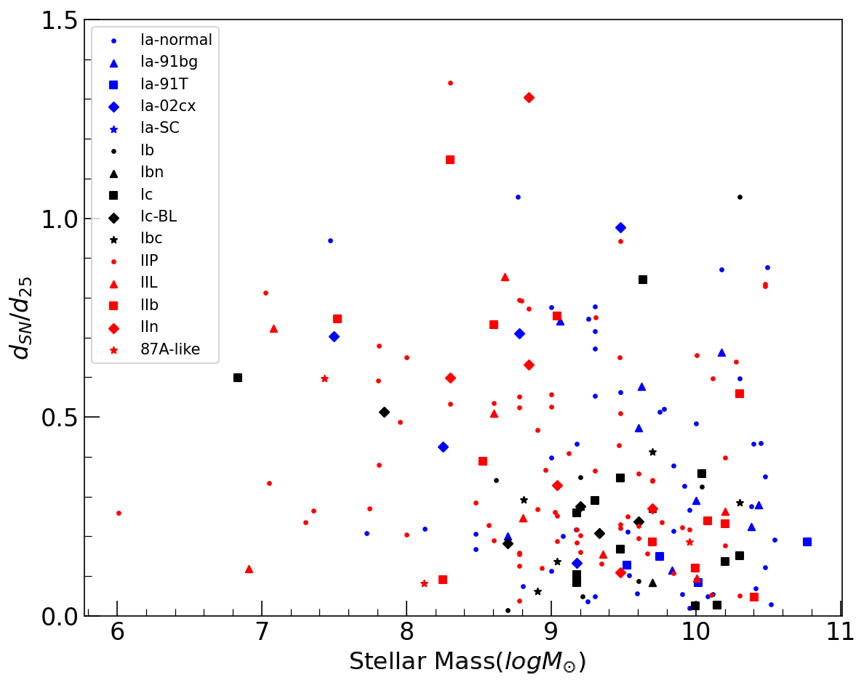}
		\caption{Stellar mass of host galaxies versus the SN offsets from the center of the hosts. SNe Ia, Ibc and II are shown by the color blue, black and red lines, respectively. The subtypes are shown with different shapes.}\label{stellarmass_radial}
    \end{figure}

In Fig.~\ref{stellarmass_radial}, we present the correlation between the stellar mass of host and the relative SN offset from the center of their hosts. For most subtypes, no correlation could be found, except for 91bg-like, IIL, IIb, and IIn that present a tendency to explode at smaller offsets when they occur in more massive hosts. 

\section{Conclusions} \label{sum}
	
We established a relatively complete nearby SN catalog based on the discoveries made over the period from year 2016 to 2023 when the SN searches enter into the wide-field survey era. After applying a distance cut of 40 Mpc, we collected a sample containing 211 SNe, of which 69 are SNe Ia, 34 are SNe Ibc, and 109 are SNe II. The main advantages of this new nearby SN sample follow as: (1) having reliable subtype classifications owing to abundant spectroscopic information and light-curve evolution; (2) suffering less observation bias due to the magnitude limit in the SN surveys; (3) obtaining relatively complete Hubble-type classifications for the host galaxies. After corrections for the possible Malmquist bias, the fractions of SNe Ia, Ibc, and II are derived as $30.4^{+3.7}_{-11.5}\%$, $16.3^{+3.7}_{-7.4}\%$, and $53.3^{+9.5}_{-18.7}\%$, respectively. 
Compared with the \citet{Li2011} result, the fraction of SNe Ia in the 40-Mpc sample shows a non-negligible increase, likely related to the change in the birth environments of SNe Ia. 

The host galaxy catalog was accordingly built for the 40-Mpc SN sample, which contains a total of 191 galaxies. Combining it with the SN catalog, we find that most of our SNe tend to reside in Sc galaxies, while they are relatively sparse in E, Sa, and Irr galaxies. This Hubble-type distribution shows significant differences from that of the full 40-Mpc nearby galaxy sample established with the GLADE+ database, in which the galaxy samples are dominated by Irr, Scd, and E/S0 types. We also compared the hosts of our SN sample with that of \citet{Li2011}, and found that this distribution is similar for CCSNe but shows a noticeable difference for SNe Ia. For the latter, a prominent double-peak structure exists in early-type S0 and late-type Sc galaxies, with an excess distribution in Sc galaxies. This structure could suggest a two-component model for the delayed time distribution of SNe Ia, with a prompt component corresponding to the young stellar progenitor population (i.e., in late-type spirals) and a delayed component corresponding to the old stellar progenitor population (i.e., in E-S0 galaxies), respectively. We further examined the hosts of each SN Ia subtype and found that 91T-like and 02cx-like events tend to occur in star-forming spirals, whereas 91bg-like events prefer in E-S0 galaxies having an old stellar population, as previously suggested. 

We further examined the stellar mass of the host galaxies and the comparison galaxy samples, and found that the average stellar mass of the GLADE+ sample is only $\sim$ 40\% of the value derived from our SN-host sample. In particular, the stellar mass of elliptical hosts is about 4 times larger than the average value of the GLADE+ elliptical galaxies, suggesting that stellar mass is the dominant factor affecting SN Ia production in those tardy galaxies. In contrast, the average stellar mass of the most SN-productive Sc-type galaxies show much less significant differences between the SN hosts and the GLADE+ sample (i.e., 0.25 vs. 0.15 in units of 10$^{10}$ M$_{\odot}$). For the 40-Mpc SN sample, the hosts of SNe Ia are more massive than those of CCSNe, while the hosts of SNe Ic-BL and 02cx-like objects tend to have lower stellar masses than other subtypes of SNe, consistent with their origins of less massive star-forming galaxies.

Analysis of the radial distribution of the 40-Mpc SNe in their host galaxies came to the following conclusions: (1) SNe Ibc tend to occur closer to the center of their host galaxies relative to SNe Ia and II; (2) Among SNe Ia, the HV subtype (and perhaps 91T-like events) reside preferentially in inner regions of their hosts compared to other subtypes of SNe Ia, especially the 02cx-like events; (3) CCSNe of different subtypes do not show significant differences in their radial distribution. Given the negative metallicity gradients in most disk galaxies, we would expect that SNe Ibc and HV SNe Ia are likely associated with metal-rich progenitors, while 02cx-like events tend to originate from metal-poorer progenitors. By considering the correlations between host-galaxy stellar mass/SN offset with stellar mass density, we find that 91bg-like, IIL, IIb, and IIn tend to explode at smaller offsets when they occur in more massive hosts.

\begin{acknowledgements}
We acknowledge the support of the staff of the Lijiang 2.4~m and Xinglong 2.16~m telescopes. This work is supported by the National Science Foundation of China (NSFC grants 12288102 and 12033003), the science research grant from the China Manned Space Project No. CMS-CSST-2021-A12 and the Tencent Xplorer prize. 
This work makes use of observations from the Las Cumbres Observatory global telescope network. The LCO group is supported by NSF grants AST-1911225 and AST-1911151. 
We acknowledge the usage of the HyperLeda database (http://leda.univ-lyon1.fr). Lasair is supported by the UKRI Science and Technology Facilities Council and is a collaboration between the University of Edinburgh (grant ST/N002512/1) and Queen’s University Belfast (grant ST/N002520/1) within the LSST:UK Science Consortium. 
ZTF is supported by National Science Foundation grant AST-1440341 and a collaboration including Caltech, IPAC, the Weizmann Institute for Science, the Oskar Klein Center at Stockholm University, the University of Maryland, the University of Washington, Deutsches Elektronen-Synchrotron and Humboldt University, Los Alamos National Laboratories, the TANGO Consortium of Taiwan, the University of Wisconsin at Milwaukee, and Lawrence Berkeley National Laboratories. Operations are conducted by COO, IPAC, and UW. This research has made use of ``Aladin sky atlas’‘ developed at CDS, Strasbourg Observatory, France 2000A\&AS..143…33B and 2014ASPC..485..277B.
This work has used data from the Asteroid Terrestrial Impact Last Alert System (ATLAS) project. The Asteroid Terrestrial Impact Last Alert System (ATLAS) project is primarily funded to search for near-earth asteroids through NASA grants NN12AR55G, 80NSSC18K0284, and 80NSSC18K1575; by-products of the NEO search include images and catalogs from the survey area. This work was partially funded by Kepler/K2 grant J1944/80NSSC19K0112 and HST GO-15889, and the STFC grants ST/T000198/1 and ST/S006109/1. The ATLAS science products have been made possible through the contributions of the University of Hawaii Institute for Astronomy, Queen's University Belfast, the Space Telescope Science Institute, the South African Astronomical Observatory, and The Millennium Institute of Astrophysics (MAS), Chile.
We acknowledge ESA Gaia, DPAC and the Photometric Science Alerts Team (http://gsaweb.ast.cam.ac.uk/alerts). H.Lin is supported by the National Natural Science Foundation of China (NSFC, Grant No. 12403061) and the innovative project of "Caiyun Post-doctoral Project" of Yunnan Province.
\end{acknowledgements}

\section*{Data Availability}
	
The data underlying this article will be shared on reasonable request with the corresponding author.

\bibliographystyle{aa} 
\bibliography{reference}

\begin{thebibliography}{124}
\expandafter\ifx\csname natexlab\endcsname\relax\def\natexlab#1{#1}\fi

\bibitem[{{Ahumada} {et~al.}(2020){Ahumada}, {Allende Prieto}, {Almeida}, {Anders}, {Anderson}, {Andrews}, {Anguiano}, {Arcodia}, {Armengaud}, {Aubert}, {Avila}, {Avila-Reese}, {Badenes}, {Balland}, {Barger}, {Barrera-Ballesteros}, {Basu}, {Bautista}, {Beaton}, {Beers}, {Benavides}, {Bender}, {Bernardi}, {Bershady}, {Beutler}, {Bidin}, {Bird}, {Bizyaev}, {Blanc}, {Blanton}, {Boquien}, {Borissova}, {Bovy}, {Brandt}, {Brinkmann}, {Brownstein}, {Bundy}, {Bureau}, {Burgasser}, {Burtin}, {Cano-D{\'\i}az}, {Capasso}, {Cappellari}, {Carrera}, {Chabanier}, {Chaplin}, {Chapman}, {Cherinka}, {Chiappini}, {Doohyun Choi}, {Chojnowski}, {Chung}, {Clerc}, {Coffey}, {Comerford}, {Comparat}, {da Costa}, {Cousinou}, {Covey}, {Crane}, {Cunha}, {Ilha}, {Dai}, {Damsted}, {Darling}, {Davidson}, {Davies}, {Dawson}, {De}, {de la Macorra}, {De Lee}, {Queiroz}, {Deconto Machado}, {de la Torre}, {Dell'Agli}, {du Mas des Bourboux}, {Diamond-Stanic}, {Dillon}, {Donor}, {Drory}, {Duckworth}, {Dwelly}, {Ebelke}, {Eftekharzadeh}, {Davis
  Eigenbrot}, {Elsworth}, {Eracleous}, {Erfanianfar}, {Escoffier}, {Fan}, {Farr}, {Fern{\'a}ndez-Trincado}, {Feuillet}, {Finoguenov}, {Fofie}, {Fraser-McKelvie}, {Frinchaboy}, {Fromenteau}, {Fu}, {Galbany}, {Garcia}, {Garc{\'\i}a-Hern{\'a}ndez}, {Garma Oehmichen}, {Ge}, {Geimba Maia}, {Geisler}, {Gelfand}, {Goddy}, {Gonzalez-Perez}, {Grabowski}, {Green}, {Grier}, {Guo}, {Guy}, {Harding}, {Hasselquist}, {Hawken}, {Hayes}, {Hearty}, {Hekker}, {Hogg}, {Holtzman}, {Horta}, {Hou}, {Hsieh}, {Huber}, {Hunt}, {Ider Chitham}, {Imig}, {Jaber}, {Jimenez Angel}, {Johnson}, {Jones}, {J{\"o}nsson}, {Jullo}, {Kim}, {Kinemuchi}, {Kirkpatrick}, {Kite}, {Klaene}, {Kneib}, {Kollmeier}, {Kong}, {Kounkel}, {Krishnarao}, {Lacerna}, {Lan}, {Lane}, {Law}, {Le Goff}, {Leung}, {Lewis}, {Li}, {Lian}, {Lin}, {Long}, {Longa-Pe{\~n}a}, {Lundgren}, {Lyke}, {Mackereth}, {MacLeod}, {Majewski}, {Manchado}, {Maraston}, {Martini}, {Masseron}, {Masters}, {Mathur}, {McDermid}, {Merloni}, {Merrifield}, {M{\'e}sz{\'a}ros}, {Miglio}, {Minniti},
  {Minsley}, {Miyaji}, {Mohammad}, {Mosser}, {Mueller}, {Muna}, {Mu{\~n}oz-Guti{\'e}rrez}, {Myers}, {Nadathur}, {Nair}, {Nandra}, {Correa do Nascimento}, {Nevin}, {Newman}, {Nidever}, {Nitschelm}, {Noterdaeme}, {O'Connell}, {Olmstead}, {Oravetz}, {Oravetz}, {Osorio}, {Pace}, {Padilla}, {Palanque-Delabrouille}, {Palicio}, {Pan}, {Pan}, {Parker}, {Paviot}, {Peirani}, {Ram{\'r}ez}, {Penny}, {Percival}, {Perez-Fournon}, {P{\'e}rez-R{\`a}fols}, {Petitjean}, {Pieri}, {Pinsonneault}, {Poovelil}, {Povick}, {Prakash}, {Price-Whelan}, {Raddick}, {Raichoor}, {Ray}, {Rembold}, {Rezaie}, {Riffel}, {Riffel}, {Rix}, {Robin}, {Roman-Lopes}, {Rom{\'a}n-Z{\'u}{\~n}iga}, {Rose}, {Ross}, {Rossi}, {Rowlands}, {Rubin}, {Salvato}, {S{\'a}nchez}, {S{\'a}nchez-Menguiano}, {S{\'a}nchez-Gallego}, {Sayres}, {Schaefer}, {Schiavon}, {Schimoia}, {Schlafly}, {Schlegel}, {Schneider}, {Schultheis}, {Schwope}, {Seo}, {Serenelli}, {Shafieloo}, {Shamsi}, {Shao}, {Shen}, {Shetrone}, {Shirley}, {Silva Aguirre}, {Simon}, {Skrutskie}, {Slosar},
  {Smethurst}, {Sobeck}, {Sodi}, {Souto}, {Stark}, {Stassun}, {Steinmetz}, {Stello}, {Stermer}, {Storchi-Bergmann}, {Streblyanska}, {Stringfellow}, {Stutz}, {Su{\'a}rez}, {Sun}, {Taghizadeh-Popp}, {Talbot}, {Tayar}, {Thakar}, {Theriault}, {Thomas}, {Thomas}, {Tinker}, {Tojeiro}, {Toledo}, {Tremonti}, {Troup}, {Tuttle}, {Unda-Sanzana}, {Valentini}, {Vargas-Gonz{\'a}lez}, {Vargas-Maga{\~n}a}, {V{\'a}zquez-Mata}, {Vivek}, {Wake}, {Wang}, {Weaver}, {Weijmans}, {Wild}, {Wilson}, {Wilson}, {Wolthuis}, {Wood-Vasey}, {Yan}, {Yang}, {Y{\`e}che}, {Zamora}, {Zarrouk}, {Zasowski}, {Zhang}, {Zhao}, {Zhao}, {Zheng}, {Zheng}, {Zhu}, \& {Zou}}]{Ahumada2020}
{Ahumada}, R., {Allende Prieto}, C., {Almeida}, A., {et~al.} 2020, \apjs, 249, 3

\bibitem[{{Ailawadhi} {et~al.}(2023){Ailawadhi}, {Dastidar}, {Misra}, {Roy}, {Hiramatsu}, {Howell}, {Brink}, {Zheng}, {Galbany}, {Shahbandeh}, {Arcavi}, {Ashall}, {Bostroem}, {Burke}, {Chapman}, {Dimple}, {Filippenko}, {Gangopadhyay}, {Ghosh}, {Hoffman}, {Hosseinzadeh}, {Jennings}, {Jha}, {Kumar}, {Karamehmetoglu}, {McCully}, {McGinness}, {M{\"u}ller-Bravo}, {Murakami}, {Pandey}, {Pellegrino}, {Piscarreta}, {Rho}, {Stritzinger}, {Sunseri}, {Van Dyk}, \& {Yadav}}]{Ailawadhi2023}
{Ailawadhi}, B., {Dastidar}, R., {Misra}, K., {et~al.} 2023, \mnras, 519, 248

\bibitem[{{Anderson} \& {James}(2009)}]{Anderson2009}
{Anderson}, J.~P. \& {James}, P.~A. 2009, \mnras, 399, 559

\bibitem[{{Andrews} {et~al.}(2022){Andrews}, {Pearson}, {Lundquist}, {Sand}, {Jencson}, {Bostroem}, {Hosseinzadeh}, {Valenti}, {Smith}, {Amaro}, {Dong}, {Janzen}, {Meza}, {Wyatt}, {Burke}, {Hiramatsu}, {Howell}, {McCully}, \& {Pellegrino}}]{Andrews2022}
{Andrews}, J.~E., {Pearson}, J., {Lundquist}, M.~J., {et~al.} 2022, \apj, 938, 19

\bibitem[{{Audcent-Ross} {et~al.}(2020){Audcent-Ross}, {Meurer}, {Audcent}, {Ryder}, {Wong}, {Phan}, {Williamson}, \& {Kim}}]{Audcent2020}
{Audcent-Ross}, F.~M., {Meurer}, G.~R., {Audcent}, J.~R., {et~al.} 2020, \mnras, 492, 848

\bibitem[{{Barna} {et~al.}(2021){Barna}, {Szalai}, {Jha}, {Camacho-Neves}, {Kwok}, {Foley}, {Kilpatrick}, {Coulter}, {Dimitriadis}, {Rest}, {Rojas-Bravo}, {Siebert}, {Brown}, {Burke}, {Padilla Gonzalez}, {Hiramatsu}, {Howell}, {McCully}, {Pellegrino}, {Dobson}, {Smartt}, {Swift}, {Stacey}, {Rahman}, {Sand}, {Andrews}, {Wyatt}, {Hsiao}, {Anderson}, {Chen}, {Della Valle}, {Galbany}, {Gromadzki}, {Inserra}, {Lyman}, {Magee}, {Maguire}, {M{\"u}ller-Bravo}, {Nicholl}, {Srivastav}, \& {Williams}}]{Barna2021}
{Barna}, B., {Szalai}, T., {Jha}, S.~W., {et~al.} 2021, \mnras, 501, 1078

\bibitem[{{Bellm} {et~al.}(2019){Bellm}, {Kulkarni}, {Graham}, {Dekany}, {Smith}, {Riddle}, {Masci}, {Helou}, {Prince}, {Adams}, {Barbarino}, {Barlow}, {Bauer}, {Beck}, {Belicki}, {Biswas}, {Blagorodnova}, {Bodewits}, {Bolin}, {Brinnel}, {Brooke}, {Bue}, {Bulla}, {Burruss}, {Cenko}, {Chang}, {Connolly}, {Coughlin}, {Cromer}, {Cunningham}, {De}, {Delacroix}, {Desai}, {Duev}, {Eadie}, {Farnham}, {Feeney}, {Feindt}, {Flynn}, {Franckowiak}, {Frederick}, {Fremling}, {Gal-Yam}, {Gezari}, {Giomi}, {Goldstein}, {Golkhou}, {Goobar}, {Groom}, {Hacopians}, {Hale}, {Henning}, {Ho}, {Hover}, {Howell}, {Hung}, {Huppenkothen}, {Imel}, {Ip}, {Ivezi{\'c}}, {Jackson}, {Jones}, {Juric}, {Kasliwal}, {Kaspi}, {Kaye}, {Kelley}, {Kowalski}, {Kramer}, {Kupfer}, {Landry}, {Laher}, {Lee}, {Lin}, {Lin}, {Lunnan}, {Giomi}, {Mahabal}, {Mao}, {Miller}, {Monkewitz}, {Murphy}, {Ngeow}, {Nordin}, {Nugent}, {Ofek}, {Patterson}, {Penprase}, {Porter}, {Rauch}, {Rebbapragada}, {Reiley}, {Rigault}, {Rodriguez}, {van Roestel}, {Rusholme}, {van
  Santen}, {Schulze}, {Shupe}, {Singer}, {Soumagnac}, {Stein}, {Surace}, {Sollerman}, {Szkody}, {Taddia}, {Terek}, {Van Sistine}, {van Velzen}, {Vestrand}, {Walters}, {Ward}, {Ye}, {Yu}, {Yan}, \& {Zolkower}}]{Bellm2019}
{Bellm}, E.~C., {Kulkarni}, S.~R., {Graham}, M.~J., {et~al.} 2019, \pasp, 131, 018002

\bibitem[{{Benetti} {et~al.}(2005){Benetti}, {Cappellaro}, {Mazzali}, {Turatto}, {Altavilla}, {Bufano}, {Elias-Rosa}, {Kotak}, {Pignata}, {Salvo}, \& {Stanishev}}]{Benetti2005}
{Benetti}, S., {Cappellaro}, E., {Mazzali}, P.~A., {et~al.} 2005, \apj, 623, 1011

\bibitem[{{Blondin} \& {Tonry}(2007)}]{Blondin2007}
{Blondin}, S. \& {Tonry}, J.~L. 2007, \apj, 666, 1024

\bibitem[{{Bouquin} {et~al.}(2018){Bouquin}, {Gil de Paz}, {Mu{\~n}oz-Mateos}, {Boissier}, {Sheth}, {Zaritsky}, {Peletier}, {Knapen}, \& {Gallego}}]{Bouquin2018}
{Bouquin}, A. Y.~K., {Gil de Paz}, A., {Mu{\~n}oz-Mateos}, J.~C., {et~al.} 2018, \apjs, 234, 18

\bibitem[{{Branch} {et~al.}(2009){Branch}, {Chau Dang}, \& {Baron}}]{Branch2009}
{Branch}, D., {Chau Dang}, L., \& {Baron}, E. 2009, \pasp, 121, 238

\bibitem[{{Branch} \& {van den Bergh}(1993)}]{Branch1993}
{Branch}, D. \& {van den Bergh}, S. 1993, \aj, 105, 2231

\bibitem[{{Brennan} {et~al.}(2024){Brennan}, {Sollerman}, {Irani}, {Schulze}, {Chen}, {Das}, {De}, {Fransson}, {Gal-Yam}, {Gkini}, {Hinds}, {Lunnan}, {Perley}, {Qin}, {Stein}, {Wise}, {Yan}, {Zimmerman}, {Anand}, {Bruch}, {Dekany}, {Drake}, {Fremling}, {Healy}, {Karambelkar}, {Kasliwal}, {Kong}, {Kulkarni}, {Masci}, {Post}, {Purdum}, {Rich}, \& {Wold}}]{Brennan2024}
{Brennan}, S.~J., {Sollerman}, J., {Irani}, I., {et~al.} 2024, \aap, 684, L18

\bibitem[{{Brown} {et~al.}(2013){Brown}, {Baliber}, {Bianco}, {Bowman}, {Burleson}, {Conway}, {Crellin}, {Depagne}, {De Vera}, {Dilday}, {Dragomir}, {Dubberley}, {Eastman}, {Elphick}, {Falarski}, {Foale}, {Ford}, {Fulton}, {Garza}, {Gomez}, {Graham}, {Greene}, {Haldeman}, {Hawkins}, {Haworth}, {Haynes}, {Hidas}, {Hjelstrom}, {Howell}, {Hygelund}, {Lister}, {Lobdill}, {Martinez}, {Mullins}, {Norbury}, {Parrent}, {Paulson}, {Petry}, {Pickles}, {Posner}, {Rosing}, {Ross}, {Sand}, {Saunders}, {Shobbrook}, {Shporer}, {Street}, {Thomas}, {Tsapras}, {Tufts}, {Valenti}, {Vander Horst}, {Walker}, {White}, \& {Willis}}]{Brown2013}
{Brown}, T.~M., {Baliber}, N., {Bianco}, F.~B., {et~al.} 2013, \pasp, 125, 1031

\bibitem[{{Byler} {et~al.}(2017){Byler}, {Dalcanton}, {Conroy}, \& {Johnson}}]{Byler2017}
{Byler}, N., {Dalcanton}, J.~J., {Conroy}, C., \& {Johnson}, B.~D. 2017, \apj, 840, 44

\bibitem[{{Cao} {et~al.}(2013){Cao}, {Kasliwal}, {Arcavi}, {Horesh}, {Hancock}, {Valenti}, {Cenko}, {Kulkarni}, {Gal-Yam}, {Gorbikov}, {Ofek}, {Sand}, {Yaron}, {Graham}, {Silverman}, {Wheeler}, {Marion}, {Walker}, {Mazzali}, {Howell}, {Li}, {Kong}, {Bloom}, {Nugent}, {Surace}, {Masci}, {Carpenter}, {Degenaar}, \& {Gelino}}]{Cao2013}
{Cao}, Y., {Kasliwal}, M.~M., {Arcavi}, I., {et~al.} 2013, \apjl, 775, L7

\bibitem[{{Chabrier}(2003)}]{Chabrier2003}
{Chabrier}, G. 2003, \pasp, 115, 763

\bibitem[{{Chen} {et~al.}(2022){Chen}, {Dong}, {Kochanek}, {Stanek}, {Post}, {Stritzinger}, {Prieto}, {Filippenko}, {Kollmeier}, {Elias-Rosa}, {Katz}, {Tomasella}, {Bose}, {Ashall}, {Benetti}, {Bersier}, {Brimacombe}, {Brink}, {Brown}, {Buckley}, {Cappellaro}, {Christie}, {Fraser}, {Gromadzki}, {Holoien}, {Hu}, {Kankare}, {Koff}, {Lundqvist}, {Mattila}, {Milne}, {Morrell}, {Mu{\~n}oz}, {Mutel}, {Natusch}, {Nicolas}, {Pastorello}, {Prentice}, {Roth}, {Shappee}, {Stone}, {Thompson}, {Villanueva}, \& {Zheng}}]{Chen2022}
{Chen}, P., {Dong}, S., {Kochanek}, C.~S., {et~al.} 2022, \apjs, 259, 53

\bibitem[{{Chen} {et~al.}(2023){Chen}, {Yan}, {Kangas}, {Lunnan}, {Schulze}, {Sollerman}, {Perley}, {Chen}, {Taggart}, {Hinds}, {Gal-Yam}, {Wang}, {Andreoni}, {Bellm}, {Bloom}, {Burdge}, {Burgos}, {Cook}, {Dahiwale}, {De}, {Dekany}, {Dugas}, {Frederik}, {Fremling}, {Graham}, {Hankins}, {Ho}, {Jencson}, {Karambelkar}, {Kasliwal}, {Kulkarni}, {Laher}, {Rusholme}, {Sharma}, {Taddia}, {Tartaglia}, {Thomas}, {Tzanidakis}, {Van Roestel}, {Walter}, {Yang}, {Yao}, \& {Yaron}}]{Chen2023a}
{Chen}, Z.~H., {Yan}, L., {Kangas}, T., {et~al.} 2023, \apj, 943, 41

\bibitem[{{Conroy} \& {Gunn}(2010)}]{Conroy2010}
{Conroy}, C. \& {Gunn}, J.~E. 2010, {FSPS: Flexible Stellar Population Synthesis}, Astrophysics Source Code Library, record ascl:1010.043

\bibitem[{{Conroy} {et~al.}(2009){Conroy}, {Gunn}, \& {White}}]{Conroy2009}
{Conroy}, C., {Gunn}, J.~E., \& {White}, M. 2009, \apj, 699, 486

\bibitem[{{Courtois} \& {Tully}(2012)}]{Courtois2012}
{Courtois}, H.~M. \& {Tully}, R.~B. 2012, \apj, 749, 174

\bibitem[{{D{\'a}lya} {et~al.}(2022){D{\'a}lya}, {D{\'\i}az}, {Bouchet}, {Frei}, {Jasche}, {Lavaux}, {Macas}, {Mukherjee}, {P{\'a}lfi}, {de Souza}, {Wandelt}, {Bilicki}, \& {Raffai}}]{Dlya2022}
{D{\'a}lya}, G., {D{\'\i}az}, R., {Bouchet}, F.~R., {et~al.} 2022, \mnras, 514, 1403

\bibitem[{{Desai} {et~al.}(2024){Desai}, {Kochanek}, {Shappee}, {Jayasinghe}, {Stanek}, {Holoien}, {Thompson}, {Ashall}, {Beacom}, {Do}, {Dong}, \& {Prieto}}]{Desai2024}
{Desai}, D.~D., {Kochanek}, C.~S., {Shappee}, B.~J., {et~al.} 2024, \mnras, 530, 5016

\bibitem[{{Falc{\'o}n-Barroso} {et~al.}(2011){Falc{\'o}n-Barroso}, {S{\'a}nchez-Bl{\'a}zquez}, {Vazdekis}, {Ricciardelli}, {Cardiel}, {Cenarro}, {Gorgas}, \& {Peletier}}]{Falcon2011}
{Falc{\'o}n-Barroso}, J., {S{\'a}nchez-Bl{\'a}zquez}, P., {Vazdekis}, A., {et~al.} 2011, \aap, 532, A95

\bibitem[{{Fan} {et~al.}(2015){Fan}, {Bai}, {Zhang}, {Wang}, {Chang}, {Xin}, \& {Zhang}}]{Fan2015}
{Fan}, Y.-F., {Bai}, J.-M., {Zhang}, J.-J., {et~al.} 2015, Research in Astronomy and Astrophysics, 15, 918

\bibitem[{{Fan} {et~al.}(2016){Fan}, {Wang}, {Jiang}, {Wu}, {Li}, {Huang}, {Xu}, {Hu}, {Zhu}, {Wang}, {Komossa}, \& {Zhang}}]{Fan2016}
{Fan}, Z., {Wang}, H., {Jiang}, X., {et~al.} 2016, \pasp, 128, 115005

\bibitem[{{Filippenko}(1997)}]{Filippenko1997}
{Filippenko}, A.~V. 1997, \araa, 35, 309

\bibitem[{{Filippenko}(2005)}]{Filippenko2005}
{Filippenko}, A.~V. 2005, in Astrophysics and Space Science Library, Vol. 332, White dwarfs: cosmological and galactic probes, ed. E.~M. {Sion}, S.~{Vennes}, \& H.~L. {Shipman}, 97--133

\bibitem[{{Filippenko} {et~al.}(1992{\natexlab{a}}){Filippenko}, {Richmond}, {Branch}, {Gaskell}, {Herbst}, {Ford}, {Treffers}, {Matheson}, {Ho}, {Dey}, {Sargent}, {Small}, \& {van Breugel}}]{Filippenko1992b}
{Filippenko}, A.~V., {Richmond}, M.~W., {Branch}, D., {et~al.} 1992{\natexlab{a}}, \aj, 104, 1543

\bibitem[{{Filippenko} {et~al.}(1992{\natexlab{b}}){Filippenko}, {Richmond}, {Matheson}, {Shields}, {Burbidge}, {Cohen}, {Dickinson}, {Malkan}, {Nelson}, {Pietz}, {Schlegel}, {Schmeer}, {Spinrad}, {Steidel}, {Tran}, \& {Wren}}]{Filippenko1992a}
{Filippenko}, A.~V., {Richmond}, M.~W., {Matheson}, T., {et~al.} 1992{\natexlab{b}}, \apjl, 384, L15

\bibitem[{{Folatelli} {et~al.}(2006){Folatelli}, {Contreras}, {Phillips}, {Woosley}, {Blinnikov}, {Morrell}, {Suntzeff}, {Lee}, {Hamuy}, {Gonz{\'a}lez}, {Krzeminski}, {Roth}, {Li}, {Filippenko}, {Foley}, {Freedman}, {Madore}, {Persson}, {Murphy}, {Boissier}, {Galaz}, {Gonz{\'a}lez}, {McCarthy}, {McWilliam}, \& {Pych}}]{Folatelli2006}
{Folatelli}, G., {Contreras}, C., {Phillips}, M.~M., {et~al.} 2006, \apj, 641, 1039

\bibitem[{{Foley} {et~al.}(2009){Foley}, {Challis}, {Groner}, {Silverman}, {Cenko}, {Filippenko}, \& {Li}}]{Foley2009}
{Foley}, R.~J., {Challis}, P., {Groner}, T., {et~al.} 2009, Central Bureau Electronic Telegrams, 1817, 2

\bibitem[{{Foley} {et~al.}(2013){Foley}, {Challis}, {Chornock}, {Ganeshalingam}, {Li}, {Marion}, {Morrell}, {Pignata}, {Stritzinger}, {Silverman}, {Wang}, {Anderson}, {Filippenko}, {Freedman}, {Hamuy}, {Jha}, {Kirshner}, {McCully}, {Persson}, {Phillips}, {Reichart}, \& {Soderberg}}]{Foley2013}
{Foley}, R.~J., {Challis}, P.~J., {Chornock}, R., {et~al.} 2013, \apj, 767, 57

\bibitem[{{Gal-Yam}(2012)}]{Gal-Yam2012}
{Gal-Yam}, A. 2012, Science, 337, 927

\bibitem[{{Gal-Yam} {et~al.}(2022){Gal-Yam}, {Bruch}, {Schulze}, {Yang}, {Perley}, {Irani}, {Sollerman}, {Kool}, {Soumagnac}, {Yaron}, {Strotjohann}, {Zimmerman}, {Barbarino}, {Kulkarni}, {Kasliwal}, {De}, {Yao}, {Fremling}, {Yan}, {Ofek}, {Fransson}, {Filippenko}, {Zheng}, {Brink}, {Copperwheat}, {Foley}, {Brown}, {Siebert}, {Leloudas}, {Cabrera-Lavers}, {Garcia-Alvarez}, {Marante-Barreto}, {Frederick}, {Hung}, {Wheeler}, {Vink{\'o}}, {Thomas}, {Graham}, {Duev}, {Drake}, {Dekany}, {Bellm}, {Rusholme}, {Shupe}, {Andreoni}, {Sharma}, {Riddle}, {van Roestel}, \& {Knezevic}}]{Gal-Yam2022}
{Gal-Yam}, A., {Bruch}, R., {Schulze}, S., {et~al.} 2022, \nat, 601, 201

\bibitem[{{Ganeshalingam} {et~al.}(2012){Ganeshalingam}, {Li}, {Filippenko}, {Silverman}, {Chornock}, {Foley}, {Matheson}, {Kirshner}, {Milne}, {Calkins}, \& {Shen}}]{Ganeshalingam2012}
{Ganeshalingam}, M., {Li}, W., {Filippenko}, A.~V., {et~al.} 2012, \apj, 751, 142

\bibitem[{{Giraud}(1987)}]{Giraud1987}
{Giraud}, E. 1987, \aap, 174, 23

\bibitem[{{Guti{\'e}rrez} {et~al.}(2020){Guti{\'e}rrez}, {Pastorello}, {Jerkstrand}, {Galbany}, {Sullivan}, {Anderson}, {Taubenberger}, {Kuncarayakti}, {Gonz{\'a}lez-Gait{\'a}n}, {Wiseman}, {Inserra}, {Fraser}, {Maguire}, {Smartt}, {M{\"u}ller-Bravo}, {Arcavi}, {Benetti}, {Bersier}, {Bose}, {Bostroem}, {Burke}, {Chen}, {Chen}, {Della Valle}, {Dong}, {Gal-Yam}, {Gromadzki}, {Hiramatsu}, {Holoien}, {Hosseinzadeh}, {Howell}, {Kankare}, {Kochanek}, {McCully}, {Nicholl}, {Pignata}, {Prieto}, {Shappee}, {Taggart}, {Tomasella}, {Valenti}, \& {Young}}]{Gutierrez2020}
{Guti{\'e}rrez}, C.~P., {Pastorello}, A., {Jerkstrand}, A., {et~al.} 2020, \mnras, 499, 974

\bibitem[{{Henry} \& {Worthey}(1999)}]{Henry1999}
{Henry}, R.~B.~C. \& {Worthey}, G. 1999, \pasp, 111, 919

\bibitem[{{Hodgkin} {et~al.}(2021){Hodgkin}, {Harrison}, {Breedt}, {Wevers}, {Rixon}, {Delgado}, {Yoldas}, {Kostrzewa-Rutkowska}, {Wyrzykowski}, {van Leeuwen}, {Blagorodnova}, {Campbell}, {Eappachen}, {Fraser}, {Ihanec}, {Koposov}, {Kruszy{\'n}ska}, {Marton}, {Rybicki}, {Brown}, {Burgess}, {Busso}, {Cowell}, {De Angeli}, {Diener}, {Evans}, {Gilmore}, {Holland}, {Jonker}, {van Leeuwen}, {Mignard}, {Osborne}, {Portell}, {Prusti}, {Richards}, {Riello}, {Seabroke}, {Walton}, {{\'A}brah{\'a}m}, {Altavilla}, {Baker}, {Bastian}, {O'Brien}, {de Bruijne}, {Butterley}, {Carrasco}, {Casta{\~n}eda}, {Clark}, {Clementini}, {Copperwheat}, {Cropper}, {Damljanovic}, {Davidson}, {Davis}, {Dennefeld}, {Dhillon}, {Dolding}, {Dominik}, {Esquej}, {Eyer}, {Fabricius}, {Fridman}, {Froebrich}, {Garralda}, {Gomboc}, {Gonz{\'a}lez-Vidal}, {Guerra}, {Hambly}, {Hardy}, {Holl}, {Hourihane}, {Japelj}, {Kann}, {Kiss}, {Knigge}, {Kolb}, {Komossa}, {K{\'o}sp{\'a}l}, {Kov{\'a}cs}, {Kun}, {Leto}, {Lewis}, {Littlefair}, {Mahabal}, {Mundell},
  {Nagy}, {Padeletti}, {Palaversa}, {Pigulski}, {Pretorius}, {van Reeven}, {Ribeiro}, {Roelens}, {Rowell}, {Schartel}, {Scholz}, {Schwope}, {Sip{\H{o}}cz}, {Smartt}, {Smith}, {Serraller}, {Steeghs}, {Sullivan}, {Szabados}, {Szegedi-Elek}, {Tisserand}, {Tomasella}, {van Velzen}, {Whitelock}, {Wilson}, \& {Young}}]{Hodgkin2021}
{Hodgkin}, S.~T., {Harrison}, D.~L., {Breedt}, E., {et~al.} 2021, \aap, 652, A76

\bibitem[{{Hosseinzadeh} {et~al.}(2017){Hosseinzadeh}, {Arcavi}, {Valenti}, {McCully}, {Howell}, {Johansson}, {Sollerman}, {Pastorello}, {Benetti}, {Cao}, {Cenko}, {Clubb}, {Corsi}, {Duggan}, {Elias-Rosa}, {Filippenko}, {Fox}, {Fremling}, {Horesh}, {Karamehmetoglu}, {Kasliwal}, {Marion}, {Ofek}, {Sand}, {Taddia}, {Zheng}, {Fraser}, {Gal-Yam}, {Inserra}, {Laher}, {Masci}, {Rebbapragada}, {Smartt}, {Smith}, {Sullivan}, {Surace}, \& {Wo{\'z}niak}}]{Hosseinzadeh2017}
{Hosseinzadeh}, G., {Arcavi}, I., {Valenti}, S., {et~al.} 2017, \apj, 836, 158

\bibitem[{{Howell}(2019)}]{Howell2019}
{Howell}, D. 2019, in American Astronomical Society Meeting Abstracts, Vol. 233, American Astronomical Society Meeting Abstracts \#233, 258.16

\bibitem[{{Howell} {et~al.}(2006){Howell}, {Sullivan}, {Nugent}, {Ellis}, {Conley}, {Le Borgne}, {Carlberg}, {Guy}, {Balam}, {Basa}, {Fouchez}, {Hook}, {Hsiao}, {Neill}, {Pain}, {Perrett}, \& {Pritchet}}]{Howell2006}
{Howell}, D.~A., {Sullivan}, M., {Nugent}, P.~E., {et~al.} 2006, \nat, 443, 308

\bibitem[{{Hsiao} {et~al.}(2020){Hsiao}, {Hoeflich}, {Ashall}, {Lu}, {Contreras}, {Burns}, {Phillips}, {Galbany}, {Anderson}, {Baltay}, {Baron}, {Castell{\'o}n}, {Davis}, {Freedman}, {Gall}, {Gonzalez}, {Graham}, {Hamuy}, {Holoien}, {Karamehmetoglu}, {Krisciunas}, {Kumar}, {Kuncarayakti}, {Morrell}, {Moriya}, {Nugent}, {Perlmutter}, {Persson}, {Piro}, {Rabinowitz}, {Roth}, {Shahbandeh}, {Shappee}, {Stritzinger}, {Suntzeff}, {Taddia}, \& {Uddin}}]{Hsiao2020}
{Hsiao}, E.~Y., {Hoeflich}, P., {Ashall}, C., {et~al.} 2020, \apj, 900, 140

\bibitem[{{Huang} {et~al.}(2018){Huang}, {Wang}, {Hosseinzadeh}, {Brown}, {Mo}, {Zhang}, {Zhang}, {Zhang}, {Howell}, {Arcavi}, {McCully}, {Valenti}, {Rui}, {Song}, {Xiang}, {Li}, {Lin}, \& {Wang}}]{Huang2018}
{Huang}, F., {Wang}, X.~F., {Hosseinzadeh}, G., {et~al.} 2018, \mnras, 475, 3959

\bibitem[{{Iben} \& {Tutukov}(1984)}]{Iben1984}
{Iben}, I., J. \& {Tutukov}, A.~V. 1984, \apj, 284, 719

\bibitem[{{Jencson} {et~al.}(2019){Jencson}, {Kasliwal}, {Adams}, {Bond}, {De}, {Johansson}, {Karambelkar}, {Lau}, {Tinyanont}, {Ryder}, {Cody}, {Masci}, {Bally}, {Blagorodnova}, {Castell{\'o}n}, {Fremling}, {Gehrz}, {Helou}, {Kilpatrick}, {Milne}, {Morrell}, {Perley}, {Phillips}, {Smith}, {van Dyk}, \& {Williams}}]{Jencson2019}
{Jencson}, J.~E., {Kasliwal}, M.~M., {Adams}, S.~M., {et~al.} 2019, \apj, 886, 40

\bibitem[{{Johnson} {et~al.}(2021){Johnson}, {Leja}, {Conroy}, \& {Speagle}}]{Johnson2021}
{Johnson}, B.~D., {Leja}, J., {Conroy}, C., \& {Speagle}, J.~S. 2021, \apjs, 254, 22

\bibitem[{{Karachentsev} {et~al.}(2006){Karachentsev}, {Kudrya}, {Karachentseva}, \& {Mitronova}}]{Karachentsev2006}
{Karachentsev}, I.~D., {Kudrya}, Y.~N., {Karachentseva}, V.~E., \& {Mitronova}, S.~N. 2006, Astrophysics, 49, 450

\bibitem[{{Karambelkar} {et~al.}(2021){Karambelkar}, {Kasliwal}, {Maguire}, {Anand}, {Andreoni}, {De}, {Drake}, {Duev}, {Graham}, {Kool}, {Laher}, {Magee}, {Mahabal}, {Medford}, {Perley}, {Rigault}, {Rusholme}, {Schulze}, {Sharma}, {Sollerman}, {Tzanidakis}, {Walters}, \& {Yao}}]{Karambelkar2021}
{Karambelkar}, V.~R., {Kasliwal}, M.~M., {Maguire}, K., {et~al.} 2021, \apjl, 921, L6

\bibitem[{{Kelly} {et~al.}(2014){Kelly}, {Filippenko}, {Modjaz}, \& {Kocevski}}]{Kelly2014}
{Kelly}, P.~L., {Filippenko}, A.~V., {Modjaz}, M., \& {Kocevski}, D. 2014, \apj, 789, 23

\bibitem[{{Kelly} \& {Kirshner}(2012)}]{Kelly2012}
{Kelly}, P.~L. \& {Kirshner}, R.~P. 2012, \apj, 759, 107

\bibitem[{{Kilpatrick} {et~al.}(2021){Kilpatrick}, {Drout}, {Auchettl}, {Dimitriadis}, {Foley}, {Jones}, {DeMarchi}, {French}, {Gall}, {Hjorth}, {Jacobson-Gal{\'a}n}, {Margutti}, {Piro}, {Ramirez-Ruiz}, {Rest}, \& {Rojas-Bravo}}]{Kilpatrick2021}
{Kilpatrick}, C.~D., {Drout}, M.~R., {Auchettl}, K., {et~al.} 2021, \mnras, 504, 2073

\bibitem[{{Kochanek} {et~al.}(2017){Kochanek}, {Shappee}, {Stanek}, {Holoien}, {Thompson}, {Prieto}, {Dong}, {Shields}, {Will}, {Britt}, {Perzanowski}, \& {Pojma{\'n}ski}}]{Kochanek2017}
{Kochanek}, C.~S., {Shappee}, B.~J., {Stanek}, K.~Z., {et~al.} 2017, \pasp, 129, 104502

\bibitem[{{Kriek} \& {Conroy}(2013)}]{Kriek2013}
{Kriek}, M. \& {Conroy}, C. 2013, \apjl, 775, L16

\bibitem[{{Kwok} {et~al.}(2024){Kwok}, {Siebert}, {Johansson}, {Jha}, {Blondin}, {Dessart}, {Foley}, {Hillier}, {Larison}, {Pakmor}, {Temim}, {Andrews}, {Auchettl}, {Badenes}, {Barnabas}, {Bostroem}, {Brenner Newman}, {Brink}, {Bustamante-Rosell}, {Camacho-Neves}, {Clocchiatti}, {Coulter}, {Davis}, {Deckers}, {Dimitriadis}, {Dong}, {Farah}, {Filippenko}, {Fl{\"o}rs}, {Fox}, {Garnavich}, {Padilla Gonzalez}, {Graur}, {Hambsch}, {Hosseinzadeh}, {Howell}, {Hughes}, {Kerzendorf}, {Saux}, {Maeda}, {Maguire}, {McCully}, {Mihalenko}, {Newsome}, {O'Brien}, {Pearson}, {Pellegrino}, {Pierel}, {Polin}, {Rest}, {Rojas-Bravo}, {Sand}, {Schwab}, {Shahbandeh}, {Shrestha}, {Smith}, {Strolger}, {Szalai}, {Taggart}, {Terreran}, {Terwel}, {Tinyanont}, {Valenti}, {Vink{\'o}}, {Wheeler}, {Yang}, {Zheng}, {Ashall}, {DerKacy}, {Galbany}, {Hoeflich}, {de Jaeger}, {Lu}, {Maund}, {Medler}, {Morell}, {Shappee}, {Stritzinger}, {Suntzeff}, {Tucker}, \& {Wang}}]{Kwok2024}
{Kwok}, L.~A., {Siebert}, M.~R., {Johansson}, J., {et~al.} 2024, \apj, 966, 135

\bibitem[{{Lagattuta} {et~al.}(2013){Lagattuta}, {Mould}, {Staveley-Smith}, {Hong}, {Springob}, {Masters}, {Koribalski}, \& {Jones}}]{Lagattuta2013}
{Lagattuta}, D.~J., {Mould}, J.~R., {Staveley-Smith}, L., {et~al.} 2013, \apj, 771, 88

\bibitem[{{Leaman} {et~al.}(2011){Leaman}, {Li}, {Chornock}, \& {Filippenko}}]{Leaman2011}
{Leaman}, J., {Li}, W., {Chornock}, R., \& {Filippenko}, A.~V. 2011, \mnras, 412, 1419

\bibitem[{{Leibundgut} {et~al.}(1993){Leibundgut}, {Kirshner}, {Phillips}, {Wells}, {Suntzeff}, {Hamuy}, {Schommer}, {Walker}, {Gonzalez}, {Ugarte}, {Williams}, {Williger}, {Gomez}, {Marzke}, {Schmidt}, {Whitney}, {Caldwell}, {Peters}, {Chaffee}, {Foltz}, {Rehner}, {Siciliano}, {Barnes}, {Cheng}, {Hintzen}, {Kim}, {Maza}, {Parker}, {Porter}, {Schmidtke}, \& {Sonneborn}}]{Leibundgut1993}
{Leibundgut}, B., {Kirshner}, R.~P., {Phillips}, M.~M., {et~al.} 1993, \aj, 105, 301

\bibitem[{{Leja} {et~al.}(2017){Leja}, {Johnson}, {Conroy}, {van Dokkum}, \& {Byler}}]{Leja2017}
{Leja}, J., {Johnson}, B.~D., {Conroy}, C., {van Dokkum}, P.~G., \& {Byler}, N. 2017, \apj, 837, 170

\bibitem[{{Li} {et~al.}(2011{\natexlab{a}}){Li}, {Chornock}, {Leaman}, {Filippenko}, {Poznanski}, {Wang}, {Ganeshalingam}, \& {Mannucci}}]{Li2011b}
{Li}, W., {Chornock}, R., {Leaman}, J., {et~al.} 2011{\natexlab{a}}, \mnras, 412, 1473

\bibitem[{{Li} {et~al.}(2003){Li}, {Filippenko}, {Chornock}, {Berger}, {Berlind}, {Calkins}, {Challis}, {Fassnacht}, {Jha}, {Kirshner}, {Matheson}, {Sargent}, {Simcoe}, {Smith}, \& {Squires}}]{Li2003}
{Li}, W., {Filippenko}, A.~V., {Chornock}, R., {et~al.} 2003, \pasp, 115, 453

\bibitem[{{Li} {et~al.}(2011{\natexlab{b}}){Li}, {Leaman}, {Chornock}, {Filippenko}, {Poznanski}, {Ganeshalingam}, {Wang}, {Modjaz}, {Jha}, {Foley}, \& {Smith}}]{Li2011}
{Li}, W., {Leaman}, J., {Chornock}, R., {et~al.} 2011{\natexlab{b}}, \mnras, 412, 1441

\bibitem[{{Makarov} {et~al.}(2014){Makarov}, {Prugniel}, {Terekhova}, {Courtois}, \& {Vauglin}}]{Makarov2014}
{Makarov}, D., {Prugniel}, P., {Terekhova}, N., {Courtois}, H., \& {Vauglin}, I. 2014, \aap, 570, A13

\bibitem[{{Maoz} {et~al.}(2014){Maoz}, {Mannucci}, \& {Nelemans}}]{Maoz2014}
{Maoz}, D., {Mannucci}, F., \& {Nelemans}, G. 2014, \araa, 52, 107

\bibitem[{{Masci} {et~al.}(2019){Masci}, {Laher}, {Rusholme}, {Shupe}, {Groom}, {Surace}, {Jackson}, {Monkewitz}, {Beck}, {Flynn}, {Terek}, {Landry}, {Hacopians}, {Desai}, {Howell}, {Brooke}, {Imel}, {Wachter}, {Ye}, {Lin}, {Cenko}, {Cunningham}, {Rebbapragada}, {Bue}, {Miller}, {Mahabal}, {Bellm}, {Patterson}, {Juri{\'c}}, {Golkhou}, {Ofek}, {Walters}, {Graham}, {Kasliwal}, {Dekany}, {Kupfer}, {Burdge}, {Cannella}, {Barlow}, {Van Sistine}, {Giomi}, {Fremling}, {Blagorodnova}, {Levitan}, {Riddle}, {Smith}, {Helou}, {Prince}, \& {Kulkarni}}]{Masci2019}
{Masci}, F.~J., {Laher}, R.~R., {Rusholme}, B., {et~al.} 2019, \pasp, 131, 018003

\bibitem[{{Matheson} {et~al.}(2000){Matheson}, {Filippenko}, {Chornock}, {Leonard}, \& {Li}}]{Matheson2000}
{Matheson}, T., {Filippenko}, A.~V., {Chornock}, R., {Leonard}, D.~C., \& {Li}, W. 2000, \aj, 119, 2303

\bibitem[{{Mattila} {et~al.}(2012){Mattila}, {Dahlen}, {Efstathiou}, {Kankare}, {Melinder}, {Alonso-Herrero}, {P{\'e}rez-Torres}, {Ryder}, {V{\"a}is{\"a}nen}, \& {{\"O}stlin}}]{Mattila2012}
{Mattila}, S., {Dahlen}, T., {Efstathiou}, A., {et~al.} 2012, \apj, 756, 111

\bibitem[{{Medler} {et~al.}(2022){Medler}, {Mazzali}, {Teffs}, {Ashall}, {Anderson}, {Arcavi}, {Benetti}, {Bostroem}, {Burke}, {Cai}, {Charalampopoulos}, {Elias-Rosa}, {Ergon}, {Galbany}, {Gromadzki}, {Hiramatsu}, {Howell}, {Inserra}, {Lundqvist}, {McCully}, {M{\"u}ller-Bravo}, {Newsome}, {Nicholl}, {Padilla Gonzalez}, {Paraskeva}, {Pastorello}, {Pellegrino}, {Pessi}, {Reguitti}, {Reynolds}, {Roy}, {Terreran}, {Tomasella}, \& {Young}}]{Medler2022}
{Medler}, K., {Mazzali}, P.~A., {Teffs}, J., {et~al.} 2022, \mnras, 513, 5540

\bibitem[{{Mould} {et~al.}(2000){Mould}, {Huchra}, {Freedman}, {Kennicutt}, {Ferrarese}, {Ford}, {Gibson}, {Graham}, {Hughes}, {Illingworth}, {Kelson}, {Macri}, {Madore}, {Sakai}, {Sebo}, {Silbermann}, \& {Stetson}}]{Mould2000}
{Mould}, J.~R., {Huchra}, J.~P., {Freedman}, W.~L., {et~al.} 2000, \apj, 529, 786

\bibitem[{{Nakaoka} {et~al.}(2018){Nakaoka}, {Kawabata}, {Maeda}, {Tanaka}, {Yamanaka}, {Moriya}, {Tominaga}, {Morokuma}, {Takaki}, {Kawabata}, {Kawahara}, {Itoh}, {Shiki}, {Mori}, {Hirochi}, {Abe}, {Uemura}, {Yoshida}, {Akitaya}, {Moritani}, {Ueno}, {Urano}, {Isogai}, {Hanayama}, \& {Nagayama}}]{Nakaoka2018}
{Nakaoka}, T., {Kawabata}, K.~S., {Maeda}, K., {et~al.} 2018, \apj, 859, 78

\bibitem[{{Nasonova} {et~al.}(2011){Nasonova}, {de Freitas Pacheco}, \& {Karachentsev}}]{Nasonova2011}
{Nasonova}, O.~G., {de Freitas Pacheco}, J.~A., \& {Karachentsev}, I.~D. 2011, \aap, 532, A104

\bibitem[{{Nomoto}(1982)}]{Nomoto1982}
{Nomoto}, K. 1982, \apj, 253, 798

\bibitem[{{Pan}(2020)}]{Pan2020}
{Pan}, Y.-C. 2020, \apjl, 895, L5

\bibitem[{{Pastorello} {et~al.}(2008){Pastorello}, {Mattila}, {Zampieri}, {Della Valle}, {Smartt}, {Valenti}, {Agnoletto}, {Benetti}, {Benn}, {Branch}, {Cappellaro}, {Dennefeld}, {Eldridge}, {Gal-Yam}, {Harutyunyan}, {Hunter}, {Kjeldsen}, {Lipkin}, {Mazzali}, {Milne}, {Navasardyan}, {Ofek}, {Pian}, {Shemmer}, {Spiro}, {Stathakis}, {Taubenberger}, {Turatto}, \& {Yamaoka}}]{Pastorello2008}
{Pastorello}, A., {Mattila}, S., {Zampieri}, L., {et~al.} 2008, \mnras, 389, 113

\bibitem[{{Paxton} {et~al.}(2018){Paxton}, {Schwab}, {Bauer}, {Bildsten}, {Blinnikov}, {Duffell}, {Farmer}, {Goldberg}, {Marchant}, {Sorokina}, {Thoul}, {Townsend}, \& {Timmes}}]{Paxton2018}
{Paxton}, B., {Schwab}, J., {Bauer}, E.~B., {et~al.} 2018, \apjs, 234, 34

\bibitem[{{Perley} {et~al.}(2022){Perley}, {Sollerman}, {Schulze}, {Yao}, {Fremling}, {Gal-Yam}, {Ho}, {Yang}, {Kool}, {Irani}, {Yan}, {Andreoni}, {Baade}, {Bellm}, {Brink}, {Chen}, {Cikota}, {Coughlin}, {Dahiwale}, {Dekany}, {Duev}, {Filippenko}, {Hoeflich}, {Kasliwal}, {Kulkarni}, {Lunnan}, {Masci}, {Maund}, {Medford}, {Riddle}, {Rosnet}, {Shupe}, {Strotjohann}, {Tzanidakis}, \& {Zheng}}]{Perley2022}
{Perley}, D.~A., {Sollerman}, J., {Schulze}, S., {et~al.} 2022, \apj, 927, 180

\bibitem[{{Perlmutter} {et~al.}(1999){Perlmutter}, {Aldering}, {Goldhaber}, {Knop}, {Nugent}, {Castro}, {Deustua}, {Fabbro}, {Goobar}, {Groom}, {Hook}, {Kim}, {Kim}, {Lee}, {Nunes}, {Pain}, {Pennypacker}, {Quimby}, {Lidman}, {Ellis}, {Irwin}, {McMahon}, {Ruiz-Lapuente}, {Walton}, {Schaefer}, {Boyle}, {Filippenko}, {Matheson}, {Fruchter}, {Panagia}, {Newberg}, {Couch}, \& {Project}}]{Perlmutter1999}
{Perlmutter}, S., {Aldering}, G., {Goldhaber}, G., {et~al.} 1999, \apj, 517, 565

\bibitem[{{Phillips} {et~al.}(1992){Phillips}, {Wells}, {Suntzeff}, {Hamuy}, {Leibundgut}, {Kirshner}, \& {Foltz}}]{Phillips1992}
{Phillips}, M.~M., {Wells}, L.~A., {Suntzeff}, N.~B., {et~al.} 1992, \aj, 103, 1632

\bibitem[{{Quimby} {et~al.}(2007){Quimby}, {Aldering}, {Wheeler}, {H{\"o}flich}, {Akerlof}, \& {Rykoff}}]{Quimby2007}
{Quimby}, R.~M., {Aldering}, G., {Wheeler}, J.~C., {et~al.} 2007, \apjl, 668, L99

\bibitem[{{Reguitti} {et~al.}(2022){Reguitti}, {Pastorello}, {Pignata}, {Fraser}, {Stritzinger}, {Brennan}, {Cai}, {Elias-Rosa}, {Fugazza}, {Gutierrez}, {Kankare}, {Kotak}, {Lundqvist}, {Mazzali}, {Moran}, {Salmaso}, {Tomasella}, {Valerin}, \& {Kuncarayakti}}]{Reguitti2022}
{Reguitti}, A., {Pastorello}, A., {Pignata}, G., {et~al.} 2022, \aap, 662, L10

\bibitem[{{Richardson} {et~al.}(2014){Richardson}, {Jenkins}, {Wright}, \& {Maddox}}]{Richardson2014}
{Richardson}, D., {Jenkins}, Robert~L., I., {Wright}, J., \& {Maddox}, L. 2014, \aj, 147, 118

\bibitem[{{Riess} {et~al.}(1998){Riess}, {Filippenko}, {Challis}, {Clocchiatti}, {Diercks}, {Garnavich}, {Gilliland}, {Hogan}, {Jha}, {Kirshner}, {Leibundgut}, {Phillips}, {Reiss}, {Schmidt}, {Schommer}, {Smith}, {Spyromilio}, {Stubbs}, {Suntzeff}, \& {Tonry}}]{Riess1998}
{Riess}, A.~G., {Filippenko}, A.~V., {Challis}, P., {et~al.} 1998, \aj, 116, 1009

\bibitem[{{Schlafly} \& {Finkbeiner}(2011)}]{Schlafly2011}
{Schlafly}, E.~F. \& {Finkbeiner}, D.~P. 2011, \apj, 737, 103

\bibitem[{{Schlegel}(1990)}]{Schlegel1990}
{Schlegel}, E.~M. 1990, \mnras, 244, 269

\bibitem[{{Schulze} {et~al.}(2021){Schulze}, {Yaron}, {Sollerman}, {Leloudas}, {Gal}, {Wright}, {Lunnan}, {Gal-Yam}, {Ofek}, {Perley}, {Filippenko}, {Kasliwal}, {Kulkarni}, {Neill}, {Nugent}, {Quimby}, {Sullivan}, {Strotjohann}, {Arcavi}, {Ben-Ami}, {Bianco}, {Bloom}, {De}, {Fraser}, {Fremling}, {Horesh}, {Johansson}, {Kelly}, {Kne{\v{z}}evi{\'c}}, {Kne{\v{z}}evi{\'c}}, {Maguire}, {Nyholm}, {Papadogiannakis}, {Petrushevska}, {Rubin}, {Yan}, {Yang}, {Adams}, {Bufano}, {Clubb}, {Foley}, {Green}, {Harmanen}, {Ho}, {Hook}, {Hosseinzadeh}, {Howell}, {Kong}, {Kotak}, {Matheson}, {McCully}, {Milisavljevic}, {Pan}, {Poznanski}, {Shivvers}, {van Velzen}, \& {Verbeek}}]{Schulze2021}
{Schulze}, S., {Yaron}, O., {Sollerman}, J., {et~al.} 2021, \apjs, 255, 29

\bibitem[{{Shappee} {et~al.}(2014){Shappee}, {Prieto}, {Stanek}, {Kochanek}, {Holoien}, {Jencson}, {Basu}, {Beacom}, {Szczygiel}, {Pojmanski}, {Brimacombe}, {Dubberley}, {Elphick}, {Foale}, {Hawkins}, {Mullins}, {Rosing}, {Ross}, \& {Walker}}]{Shappee2014}
{Shappee}, B., {Prieto}, J., {Stanek}, K.~Z., {et~al.} 2014, in American Astronomical Society Meeting Abstracts, Vol. 223, American Astronomical Society Meeting Abstracts \#223, 236.03

\bibitem[{{Shingles} {et~al.}(2021){Shingles}, {Smith}, {Young}, {Smartt}, {Tonry}, {Denneau}, {Heinze}, {Weiland}, {Flewelling}, {Stalder}, {Clocchiatti}, {F{\"o}rster}, {Pignata}, {Rest}, {Anderson}, {Stubbs}, \& {Erasmus}}]{Shingles2021}
{Shingles}, L., {Smith}, K.~W., {Young}, D.~R., {et~al.} 2021, Transient Name Server AstroNote, 7, 1

\bibitem[{{Shporer} {et~al.}(2011){Shporer}, {Brown}, {Lister}, {Street}, {Tsapras}, {Bianco}, {Fulton}, \& {Howell}}]{Shporer2011}
{Shporer}, A., {Brown}, T., {Lister}, T., {et~al.} 2011, in The Astrophysics of Planetary Systems: Formation, Structure, and Dynamical Evolution, ed. A.~{Sozzetti}, M.~G. {Lattanzi}, \& A.~P. {Boss}, Vol. 276, 553--555

\bibitem[{{Siebert} {et~al.}(2024){Siebert}, {Kwok}, {Johansson}, {Jha}, {Blondin}, {Dessart}, {Foley}, {Hillier}, {Larison}, {Pakmor}, {Temim}, {Andrews}, {Auchettl}, {Badenes}, {Barna}, {Bostroem}, {Brenner Newman}, {Brink}, {Bustamante-Rosell}, {Camacho-Neves}, {Clocchiatti}, {Coulter}, {Davis}, {Deckers}, {Dimitriadis}, {Dong}, {Farah}, {Filippenko}, {Fl{\"o}rs}, {Fox}, {Garnavich}, {Padilla Gonzalez}, {Graur}, {Hambsch}, {Hosseinzadeh}, {Howell}, {Hughes}, {Kerzendorf}, {Le Saux}, {Maeda}, {Maguire}, {McCully}, {Mihalenko}, {Newsome}, {O'Brien}, {Pearson}, {Pellegrino}, {Pierel}, {Polin}, {Rest}, {Rojas-Bravo}, {Sand}, {Schwab}, {Shahbandeh}, {Shrestha}, {Smith}, {Strolger}, {Szalai}, {Taggart}, {Terreran}, {Terwel}, {Tinyanont}, {Valenti}, {Vink{\'o}}, {Wheeler}, {Yang}, {Zheng}, {Ashall}, {DerKacy}, {Galbany}, {Hoeflich}, {Hsiao}, {de Jaeger}, {Lu}, {Maund}, {Medler}, {Morrell}, {Shappee}, {Stritzinger}, {Suntzeff}, {Tucker}, \& {Wang}}]{Siebert2024}
{Siebert}, M.~R., {Kwok}, L.~A., {Johansson}, J., {et~al.} 2024, \apj, 960, 88

\bibitem[{{Singh} {et~al.}(2019){Singh}, {Sahu}, {Anupama}, {Kumar}, {Kumar}, {Yamanaka}, {Baklanov}, {Tominaga}, {Blinnikov}, {Maeda}, {Dutta}, {Bhalerao}, {Anche}, {Barway}, {Akitaya}, {Nakaoka}, {Kawabata}, {Kawabata}, {Sasada}, {Takagi}, {Maehara}, {Isogai}, {Kino}, {Taguchi}, \& {Nagao}}]{Singh2019}
{Singh}, A., {Sahu}, D.~K., {Anupama}, G.~C., {et~al.} 2019, \apjl, 882, L15

\bibitem[{{Skrutskie} {et~al.}(2006){Skrutskie}, {Cutri}, {Stiening}, {Weinberg}, {Schneider}, {Carpenter}, {Beichman}, {Capps}, {Chester}, {Elias}, {Huchra}, {Liebert}, {Lonsdale}, {Monet}, {Price}, {Seitzer}, {Jarrett}, {Kirkpatrick}, {Gizis}, {Howard}, {Evans}, {Fowler}, {Fullmer}, {Hurt}, {Light}, {Kopan}, {Marsh}, {McCallon}, {Tam}, {Van Dyk}, \& {Wheelock}}]{Skrutskie2006}
{Skrutskie}, M.~F., {Cutri}, R.~M., {Stiening}, R., {et~al.} 2006, \aj, 131, 1163

\bibitem[{{Smartt} {et~al.}(2009){Smartt}, {Eldridge}, {Crockett}, \& {Maund}}]{Smartt2009}
{Smartt}, S.~J., {Eldridge}, J.~J., {Crockett}, R.~M., \& {Maund}, J.~R. 2009, \mnras, 395, 1409

\bibitem[{{Smith} {et~al.}(2020){Smith}, {Smartt}, {Young}, {Tonry}, {Denneau}, {Flewelling}, {Heinze}, {Weiland}, {Stalder}, {Rest}, {Stubbs}, {Anderson}, {Chen}, {Clark}, {Do}, {F{\"o}rster}, {Fulton}, {Gillanders}, {McBrien}, {O'Neill}, {Srivastav}, \& {Wright}}]{Smith2020}
{Smith}, K.~W., {Smartt}, S.~J., {Young}, D.~R., {et~al.} 2020, \pasp, 132, 085002

\bibitem[{{Smith} {et~al.}(2019){Smith}, {Williams}, {Young}, {Ibsen}, {Smartt}, {Lawrence}, {Morris}, {Voutsinas}, \& {Nicholl}}]{Smith2019}
{Smith}, K.~W., {Williams}, R.~D., {Young}, D.~R., {et~al.} 2019, Research Notes of the American Astronomical Society, 3, 26

\bibitem[{{Sorce} {et~al.}(2014){Sorce}, {Tully}, {Courtois}, {Jarrett}, {Neill}, \& {Shaya}}]{Sorce2014}
{Sorce}, J.~G., {Tully}, R.~B., {Courtois}, H.~M., {et~al.} 2014, \mnras, 444, 527

\bibitem[{{Speagle}(2020)}]{Speagle2020}
{Speagle}, J.~S. 2020, \mnras, 493, 3132

\bibitem[{{Tartaglia} {et~al.}(2018){Tartaglia}, {Sand}, {Valenti}, {Wyatt}, {Anderson}, {Arcavi}, {Ashall}, {Botticella}, {Cartier}, {Chen}, {Cikota}, {Coulter}, {Della Valle}, {Foley}, {Gal-Yam}, {Galbany}, {Gall}, {Haislip}, {Harmanen}, {Hosseinzadeh}, {Howell}, {Hsiao}, {Inserra}, {Jha}, {Kankare}, {Kilpatrick}, {Kouprianov}, {Kuncarayakti}, {Maccarone}, {Maguire}, {Mattila}, {Mazzali}, {McCully}, {Melandri}, {Morrell}, {Phillips}, {Pignata}, {Piro}, {Prentice}, {Reichart}, {Rojas-Bravo}, {Smartt}, {Smith}, {Sollerman}, {Stritzinger}, {Sullivan}, {Taddia}, \& {Young}}]{Tartaglia2018}
{Tartaglia}, L., {Sand}, D.~J., {Valenti}, S., {et~al.} 2018, \apj, 853, 62

\bibitem[{{Teerikorpi}(1984)}]{Teerikorpi1984}
{Teerikorpi}, P. 1984, \aap, 141, 407

\bibitem[{{Tonry} {et~al.}(2018){Tonry}, {Denneau}, {Flewelling}, {Heinze}, {Onken}, {Smartt}, {Stalder}, {Weiland}, \& {Wolf}}]{Tonry2018}
{Tonry}, J.~L., {Denneau}, L., {Flewelling}, H., {et~al.} 2018, \apj, 867, 105

\bibitem[{{Tsvetkov} {et~al.}(2019){Tsvetkov}, {Baklanov}, {Potashov}, {Oknyansky}, {Mikailov}, {Huseynov}, {Alekberov}, {Khalilov}, {Pavlyuk}, {Metlov}, {Volkov}, \& {Shugarov}}]{Tsvetkov2019}
{Tsvetkov}, D.~Y., {Baklanov}, P.~V., {Potashov}, M.~S., {et~al.} 2019, \mnras, 487, 3001

\bibitem[{{Tsvetkov} {et~al.}(2018){Tsvetkov}, {Shugarov}, {Volkov}, {Pavlyuk}, {Vozyakova}, {Shatsky}, {Nikiforova}, {Troitsky}, {Troitskaya}, \& {Baklanov}}]{Tsvetkov2018}
{Tsvetkov}, D.~Y., {Shugarov}, S.~Y., {Volkov}, I.~M., {et~al.} 2018, Astronomy Letters, 44, 315

\bibitem[{{Tully} {et~al.}(2013){Tully}, {Courtois}, {Dolphin}, {Fisher}, {H{\'e}raudeau}, {Jacobs}, {Karachentsev}, {Makarov}, {Makarova}, {Mitronova}, {Rizzi}, {Shaya}, {Sorce}, \& {Wu}}]{Tully2013}
{Tully}, R.~B., {Courtois}, H.~M., {Dolphin}, A.~E., {et~al.} 2013, \aj, 146, 86

\bibitem[{{Tully} {et~al.}(2016){Tully}, {Courtois}, \& {Sorce}}]{Tully2016}
{Tully}, R.~B., {Courtois}, H.~M., \& {Sorce}, J.~G. 2016, \aj, 152, 50

\bibitem[{{Tully} \& {Fisher}(1988)}]{Tully1988}
{Tully}, R.~B. \& {Fisher}, J.~R. 1988, {Catalog of Nearby Galaxies}

\bibitem[{{Turatto} {et~al.}(2003){Turatto}, {Benetti}, \& {Cappellaro}}]{Turatto2003}
{Turatto}, M., {Benetti}, S., \& {Cappellaro}, E. 2003, in From Twilight to Highlight: The Physics of Supernovae, ed. W.~{Hillebrandt} \& B.~{Leibundgut}, 200

\bibitem[{{Tutukov} \& {Yungelson}(1981)}]{Tutukov1981}
{Tutukov}, A.~V. \& {Yungelson}, L.~R. 1981, Nauchnye Informatsii, 49, 3

\bibitem[{{Valerin} {et~al.}(2022){Valerin}, {Pumo}, {Pastorello}, {Reguitti}, {Elias-Rosa}, {G{\'u}tierrez}, {Kankare}, {Fraser}, {Mazzali}, {Howell}, {Kotak}, {Galbany}, {Williams}, {Cai}, {Salmaso}, {Pinter}, {M{\"u}ller-Bravo}, {Burke}, {Padilla Gonzalez}, {Hiramatsu}, {McCully}, {Newsome}, \& {Pellegrino}}]{Valerin2022}
{Valerin}, G., {Pumo}, M.~L., {Pastorello}, A., {et~al.} 2022, \mnras, 513, 4983

\bibitem[{{Van Dyk} {et~al.}(2018){Van Dyk}, {Zheng}, {Brink}, {Filippenko}, {Milisavljevic}, {Andrews}, {Smith}, {Cignoni}, {Fox}, {Kelly}, {Adamo}, {Yunus}, {Zhang}, \& {Kumar}}]{Van2018}
{Van Dyk}, S.~D., {Zheng}, W., {Brink}, T.~G., {et~al.} 2018, \apj, 860, 90

\bibitem[{{Wang} {et~al.}(2019){Wang}, {Chen}, {Wang}, {Hu}, {Xi}, {Yang}, {Zhao}, \& {Li}}]{Wang2019}
{Wang}, X., {Chen}, J., {Wang}, L., {et~al.} 2019, \apj, 882, 120

\bibitem[{{Wang} {et~al.}(2009){Wang}, {Filippenko}, {Ganeshalingam}, {Li}, {Silverman}, {Wang}, {Chornock}, {Foley}, {Gates}, {Macomber}, {Serduke}, {Steele}, \& {Wong}}]{Wang2009}
{Wang}, X., {Filippenko}, A.~V., {Ganeshalingam}, M., {et~al.} 2009, \apjl, 699, L139

\bibitem[{{Wang} {et~al.}(2013){Wang}, {Wang}, {Filippenko}, {Zhang}, \& {Zhao}}]{Wang2013}
{Wang}, X., {Wang}, L., {Filippenko}, A.~V., {Zhang}, T., \& {Zhao}, X. 2013, Science, 340, 170

\bibitem[{{Webbink}(1984)}]{Webbink1984}
{Webbink}, R.~F. 1984, \apj, 277, 355

\bibitem[{{Whelan} \& {Iben}(1973)}]{Whelan1973}
{Whelan}, J. \& {Iben}, Icko, J. 1973, \apj, 186, 1007

\bibitem[{{Willick} {et~al.}(1997){Willick}, {Courteau}, {Faber}, {Burstein}, {Dekel}, \& {Strauss}}]{Willick1997}
{Willick}, J.~A., {Courteau}, S., {Faber}, S.~M., {et~al.} 1997, \apjs, 109, 333

\bibitem[{{Wright} {et~al.}(2010){Wright}, {Eisenhardt}, {Mainzer}, {Ressler}, {Cutri}, {Jarrett}, {Kirkpatrick}, {Padgett}, {McMillan}, {Skrutskie}, {Stanford}, {Cohen}, {Walker}, {Mather}, {Leisawitz}, {Gautier}, {McLean}, {Benford}, {Lonsdale}, {Blain}, {Mendez}, {Irace}, {Duval}, {Liu}, {Royer}, {Heinrichsen}, {Howard}, {Shannon}, {Kendall}, {Walsh}, {Larsen}, {Cardon}, {Schick}, {Schwalm}, {Abid}, {Fabinsky}, {Naes}, \& {Tsai}}]{Wright2010}
{Wright}, E.~L., {Eisenhardt}, P. R.~M., {Mainzer}, A.~K., {et~al.} 2010, \aj, 140, 1868

\bibitem[{{Xi} {et~al.}(2024){Xi}, {Wang}, {Li}, {Liu}, {Yan}, {Lin}, {Zhao}, {Filippenko}, {Zheng}, {Brink}, {Yang}, {Ehgamberdiev}, {Mirzaqulov}, {Reguitti}, {Pastorello}, {Tomasella}, {Cai}, {Zhang}, {Li}, {Zhang}, {Sai}, {Chen}, {Liu}, {Ma}, \& {Xiang}}]{Xi2024}
{Xi}, G., {Wang}, X., {Li}, G., {et~al.} 2024, \mnras, 527, 9957

\bibitem[{{Xiang} {et~al.}(2019){Xiang}, {Wang}, {Mo}, {Wang}, {Smartt}, {Fraser}, {Ehgamberdiev}, {Mirzaqulov}, {Zhang}, {Zhang}, {Vinko}, {Wheeler}, {Hosseinzadeh}, {Howell}, {McCully}, {DerKacy}, {Baron}, {Brown}, {Zhang}, {Bi}, {Song}, {Zhang}, {Rest}, {Nomoto}, {Tolstov}, \& {Blinnikov}}]{Xiang2019}
{Xiang}, D., {Wang}, X., {Mo}, J., {et~al.} 2019, \apj, 871, 176

\bibitem[{{Xiang} {et~al.}(2023){Xiang}, {Wang}, {Zhang}, {Sai}, {Zhang}, {Brink}, {Filippenko}, {Mo}, {Zhang}, {Chen}, {Dessart}, {Li}, {Yan}, {Blinnikov}, {Rui}, {Baron}, \& {DerKacy}}]{Xiang2023}
{Xiang}, D., {Wang}, X., {Zhang}, X., {et~al.} 2023, \mnras, 520, 2965

\bibitem[{{Yang} {et~al.}(2021){Yang}, {Sollerman}, {Strotjohann}, {Schulze}, {Lunnan}, {Kool}, {Fremling}, {Perley}, {Ofek}, {Schweyer}, {Bellm}, {Kasliwal}, {Masci}, {Rigault}, \& {Yang}}]{Yang2021}
{Yang}, S., {Sollerman}, J., {Strotjohann}, N.~L., {et~al.} 2021, \aap, 655, A90

\bibitem[{{Yoon}(2015)}]{Yoon2015}
{Yoon}, S.-C. 2015, \pasa, 32, e015

\bibitem[{{Zhang} {et~al.}(2020){Zhang}, {Wang}, {J{\'o}zsef}, {Zhai}, {Zhang}, {Filippenko}, {Brink}, {Zheng}, {Wyrzykowski}, {Miko{\l}ajczyk}, {Huang}, {Rui}, {Mo}, {Sai}, {Zhang}, {Wang}, {DerKacy}, {Baron}, {S{\'a}rneczky}, {B{\'o}di}, {Cs{\"o}rnyei}, {Hanyecz}, {Ign{\'a}cz}, {Kalup}, {Kriskovics}, {K{\"o}nyves-T{\'o}th}, {Ordasi}, {P{\'a}l}, {S{\'o}dor}, {Szak{\'a}ts}, {Vida}, \& {Zsidi}}]{Zhang2020}
{Zhang}, J., {Wang}, X., {J{\'o}zsef}, V., {et~al.} 2020, \mnras, 498, 84

\bibitem[{{Zheng} {et~al.}(2017){Zheng}, {Filippenko}, {Mauerhan}, {Graham}, {Yuk}, {Hosseinzadeh}, {Silverman}, {Rui}, {Arbour}, {Foley}, {Abolfathi}, {Abramson}, {Arcavi}, {Barth}, {Bennert}, {Brandel}, {Cooper}, {Cosens}, {Fillingham}, {Fulton}, {Halevi}, {Howell}, {Hsyu}, {Kelly}, {Kumar}, {Li}, {Li}, {Malkan}, {Manzano-King}, {McCully}, {Nugent}, {Pan}, {Pei}, {Scott}, {Sexton}, {Shivvers}, {Stahl}, {Treu}, {Valenti}, {Vogler}, {Walsh}, \& {Wang}}]{Zheng2017}
{Zheng}, W., {Filippenko}, A.~V., {Mauerhan}, J., {et~al.} 2017, \apj, 841, 64

\end{thebibliography}

\begin{appendix}

\begin{landscape}

\section{The SNe sample}\label{SNesample}

		\begin{table}[!h]
			\centering
		    \caption{The SNe Ia sample\tablefootmark{a}}\label{Ia_info}
		    \begin{threeparttable}
			
			    \resizebox{\columnwidth}{!}{
				    \begin{tabular}{cccccccccccccccc}
				    	\toprule
					    SN & Type\tablefootmark{b} & Discoverer & Disc. date\tablefootmark{c} & Spec. ID\tablefootmark{d} & $\mathrm{M_r}$\tablefootmark{e} & Galactic extinction(r)/mag & E(B-V)/mag\tablefootmark{f} & ra & dec & Redshift & Distance/Mpc & Method\tablefootmark{g} & Reference\tablefootmark{h} & Host & Offset\tablefootmark{i}\\
      					\midrule
	    				SN 2016coj & IaH & LOSS & 2016/05/28 &  & -19.0$\pm$0.2 & 0.043 & $\lesssim$ 0.020 & 12:08:06.800 & +65:10:38.24 & 0.004483 & 24.90$\pm$1.75 & Tully est & (7) & NGC 4125 & 12.38 \\
		    			SN 2016eiy & IaH & ASAS-SN & 2016/07/26 & 25095 & -18.5$\pm$0.1 & 0.289 && 13:34:38.640 & -23:40:53.04 & 0.002 & 35.00$\pm$2.45 & Virgo Infall & (4) & ESO 509-IG 064 & 9.56 \\
			    		SN 2016gwl & IaN & GaiaAlerts & 2016/10/02 & 15794 &  & 0.192 && 09:23:28.008 & -23:10:10.92 & 0.009 & 39.55$\pm$2.77 & Tully est & (7) & NGC 2865 & 42.32 \\
				    	SN 2016hsc & IaN & GaiaAlerts & 2016/10/01 & 25009 & -17.2$\pm$0.1 & 0.218 && 06:32:25.152 & -71:34:04.08 & 0.006 & 28.70$\pm$2.00 & Virgo Infall & (4) & PGC 179577 & 3.09 \\
					    SN 2017bzc & IaN & BOSS & 2017/03/07 &  & -19.4$\pm$0.1 & 0.032 && 23:16:14.690 & -42:34:10.90 & 0.005365 & 23.43$\pm$1.64 & Tully-Fisher & (7) & NGC 7552 & 69.62 \\
					    SN 2017cbv & IaN & DLT40 & 2017/03/10 & 75276 & -19.1$\pm$0.1 & 0.386 && 14:32:34.420 & -44:08:02.74 & 0.003999 & 15.02$\pm$1.05 & Tully-Fisher & (7) & NGC 5643 & 161.11 \\
					    \bottomrule
				    \end{tabular}
			    }
			
			\end{threeparttable}

            \tablefoot{
            \tablefoottext{a}{Only the first 6 entries are shown, a full table is available online.}
            \tablefoottext{b}{For normal Ia, we further divide them into high velocity (IaH) and normal velocity (IaN) according to the classification scheme of \citet{Wang2009}.}
            \tablefoottext{c}{The date of discovery.}
            \tablefoottext{d}{Spectrum ID in WISeREP of the spectrum used.}
            \tablefoottext{e}{The r-band peak magnitude after extinction correction.}
            \tablefoottext{f}{Host galaxy extinction, for SNe given in Table~\ref{reference}, the adopted values are taken from the corresponding references in Table~\ref{reference}, for the rest SNe, the extinction values are estimated from Na I D absorption lines if high resolution spectra are available.}
            \tablefoottext{g}{Methods for the distance measurements.}
            \tablefoottext{h}{References for the distance measurements.}
            \tablefoottext{i}{The offset angle from host galaxy nucleus (in arcsec).}
            }
            
            \tablebib{(1) \citet{Courtois2012}; (2) \citet{Karachentsev2006}; (3) \citet{Lagattuta2013}; (4) \citet{Mould2000}; (5) \citet{Nasonova2011}; (6) \citet{Sorce2014}; (7) \citet{Tully1988}; (8) \citet{Tully2013}; (9) \citet{Tully2016}; (10) \citet{Willick1997}.}
            
		\end{table}
		
		\begin{table}[!h]
			\centering
			\caption{The SNe Ibc sample}\label{Ibc_info}
			\begin{threeparttable}
				
				\resizebox{\columnwidth}{!}{
					\begin{tabular}{cccccccccccccccc}
						\toprule
						SN & Type & Discoverer & Disc. date & Spec. ID & $\mathrm{M_r}$ & Galactic extinction(r)/mag & E(B-V)/mag & ra & dec & Redshift & Distance/Mpc & Method & Reference & Host & Offset\\
						\midrule
						SN2016bau & Ib & Amateur & 2016/03/13 &  & -17.5$\pm$0.3 & 0.038 & 0.566$\pm$0.046 & 11:20:59.020 & +53:10:25.60 & 0.003856 & 23.16$\pm$1.62 & Tully-Fisher & (7) & NGC 3631 & 37.94 \\
						SN2016cdd & Ib & ASAS-SN & 2016/05/14 &  &  & 0.421 &  & 12:31:53.636 & -51:44:50.83 & 0.008731 & 29.92$\pm$2.10 & Tully est & (7) & ESO 218-8 & 1.15 \\
						SN2016coi & Ic-BL & ASAS-SN & 2016/05/27 &  & -18.2$\pm$0.2 & 0.194 &  & 21:59:04.100 & +18:11:10.70 & 0.0036 & 18.83$\pm$1.32 & Tully est & (7) & UGC 11868 & 32.89 \\
						SN2016dsl & Ibc & GaiaAlerts & 2016/02/24 & 27740 & & 0.186 && 13:17:13.200 & -17:15:06.50 & 0.008 & 36.38$\pm$2.55 & Tully-Fisher & (3) & IC 863 & 14.89 \\
						SN2016G & Ic-BL & Amateur & 2016/01/09 &  & -17.7$\pm$0.2 & 0.359 && 03:03:57.700 & +43:24:03.60 & 0.009 & 39.25$\pm$2.75 & Tully-Fisher & (5) & NGC 1171 & 16.52 \\
						SN2016iae & Ic & ATLAS & 2016/11/07 &  & -17.1$\pm$0.2 & 0.035 && 04:12:05.530 & -32:51:44.53 & 0.004 & 13.62$\pm$0.96 & Tully-Fisher & (7) & NGC 1532 & 44.91 \\
						\bottomrule
					\end{tabular}
				}    
				
			\end{threeparttable}

            \tablefoot{The meaning of each column is the same as Table~\ref{Ia_info}}
            
		\end{table}
		
		\begin{table}[!h]
			\centering
			\caption{The SNe II sample\tablefootmark{a}}\label{II_info}
			\begin{threeparttable}
				
				\resizebox{\columnwidth}{!}{
					\begin{tabular}{cccccccccccccccccc}
						\toprule
						SN & Type & Discoverer & Disc. date & Spec. ID & $\mathrm{M_r}$\tablefootmark{b} & Galactic extinction(r)/mag & E(B-V)/mag & ra & dec & Redshift & Distance/Mpc & Method & Reference & Host & Offset\\
						\midrule
						SN2016aai & IIP & Pan-STARRS & 2016/02/04 &  & -16.4$\pm$0.3 & 0.103 & & 12:50:28.032 & -10:50:29.02 & 0.007972 & 36.74$\pm$2.57 & Tully-Fisher & (9) & MCG -02-33-020 & 51.10 \\
						SN2016adj & IIb & BOSS & 2016/02/08 &  & -15.5$\pm$0.1 & 0.263 & 0.600$\pm$0.060 & 13:25:24.110 & -43:00:57.50 & 0.001825 & 4.46$\pm$0.32 & Tully est & (7) & NGC 5128 & 40.07 \\
						SN2016adl & IIP & ASAS-SN & 2016/02/08 & 20764 & -15.9$\pm$0.2 & 0.098 & & 11:51:56.200 & -13:25:03.10 & 0.00652 & 31.50$\pm$2.2 & Virgo Infall & (4) & GALEXASC J115155.68-132459.3 & 8.49 \\
						SN2016afa & II-87A & XOSS & 2016/02/12 & 22102 & -16.9$\pm$0.2 & 0.124 && 15:36:32.470 & +16:36:36.70 & 0.007 & 35.71$\pm$2.50 & Tully-Fisher & (9) & NGC 5962 & 14.33 \\
						SN2016aqf & IIP & ASAS-SN & 2016/02/26 & 57354 & -16.2$\pm$1.2 & 0.126 & $\lesssim$ 0.031 & 05:46:23.907 & -52:05:18.86 & 0.004 & 15.50$\pm$1.09 & Tully est & (7) & NGC 2101 & 7.74 \\
						SN2016B & IIP & ASAS-SN & 2016/01/03 &  & -17.0$\pm$0.8 & 0.048 && 11:55:04.246 & +01:43:06.78 & 0.004293 & 26.29$\pm$1.84 & Virgo Infall & (4) & PGC 037392 & 11.21 \\
						\bottomrule
					\end{tabular}
				}    
			
			\end{threeparttable}

            \tablefoot{
            \tablefoottext{a}{The meaning of each column is the same as Table~\ref{Ia_info}.}
            \tablefoottext{b}{For type IIP, we use their plateau phase maximum instead.}
            }
            
		\end{table}
		
\end{landscape}

    \onecolumn
    \section{The host galaxy sample}\label{Hostsample}

		\begin{table*}[!h]
			\centering
			\caption{The galaxy sample for SNe Ia\tablefootmark{a}}\label{Ia_host}
			\begin{threeparttable}
				
				\resizebox{\textwidth}{!}{
					\begin{tabular}{cccccccccc}
						\toprule
						SN & Host & ra\tablefootmark{b} & dec\tablefootmark{c} & $log(\mathrm{d_{25}/0.1 arcmin})$\tablefootmark{d} & $\mathrm{d_{SN}/d_{25}}$\tablefootmark{e} & Type\tablefootmark{f} & Note & t\tablefootmark{g} & M$_{\ast}$\tablefootmark{h} \\
						\midrule
						SN2016coj & NGC 4125 & 12:08:06.017 & +65:10:26.88 & $1.77\pm0.04$ & 0.069 & E &  & $-4.8\pm0.6$ & $2.600\pm0.300$ \\
						SN2016eiy & ESO 509-IG 064 & 13:34:39.3 & -23:40:50 & $1.16\pm0.06$ & 0.206 & Sab & Multiple\tablefootmark{i} & $1.5\pm3.3$ & $0.030\pm0.020$ \\
						SN2016gwl & NGC 2865 & 09:23:30.205 & -23:09:41.37 & $1.39\pm0.03$ & 0.512 & E & Bar & $-4.2\pm1.3$ & $0.559\pm0.083$ \\
						SN2016hsc & PGC 179577 & 06:32:24.5 & -71:34:04 & & & & & &$0.004\pm0.004$ \\
						SN2017bzc & NGC 7552 & 23:16:10.767 & -42:35:05.39 & $1.59\pm0.02$ & 0.596 & Sab & Bar & $2.4\pm0.7$ & $2.000\pm1.000$ \\
						SN2017cbv & NGC 5643 & 14:32:40.778 & -44:10:28.60 & $1.72\pm0.02$ & 0.871 & Sc & Bar, Ring & $5.0\pm0.3$ & $1.500\pm0.900$ \\
						\bottomrule
					\end{tabular}
				}
				
			\end{threeparttable}
            
            \tablefoot{
            \tablefoottext{a}{Only the first 6 entries are shown, full table is available online.}
            \tablefoottext{b}{Right ascension of the core of the galaxy.}
            \tablefoottext{c}{Declination of the core of the galaxy.}
            \tablefoottext{d}{The decimal logarithm of the length the projected major axis of the galaxy at the isophotal level 25 mag/arcsec$^2$ in B-band corrected for galactic extinction and inclination effect.}
            \tablefoottext{e}{The ratio of the offset angle and d$_{25}$.}
            \tablefoottext{f}{The morphological type of the galaxy.}
            \tablefoottext{g}{The morphological type code.}
            \tablefoottext{h}{Stellar mass in unit of $10^{10} \mathrm{M}_{\odot}$.}
            \tablefoottext{i}{Multiple galaxies.}
            }
		\end{table*}
		
		\begin{table}[!h]
			\centering
			\caption{The galaxy sample for SNe Ibc}\label{Ibc_host}
			\begin{threeparttable}
				
				\resizebox{\textwidth}{!}{
					\begin{tabular}{cccccccccc}
						\toprule
						SN & Host & ra & dec & $log(\mathrm{d_{25}/0.1 arcmin})$ & $\mathrm{d_{SN}/d_{25}}$ & Type & Note & t & M$_{\ast}$ \\
						\midrule
						SN2016bau & NGC 3631 & 11:21:02.887 & +53:10:10.415 & $1.57\pm0.03$ & 0.340 & Sc &  & $5.1\pm0.5$ & $0.042\pm0.007$ \\
						SN2016cdd & ESO 218-8 & 12:31:53.679 & -51:44:51.91 & $1.43\pm0.03$ & 0.013 & SBbc & Bar & $4.2\pm1.3$ & $0.050\pm0.030$ \\
						SN2016coi & UGC 11868 & 21:59:04.702 & +18:10:38.95 & $1.30\pm0.05$ & 0.513 & Sm & Bar, Ring & $9.0\pm0.5$ & $0.007\pm0.004$ \\
						SN2016dsl & IC 863 & 13:17:12.403 & -17:15:16.054 & $1.06\pm0.08$ & 0.413 & S0-a & Bar & $0.4\pm1.4$ & $0.500\pm0.200$ \\
						SN2016G & NGC 1171 & 03:03:58.959 & +43:23:54.404 & $1.25\pm0.06$ & 0.276 & Sc & & $5.7\pm0.7$ & $0.160\pm0.080$ \\
						SN2016iae & NGC 1532 & 04:12:04.361 & -32:52:26.960 & $2.05\pm0.02$ & 0.137 & SBb & Bar & $3.2\pm0.9$ & $1.600\pm0.700$ \\
						\bottomrule
					\end{tabular}
				}    
				
			\end{threeparttable}
            
			\tablefoot{The meaning of each column is the same as Table~\ref{Ia_host}}
            
		\end{table}
  
		\begin{table}[!h]
			\centering
			\caption{The galaxy sample for SNe II}\label{II_host}
			\begin{threeparttable}
				
				\resizebox{\textwidth}{!}{
					\begin{tabular}{cccccccccc}
						\toprule
						SN & Host & ra & dec & $log(\mathrm{d_{25}/0.1 arcmin})$ & $\mathrm{d_{SN}/d_{25}}$ & Type & Note & t & M$_{\ast}$ \\
						\midrule
						SN2016aai & MCG -02-33-020 & 12:50:29.414 & -10:51:15.89 & $1.52\pm0.03$ & 0.526 & Sm & Bar & $9.0\pm1.1$ & $0.100\pm0.060$ \\
						SN2016adj & NGC 5128 & 13:25:27.615 & -43:01:08.805 & $2.41\pm0.01$ & 0.046 & S0 & Multiple & $-2.1\pm0.6$ & $2.526\pm0.088$ \\
						SN2016adl & GALEXASC J115155.68-132459.3 & 11:51:55.68 & -13:24:59.3 & & & & & & \\
						SN2016afa & NGC 5962 & 15:36:31.681 & +16:36:27.933 & $1.40\pm0.03$ & 0.186 & Sc & Ring & $5.1\pm0.7$ & $0.900\pm0.400$ \\
						SN2016aqf & NGC 2101 & 05:46:23.1 & -52:05:21 & $1.09\pm0.06$ & 0.205 & IB & Bar & $9.9\pm0.5$ & $0.010\pm0.005$ \\
						SN2016B & PGC 037392 & 11:55:04.926 & +01:43:11.44 & $0.76\pm0.06$ & 0.650 & Scd & & $7.0\pm2.0$ & $0.010\pm0.005$ \\
						\bottomrule
					\end{tabular}
				}    
				
			\end{threeparttable}
            
			\tablefoot{The meaning of each column is the same as Table~\ref{Ia_host}}
            
		\end{table}

    \clearpage

\section{References for photometry data}

    \begin{table}[!h]
			\centering
			\caption{References for photometry data.}\label{reference}
			\begin{threeparttable}
				
				\resizebox{\textwidth}{!}{
					\begin{tabular}{cc}
						\toprule
						Reference & Supernovae  \\
						\midrule
                        (1) & SN2020jfo \\
                        (2) & SN2019esa \\
                        (3) & SN2019muj  \\
						\multirow{3}*{(4)} & SN2016eiy, SN2016hsc, SN2017bzc, SN2017cbv, SN2017dhr, SN2017drh, SN2017emq, SN2017erp \\ 
                        ~ & SN2017fgc, SN2017fzw, SN2017hjy, SN2017igf, SN2018aoz, SN2018ghb, SN2018gv \\
                        ~ & SN2018htt, SM2018imd, SN2018isq, SN2018pv, SN2018yu, SN2019np, SN2020ue\\
						(5) & SN2017ivv \\ 
						(6) & SN2016X \\
						(7) & SN2021fcg  \\
                        (8) & SN2020acat \\
						(9) & SN2016bkv \\
                        (10) & SN2017eaw \\
                        (11) & SN2018aoq \\
                        (12) & SN2021aai \\
					  (13)	& SN2017ein \\
                        (14) & SN2020cxd \\
                        (15) & SN2018zd \\
                        (16) & SN2016coj  \\
						\bottomrule
					\end{tabular}
				}

		\end{threeparttable}
        
        \tablebib{(1) \citet{Ailawadhi2023}; (2) \citet{Andrews2022}; (3) \citet{Barna2021}; (4) \citet{Chen2022}; (5) \citet{Gutierrez2020}; (6) \citet{Huang2018}; (7) \citet{Karambelkar2021}; (8) \citet{Medler2022}; (9) \citet{Nakaoka2018}; (10) \citet{Tsvetkov2018}; (11) \citet{Tsvetkov2019}; (12) \citet{Valerin2022}; (13) \citet{Xiang2019}; (14) \citet{Yang2021}; (15) \citet{Zhang2020}; (16) \citet{Zheng2017}.}
        
    \end{table}

\clearpage
\twocolumn
\section{The Miller diagrams}\label{Miller}

In this section, we present all Miller diagrams for each subtype of SNe after bias correction, except for 91T/02cx/87A-like events and Ia-SC for their small sample sizes. The black dots in each figure are the original data and the red dots represent SNe that were added to account for the Malmquist bias. 

    \begin{figure}[h!]
        \centering
		\includegraphics[width=0.9\columnwidth]{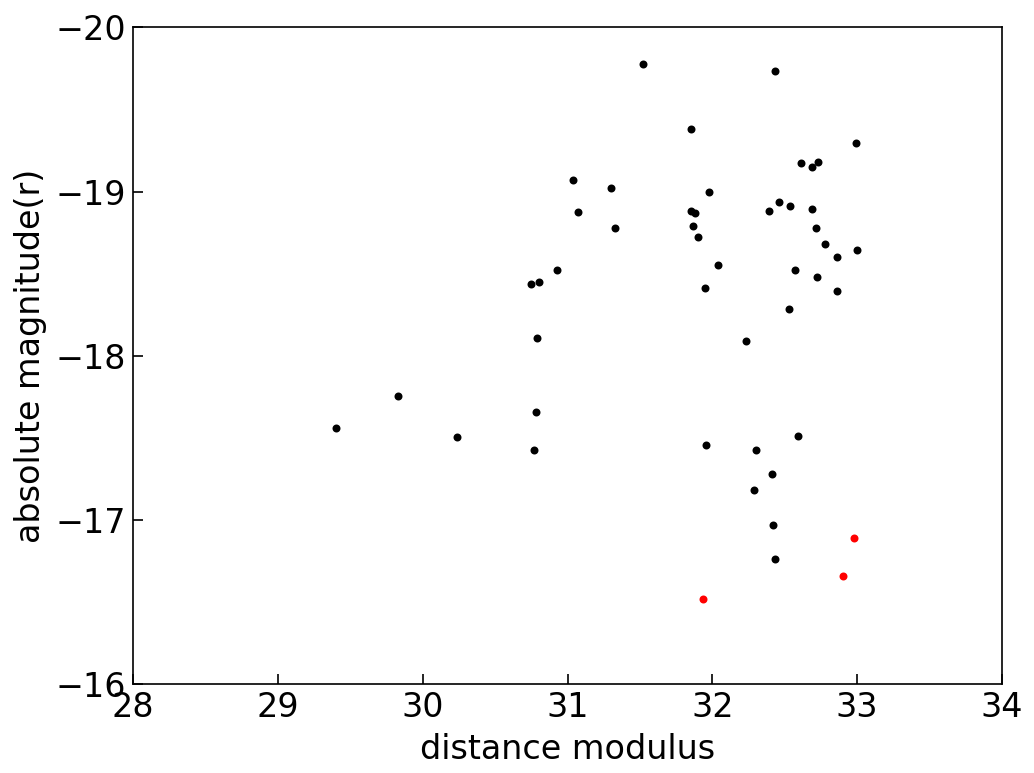}
		\caption{Miller diagrams for the normal Ia sample after the bias–correction process. The black dots are the original data while the red dots are SNe that were added to account for the Malmquist bias. The whole process stops at $\mu$ = 33.01 (40 Mpc).}\label{Ia}
    \end{figure}

    \begin{figure}[h!]
        \centering
		\includegraphics[width=0.9\columnwidth]{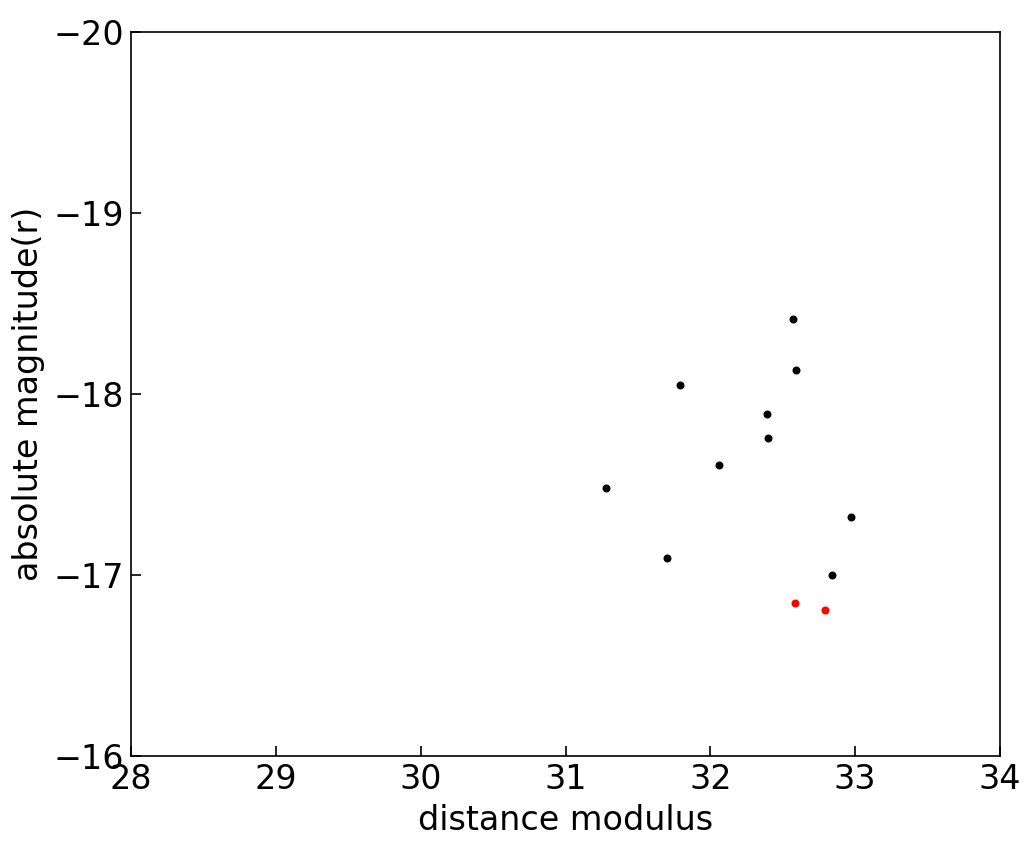}
		\caption{The same as Fig.~\ref{Ia} but for the 91bg-like sample.}
    \end{figure}

    \begin{figure}[h!]
        \centering
		\includegraphics[width=0.9\columnwidth]{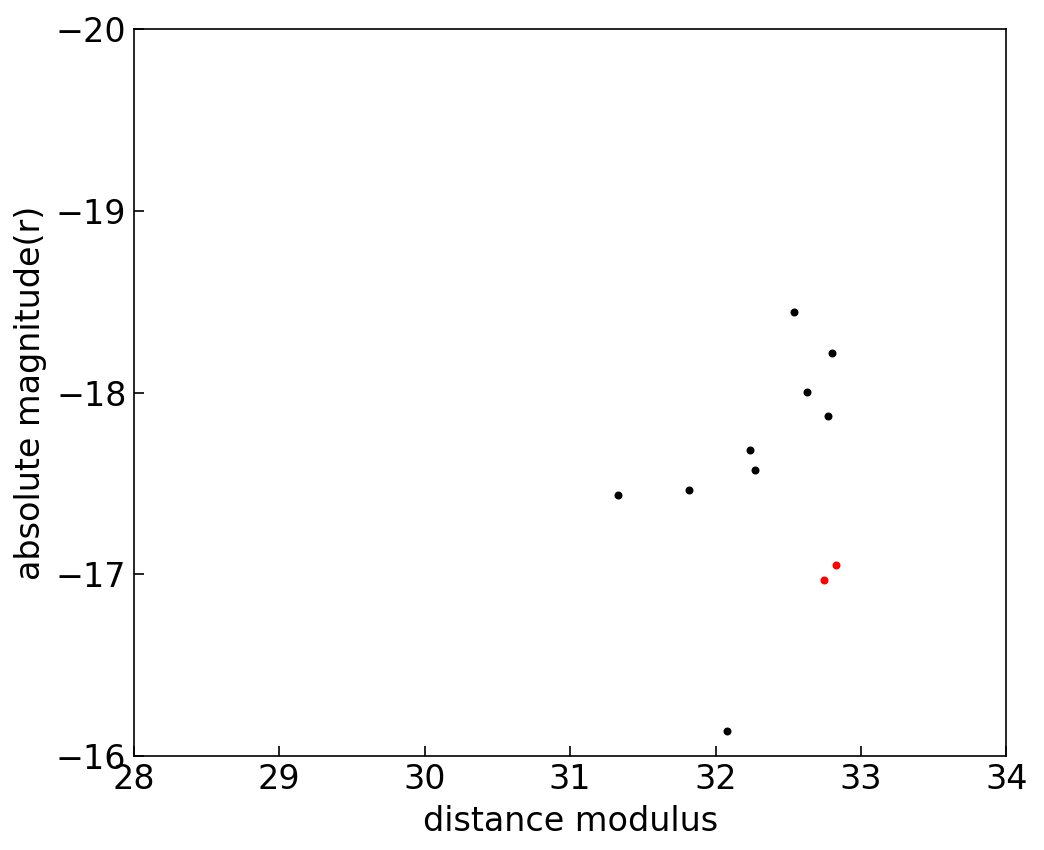}
		\caption{The same as Fig.~\ref{Ia} but for the SNe Ib sample.}
    \end{figure}

    \begin{figure}[h!]
        \centering
		\includegraphics[width=0.9\columnwidth]{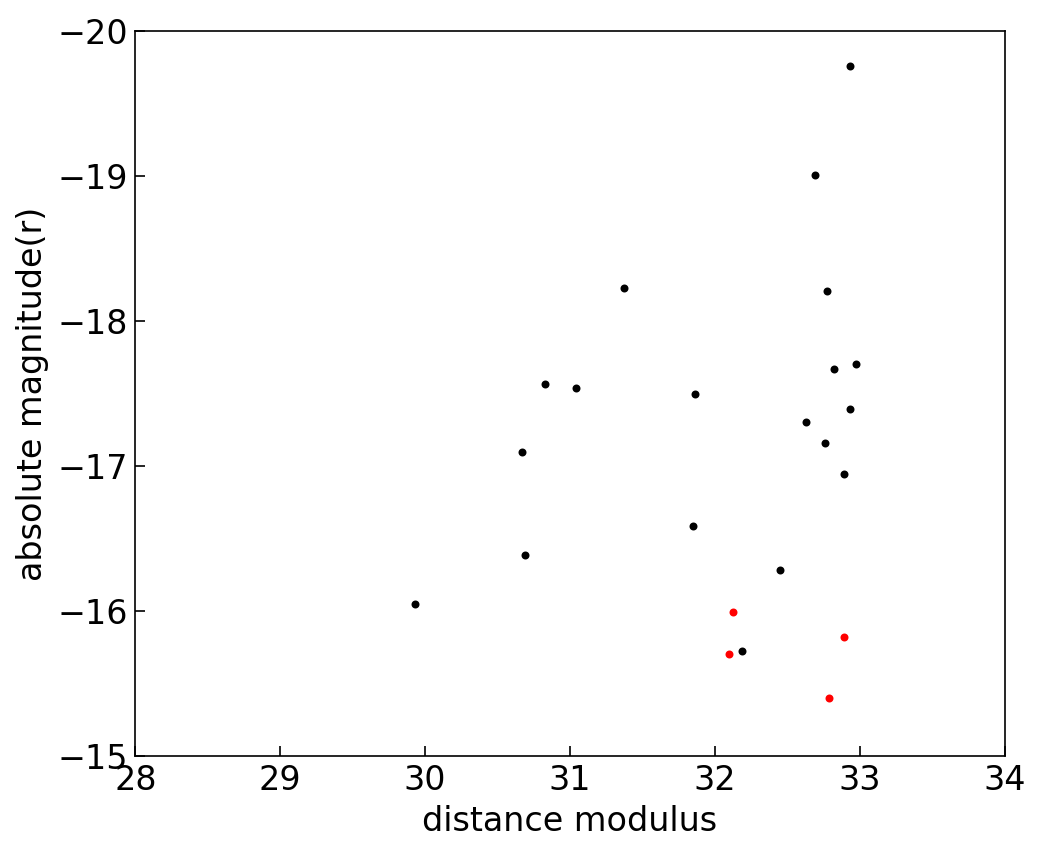}
		\caption{The same as Fig.~\ref{Ia} but for the SNe Ic sample.}
    \end{figure}

    \begin{figure}[h!]
        \centering
		\includegraphics[width=0.9\columnwidth]{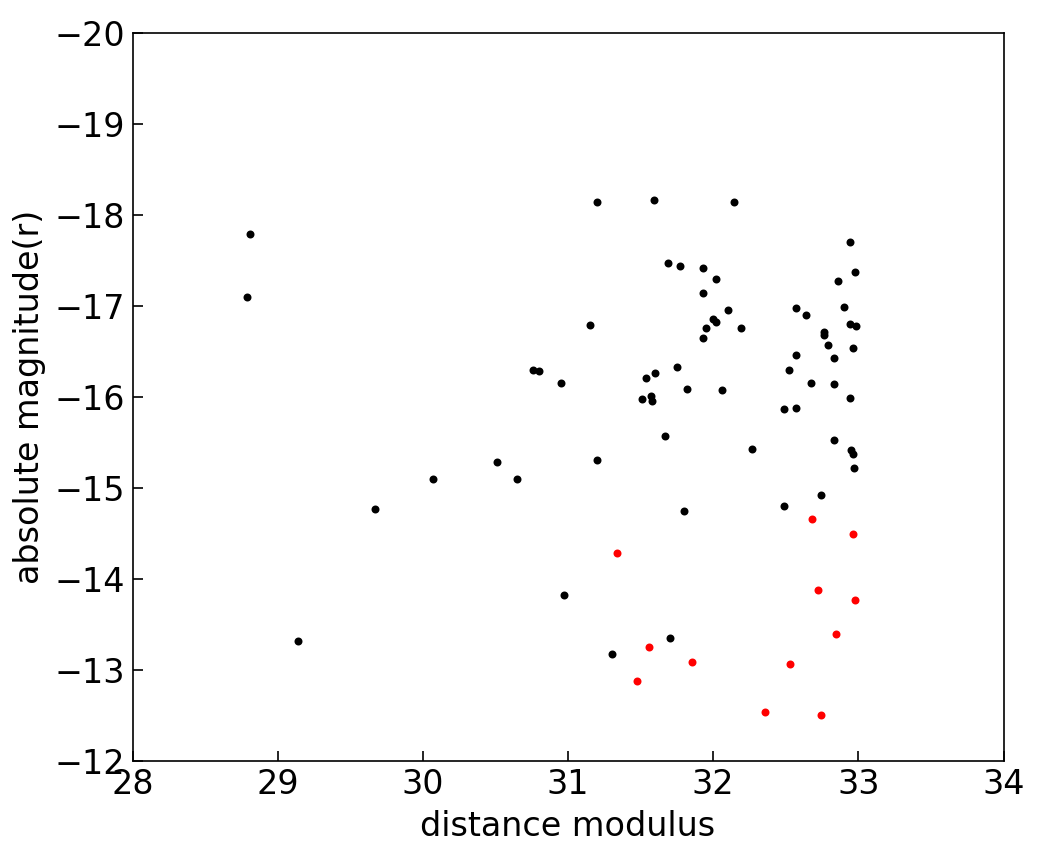}
		\caption{The same as Fig.~\ref{Ia} but for the SNe IIP sample.}
    \end{figure}

    \begin{figure}[h!]
        \centering
		\includegraphics[width=0.9\columnwidth]{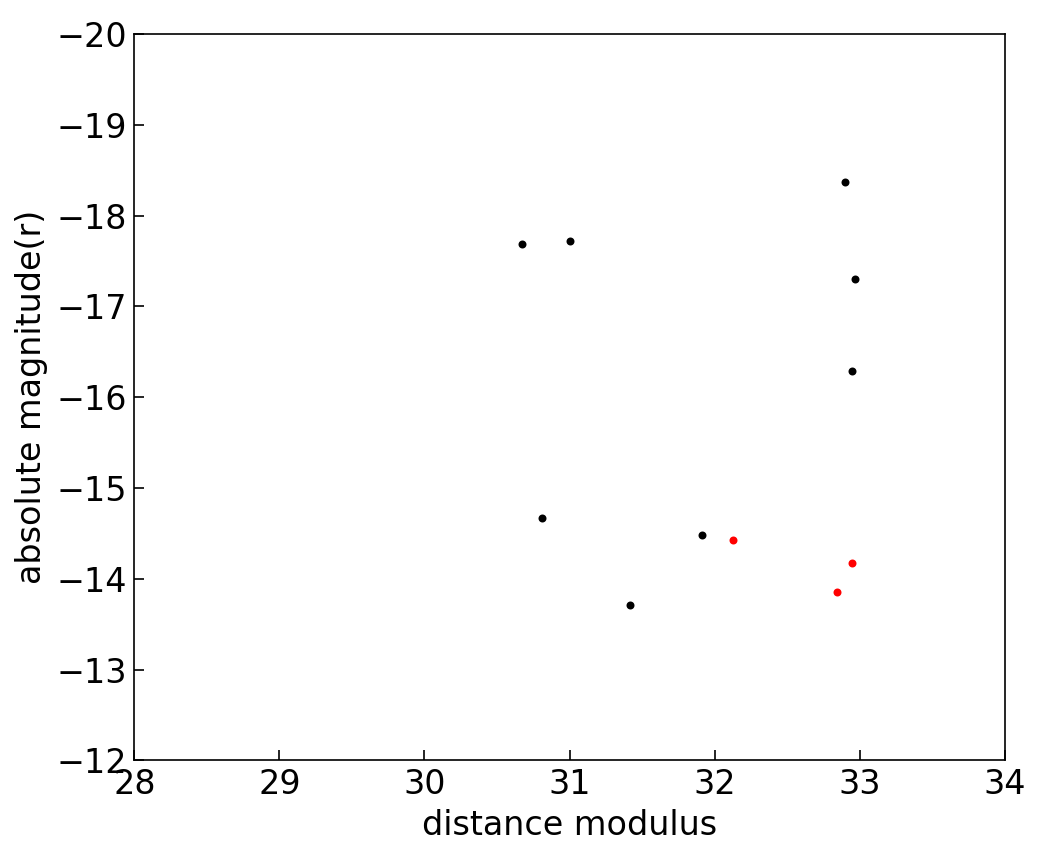}
		\caption{The same as Fig.~\ref{Ia} but for the SNe IIL sample.}
    \end{figure}

    \begin{figure}[h!]
        \centering
		\includegraphics[width=0.9\columnwidth]{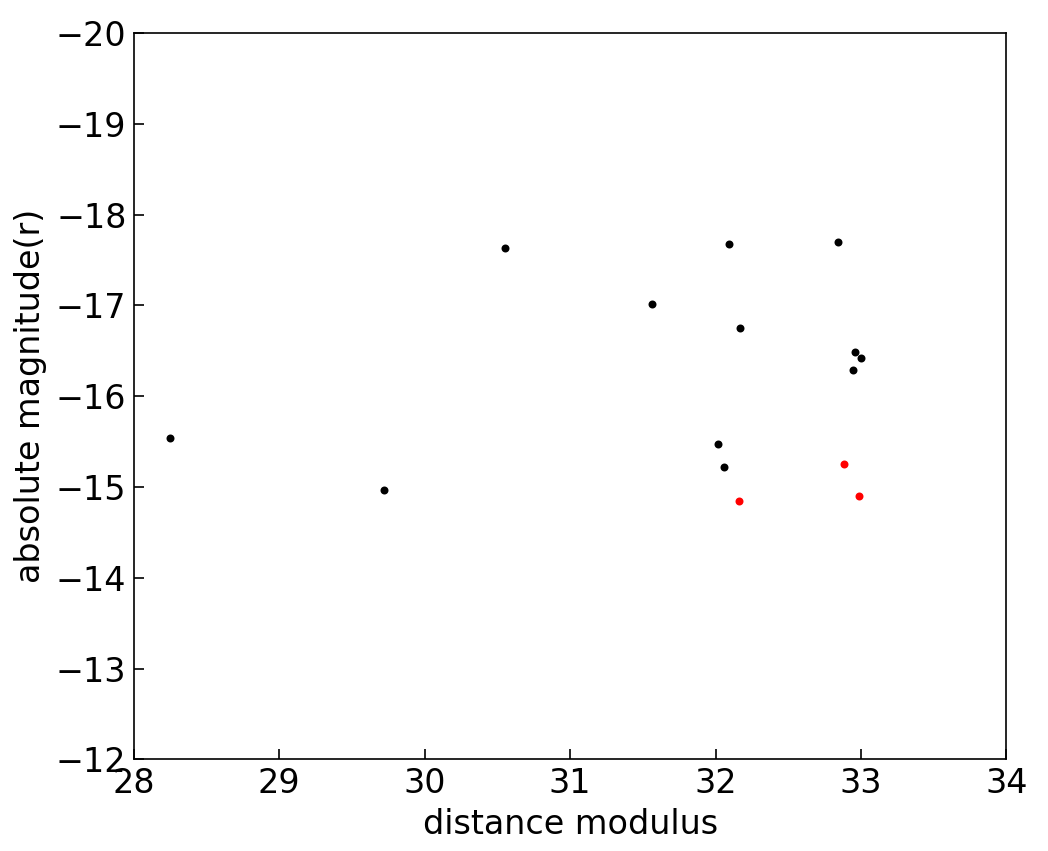}
		\caption{The same as Fig.~\ref{Ia} but for the SNe IIb sample.}
    \end{figure}

    \begin{figure}[h!]
        \centering
		\includegraphics[width=0.9\columnwidth]{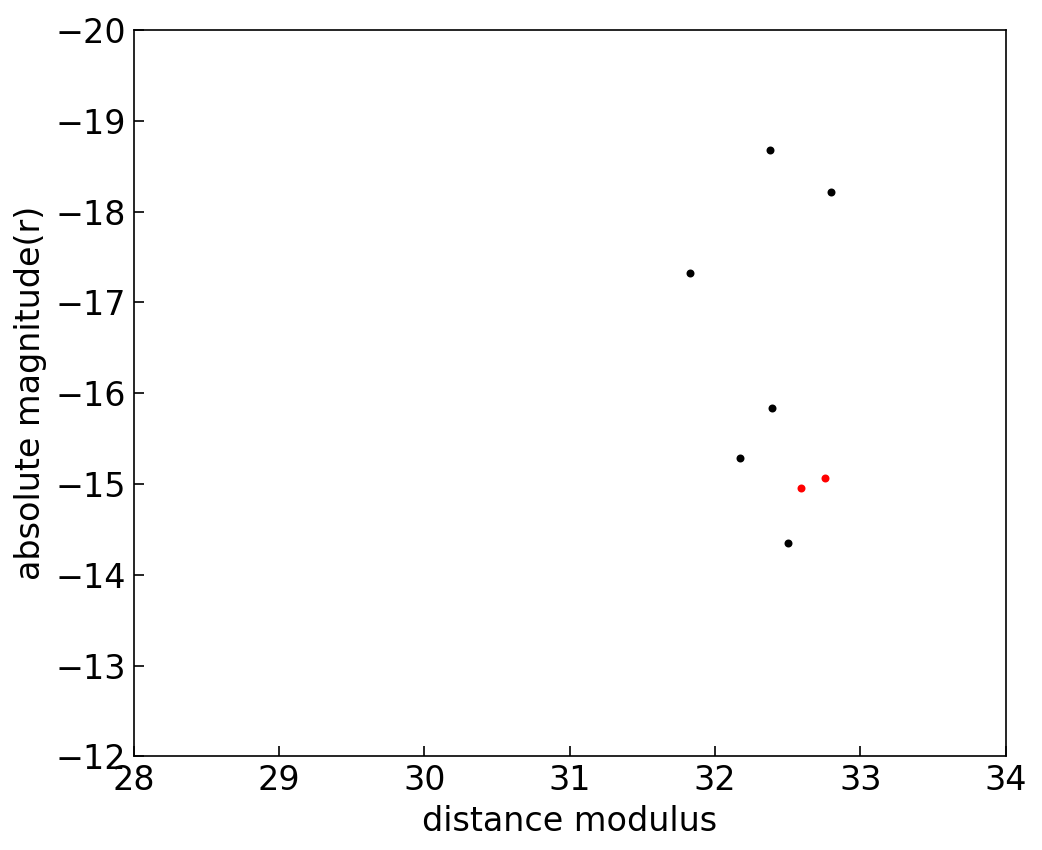}
		\caption{The same as Fig.~\ref{Ia} but for the SNe IIn sample.}
    \end{figure}

\FloatBarrier

\onecolumn

\section{Prospector fitting exapmle}

    \begin{figure}[h!]
        \centering
		\includegraphics[width=0.5\columnwidth]{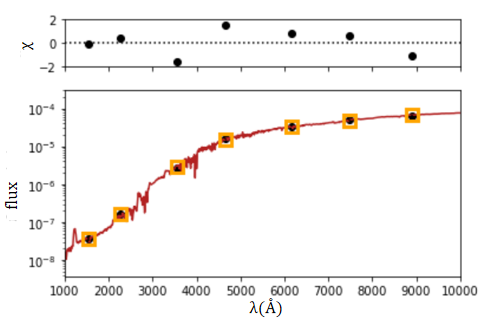}
		\caption{Prospector fitted SED for galaxy NGC 1209.}\label{SED}
    \end{figure}

    \begin{figure*}[h!]
        \centering
		\includegraphics[width=0.8\columnwidth]{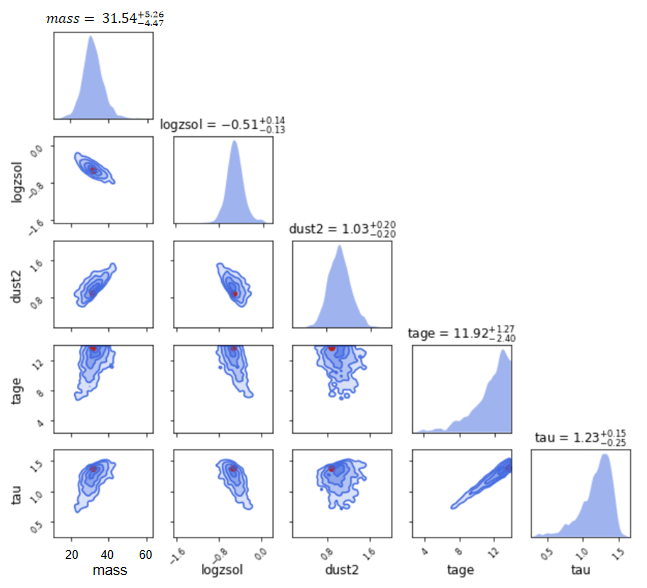}
		\caption{Prospector fitted parameters for the SED of NGC 1209. Five parameters fitted stand for total mass formed (mass, in unit of 10$^{10}$ M$_{\odot}$), stellar metallicity (logzsol = $log(Z_{\ast}/Z_{\odot})$), V-band optical depth (dust2), stellar age (tage, in unit of Gyr) and the SFH parameter ($\tau$, in unit of Gyr),respectively.}\label{prospector}
    \end{figure*}

\clearpage

\section{Spectra}    

    In this section we present part of the spectra used in our work, full data would be available on reasonable request. 
    
    \begin{figure*}[h!]
        \centering
		\includegraphics[width=0.9\textwidth]{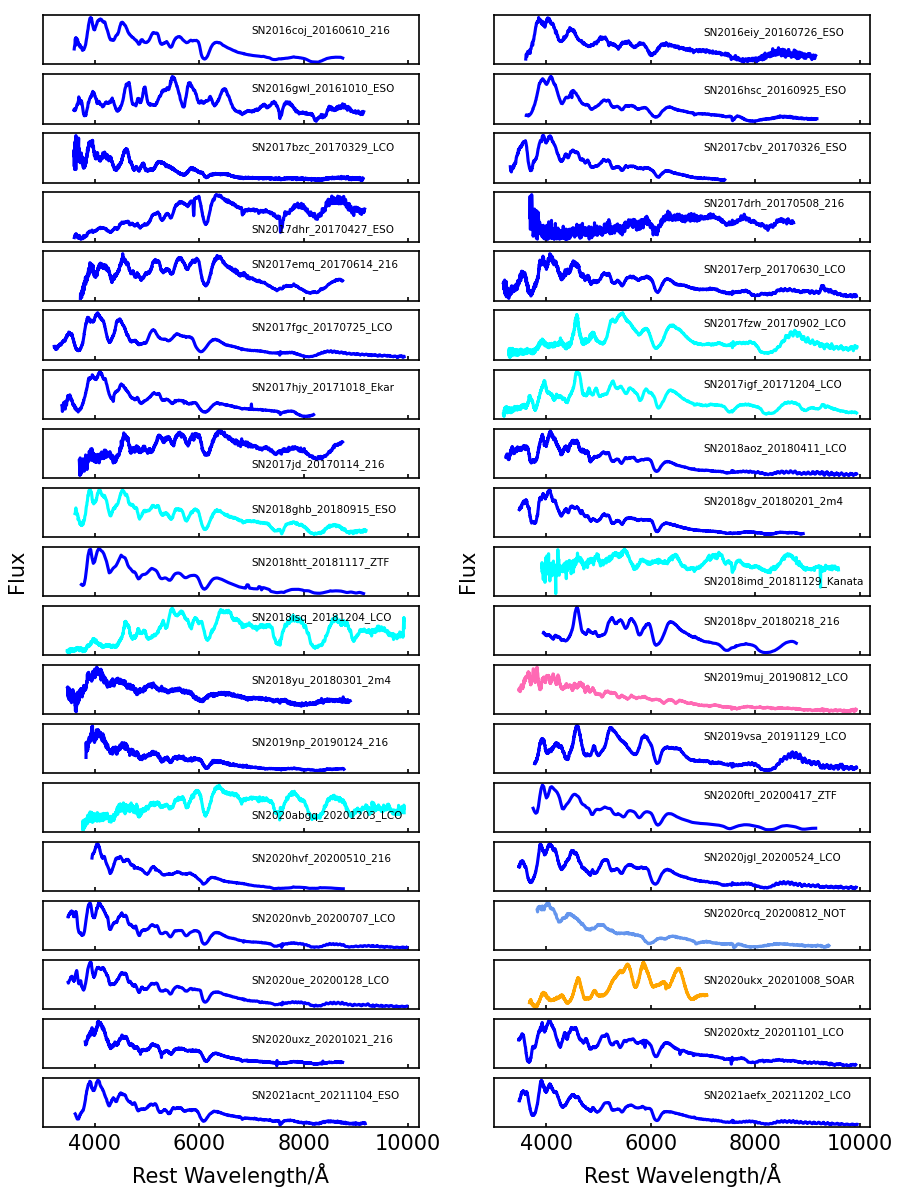}
		\caption{The spectra used for classification of our SNe Ia sample (part one), Normal Ia, 91bg-like, 91T-like and 02cx-like events are represented by the color blue, cyan, orange and pink, respectively.}\label{Iaspectra1}
    \end{figure*}

    \FloatBarrier

\end{appendix}

\end{document}